\documentclass{emulateapj}

\usepackage{apjfonts}
\usepackage{amsmath}
\usepackage{amssymb}
\usepackage{natbib}
\usepackage{xspace}
\usepackage{graphicx}
\usepackage{rotate} 
\usepackage{subfigure}
\usepackage{float}

\newcommand{\err}[2]{\ensuremath{^{+#1}_{-#2}}\xspace}

\newcommand{\NH}{$N_{\text{H}}$}
\newcommand{\NHgal}{$N_{\text{H, Gal}}$}
\newcommand{\xte}{\textsl{RXTE}\xspace}
\newcommand{\rxte}{\textsl{RXTE}\xspace}
\newcommand{\integral}{\textsl{INTEGRAL}\xspace}
\newcommand{\sax}{\textsl{BeppoSAX}\xspace}
\newcommand{\suzaku}{\textsl{Suzaku}\xspace}
\newcommand{\swiftbat}{\textsl{Swift}-BAT\xspace}
\newcommand{\xmm}{\textsl{XMM-Newton}\xspace}
\newcommand{\chandra}{\textsl{Chandra}\xspace}
\newcommand{\mr}{MR\,2251--178}

\newcommand{\pexrav}{\textsc{pexrav}\xspace}
\newcommand{\eroll}{$E_{\mathrm{roll}}$\xspace}
\newcommand{\etal}{et al.\xspace}

\shorttitle{Spectral Survey of X-Ray Bright AGN from \xte}
\shortauthors{Rivers et al.}

\begin{document}

\title{Spectral Survey of X-Ray Bright Active Galactic Nuclei from the \textsl{Rossi~X-ray~Timing~Explorer}} 

\author{Elizabeth~Rivers\altaffilmark{1}, Alex~Markowitz\altaffilmark{1}, and Richard~Rothschild\altaffilmark{1}}
\altaffiltext{1}{University of California, San Diego, Center for
  Astrophysics and Space Sciences, 9500 Gilman Dr., La Jolla, CA
  92093-0424, USA} 
\email{erivers@ucsd.edu}

%==================================================================%

\begin{abstract}

Using long-term monitoring data from the \textsl{Rossi X-ray Timing Explorer} (\xte), we have selected 23 active galactic nuclei (AGN)
with sufficient brightness and overall observation time to derive broadband X-ray spectra from 3 to $\gtrsim$100 keV.
Our sample includes mainly radio-quiet Seyferts, as well as seven radio-loud sources.
Given the longevity of the \xte mission, the greater part of our data is spread out over more than a decade, providing truly long-term average spectra
and eliminating inconsistencies arising from variability.
We present long-term average values of absorption, Fe line parameters, Compton reflection 
strengths and photon indices, as well as fluxes and luminosities for the hard and very hard energy bands, 2--10 keV and 20--100 keV respectively.
We find tentative evidence for high-energy rollovers in three of our objects.
We improve upon previous surveys of the very hard X-ray energy band in terms of accuracy and sensitivity, particularly 
with respect to confirming and quantifying the Compton reflection component.  
This survey is meant to provide a baseline for future analysis with respect to the long-term averages
for these sources and to cement the legacy of \xte, and especially its 
High Energy X-ray Timing Experiment, as a contributor to AGN spectral science.

\end{abstract}


%==================================================================%

\section{Introduction}

X-ray emission is seen universally in active galactic nuclei (AGN) and offers a wealth of information
about the geometrical and physical make-up of the accreting supermassive black holes at the hearts of distant galaxies. 
The primary X-ray continuum is thought to arise from the hot, Comptonizing electron or electron-positron pair corona close to the 
black hole (Haardt \etal 1994).  In many galaxies and especially in Seyferts the continuum is reprocessed by the accretion disk or other 
circumnuclear material creating commonly seen reflection features: the Fe K emission around 6--7 keV and the so-called 
Compton reflection hump peaking around 20--30 keV.
Nandra \& Pounds (1994) performed a spectral survey of 27 Seyferts observed with \textsl{Ginga}
in the 1.5--37 keV range and confirmed that Fe lines and Compton reflection humps were indeed
common.  Further work using the high resolution of the \textsl{Advanced Satellite for Cosmology and Astrophysics} (\textsl{ASCA}), 
\xmm and \chandra in the 2--10 keV range by Reynolds (1997), Yaqoob \& Padmanabhan (2004), 
Nandra (2006), and Nandra \etal (2007) has yielded detailed Fe K complex parameters.  
However each of these instruments lacks broadband coverage above about 10 keV, which 
complicates determination of the underlying continuum.

In the last decade the 20--100 keV sky has become more accessible, a development which is especially interesting for 
AGN because 20--100 keV AGN fluxes do not suffer from absorption by gas along the line of sight to the 
nucleus. \integral and \swiftbat surveys are turning up large numbers of obscured AGN (including some that are rather 
bright above 20 keV) which are difficult to study in lower energy bands (e.g. Winter \etal 2009, Bird \etal 2007).
These surveys yield detections and fluxes, and constrain AGN number densities and luminosity
functions.  However, for energy spectra in this band, the community has relied mainly on
\textsl{Rossi X-ray Timing Explorer} (\xte), \sax and \suzaku, and less recently, 
\textsl{High Energy Astronomy Observatory-1 (HEAO-1)} and the \textsl{Compton Gamma Ray Observatory's} 
Oriented Scintillation Spectrometer Experiment (\textit{CGRO}-OSSE).  
Of these, \xte, launched at the end of 1995, is the 
longest currently running X-ray mission and the public archive 
contains over 14 years' worth of data.

In this paper, we have taken advantage of the large quantity of AGN data 
and \xte's broad bandpass to investigate the form of AGN X-ray spectral data. 
We concentrate on quantifying the strengths of Compton reflection components
which will tighten constraints on the quantities of Compton-thick circumnuclear gas,
and searching for cutoffs or rollovers in the high-energy power-law continuum, which could
place constraints on thermal Comptonization processes in the corona (Haardt et al.\ 1994).
We have closely examined \xte archival data for 23 of the
X-ray brightest AGN and we were able to construct spectra summed over the last 14 years
covering from 3 to above 100 keV.
Our analysis takes advantage of the fact that the two pointed-observation instruments, 
the Proportional Counter Array (PCA) and the High Energy X-ray Timing Experiment (HEXTE),
are always operating simultaneously, removing ambiguity due to 
source variability inherent in splicing non-simultaneous 
~$<$10 and $>$10 keV data sets from different missions as is commonly done
to achieve broadband coverage.

\xte has been commonly used for continuous and sustained
long-term monitoring and many of the X-ray brightest AGN have been monitored
for at least 5--10 years. For many of the objects in
our sample, the spectra are time-averaged 
over timescales of years, and thus this paper serves 
as a reference for long-term average spectral properties in the hard and very hard X-ray bands.
For these sources, this complements the information obtained from
individual ``long-look'' spectra, as are commonly obtained with
many other X-ray observatories for which monitoring
is difficult (e.g., \xmm, \chandra).
This investigation also serves to 
support the legacy of \rxte by
maximizing the AGN science return from HEXTE. 

%--------------------------------------------------------------------------------------------------%
\small

\begin{deluxetable*}{lllllllll}
%   \tablewidth{0pc}
   \tablecaption{Target List \label{tabsample}}
   \tablecolumns{8}
\startdata
\hline
\hline
Source Name  &   Type &   $z$  &  Galactic \NH\tablenotemark{1} & PCA & HEXTE-A, ~B &  Lightcurve References  \\[1mm]
&&& ($10^{20}$ cm$^{-2}$) & Expo\tablenotemark{2} (ks) & Expo\tablenotemark{2} (ks, ~ks) &  &\\[1mm]
\hline
Type 1 Seyfert Galaxies \\[1mm]
\hline
NGC~4151   &   Sy1.5   &   0.003319   &   2.53    &   536   &   180, ~~~180   
& Markowitz et al.\ (2003b)\\[1mm]
      
IC~4329a   &   Sy1.2   &   0.016054   &   5.04    &   582   &   145, ~~~175    
& Markowitz (2009)\\[1mm]
 
NGC~3783   &   Sy1   &   0.009730   &   10.3    &   1297   &   204, ~~~365    
& Ar\'{e}valo et al.\ (2009), Summons et al.\ (2007)\\[1mm]
     
NGC~5548   &   Sy1.5   &   0.017175   &   1.52    &   927   &   294, ~~~312    
& Markowitz et al.\ (2003b), Uttley et al.\ (2003)\\[1mm]

Mkn~509   &   Sy1.2   &   0.034397   &   4.28    &   739   &   197, ~~~225  
& Marshall et al.\ (2008)         \\ [1mm]

MR~2251--178   &   Sy1   &   0.063980   &   2.54    &   380   &   ~58, ~~~~120  
& Ar\'{e}valo et al.\ (2008)\\[1mm]

NGC~3516   &   Sy1.5   &   0.008836   &   3.12    &   947   &   293, ~~~292   
& Markowitz et al.\ (2003b),  Maoz et al.\ (2002)   \\[1mm]
      
NGC~3227   &   Sy1.5   &   0.003859   &   1.84    &   1050   &   283, ~~~284  
& Uttley \& M$^{\textrm c}$Hardy (2005)   \\ [1mm]  
      
NGC~4593   &   Sy1   &   0.009000   &   1.55    &   960   &   168, ~~~282  
& Summons et al., in prep.,  Markowitz \& Reeves (2009)  \\ [1mm] 
   
NGC~7469   &   Sy1.2   &   0.016317   &   4.72    &   1065   &   243, ~~~305  
& Markowitz (2010), Nandra \& Papadakis (2001)  \\ [1mm]
         
\hline
Radio-Loud Seyfert 1s\\[1mm]
\hline

3C~111   &   BLRG   &   0.048500   &   119.1\tablenotemark{3}  &   808   &   127, ~~~238    
& Chatterjee et al.\ (2011),  Markowitz \& Edelson (2004)  \\[1mm]
  
3C~120   &   BLRG   &   0.033010   &   10.2   &   2102   &   505, ~~~629        
& Chatterjee et al.\ (2009), Marshall et al.\ (2009)\\[1mm]

\hline
Type 2 Seyfert Galaxies:\\[1mm]
Compton-thin\\[1mm]
\hline

Cen~A   &   NLRG   &   0.001825   &   9.07    &   632   &   110, ~~~200       
& Rothschild et al.\ (2011), Rothschild et al.\ (2006)\\[1mm]

NGC~5506   &   Sy1.9   &   0.006181   &   4.64    &   698   &   202, ~~~201        
& Uttley \& M$^{\textrm c}$Hardy (2005)\\[1mm]

MCG\,--5-23-16   &   Sy2   &   0.008486   &   8.28    &   180   &   ~55,\,~~~~55        
& this work\\[1mm]

NGC~4507   &   Sy2   &   0.011801   &   7.04    &   145   &   ~47, ~~~~47        
& this work\\[1mm]

\hline
Compton-thick\\[1mm]
\hline

Circinus   &   Sy2   &   0.001448   &   50.4   &   97   &   ~33, ~~~~32        
& this work\\[1mm]

NGC~7582   &   Sy2   &   0.005254   &   1.23    &   185   &   ~43, ~~~~43        
& this work\\[1mm]

NGC~4945   &   Sy2   &   0.001878   &   13.9   &   999   &   222, ~~~320        
&Mueller et al.\ (2004)\\[1mm]
%Mueller+in prep., Mueller+ 2004, AIPC, 714, 190

\hline
Blazars\\[1mm]
\hline

3C~273   &   FSRQ   &   0.158339   &   1.62    &   1810   &   430, ~~~530        
& Kataoka et al.\ (2000)\\[1mm]

3C~454.3   &   FSRQ   &   0.859000   &   7.24    &  37    &  ~13, ~~~~13         
& Jorstad \etal (2010)\\[1mm]

Mkn~421   &   BL~Lac  &   0.030021   &   1.53    &   1501   &   476, ~~~433        
& Giebels et al.\ (2007), Cui (2004)\\[1mm]

1ES~1959+650   &   BL~Lac   &   0.047000   &   10.1   &   199   &   ~65, ~~~~64        
& Gutierrez et al.\ (2006), Krawczynski et al.\ (2004)\\

\enddata
\tablecomments{The 23 targets selected for our sample due to reasonably bright hard (20--100 keV) 
X-ray spectra and sufficient exposure time to achieve good statistics with HEXTE.  Targets are listed
by 2--10 keV flux within each object class (flux values given in Table \ref{tabflux}).}
\tablenotetext{1}{Values taken from the LAB survey (Kalbera \etal 2005).}
\tablenotetext{2}{Good exposure time after screening.}
\tablenotetext{3}{73\% of the \NH~ column for 3C~111 is  from a molecular cloud along the line of sight.}

\end{deluxetable*}

%--------------------------------------------------------------------------------------------------%

This paper is structured in the following way: Section 2 describes the
archive, instruments, and data reduction process.  Section 3 details
the methods and analysis.  Section 4 discusses our results, including the Compton
reflection hump and the high-energy rollover.
Section 5 contains a summary of the main results of the paper.
We have included notes on individual sources with details of modeling 
and comparisons to previous work unique to each source in Appendix A.

%==================================================================%
\section{Archival Observations and Data Reduction}

\subsection{The Archive}

\xte includes two pointed observation instruments, the PCA 
(Jahoda et al.\ 2006) and HEXTE (Rothschild et al.\ 1998).
The PCA is usually the primary instrument when observing; HEXTE is not 
as sensitive as the PCA below 20 keV and is commonly overlooked for
faint sources such as AGN, especially for short observations.  
This investigation demonstrates that 
accumulation of data over long timescales can indeed
yield high quality PCA + HEXTE spectra for the brightest AGN. 

\begin{figure*}
 \epsscale{0.9}
 \plotone{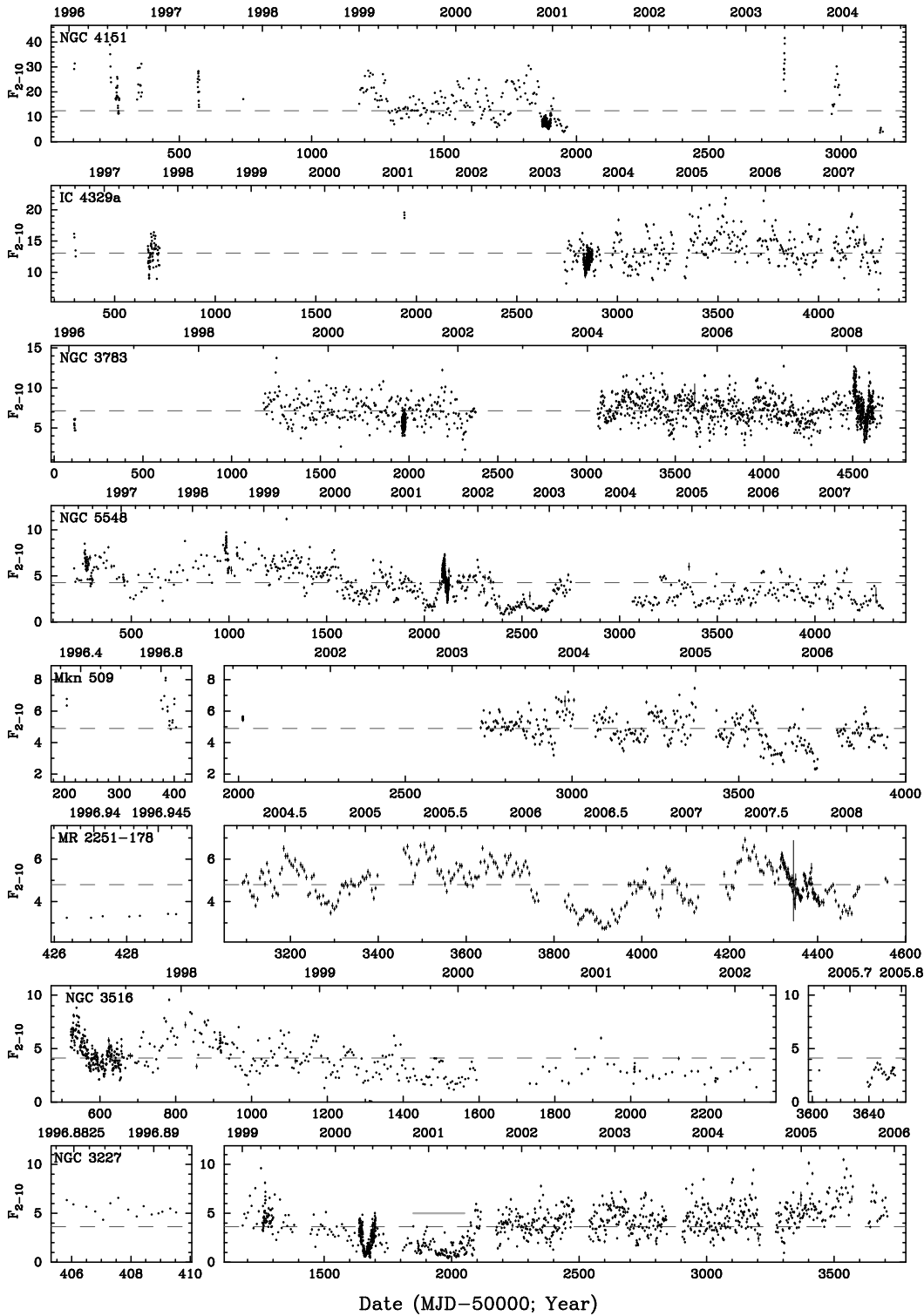}
 \caption{2--10 keV PCA flux light curves for the targets in our sample, with one flux point
for each observation ($F_{2-10}$ in units of erg cm$^{-2}$ s$^{-1}$). In many cases, error bars are smaller than the size of the data point.
Dashed lines indicate the average long-term flux for each object.}
   \label{lc1}
\end{figure*}

\begin{figure*}
 \epsscale{0.9}
 \plotone{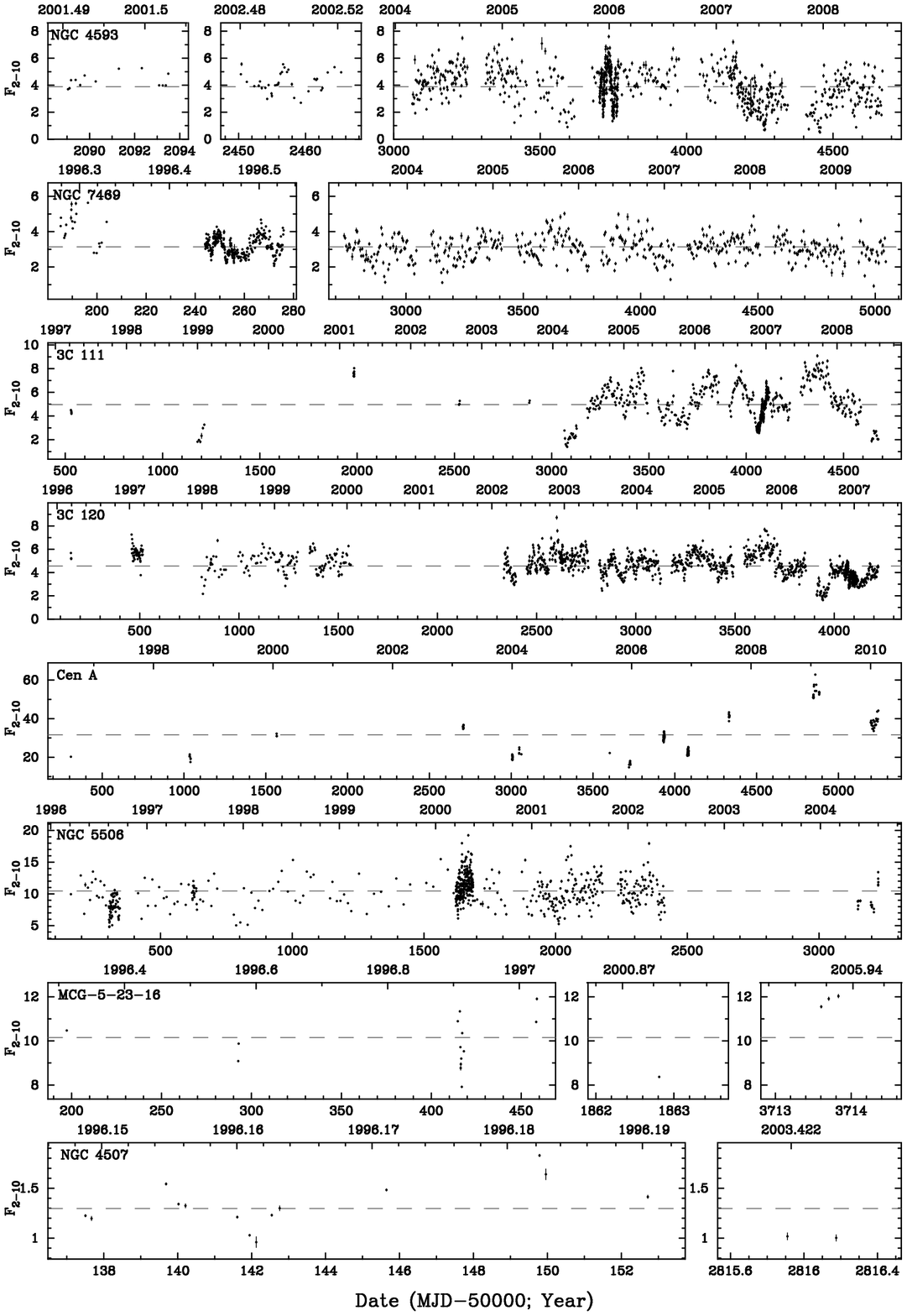}
 \caption{2--10 keV PCA flux light curves for the targets in our sample, with one flux point
for each observation ($F_{2-10}$ in units of erg cm$^{-2}$ s$^{-1}$). In many cases, error bars are smaller than the size of the data point.
Dashed lines indicate the average long-term flux for each object.}
   \label{lc2}
\end{figure*}

\begin{figure*}
 \epsscale{0.9}
 \plotone{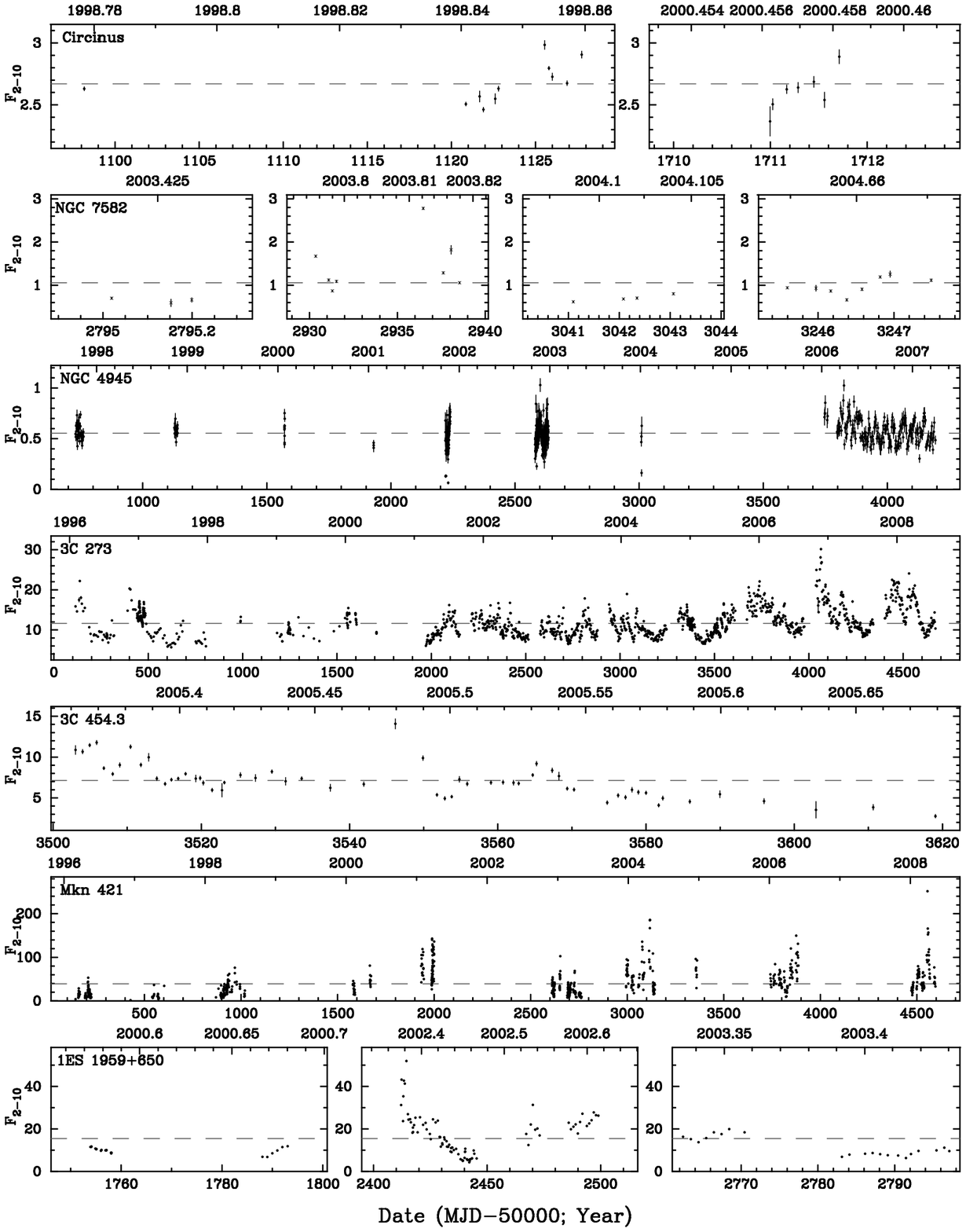}
 \caption{2--10 keV PCA flux light curves for the targets in our sample, with one flux point
for each observation ($F_{2-10}$ in units of erg cm$^{-2}$ s$^{-1}$). In many cases, error bars are smaller than the size of the data point.
Dashed lines indicate the average long-term flux for each object. For NGC~7582 data were ignored between MJD 51,850--52,050, see text.} 
   \label{lc3}
\end{figure*}

The sampling of the AGN in the \xte public archive is highly inhomogeneous, 
as targets were proposed by various groups for a variety of reasons. This 
includes, for example, multi-timescale, continuous monitoring
%%%% (numerous, 1 ks snapshots evenly spaced hours or days apart for weeks to years).
of Seyferts for power spectral density (PSD) measurements, long-look style observations
for tens of ks at a time, and target of opportunity observations on 
flaring blazars as part of multi-wavelength campaigns.
For each target, we simply used all data available in the public archive
regardless of sampling. We used data up to 2008 July (plus
proprietary data up to 2009 July for NGC\,7469 and proprietary data in 2009 January
and February for Cen~A).
Light curves in Figures \ref{lc1}--\ref{lc3} show the distribution of observations over the past
14 years.

%--------------------------------------------------------------------------------------------------%

%--------------------------------------------------------------------------------------------------%

We did not make any effort to isolate a particular class of AGN.
Target selection was guided by HEXTE data quality alone.
Our goal was to obtain high-quality spectra out to at least 100
keV for each case. This required the combination of 
the source being sufficiently X-ray bright in the 20--100 keV range
and having been observed with a sufficiently long exposure time.
The limiting 20--100 keV flux was $\gtrsim 5 \times 10^{-11}$ erg cm$^{-2}$ s$^{-1}$
depending on exposure time and the long-term trends in the HEXTE background;
see Section 2.4 for details. 

These criteria for target selection bias our sample towards objects 
with flat spectra except in the case of objects which happen to have high fluxes and/or a large amount of exposure time.
Hence this sample is not statistically complete, for example we did not have any Narrow Line Seyfert 1's in our sample
since they tend to be faint and have steeper spectra than typical Broad Line Seyfert 1's (Grupe \etal 1999, Vaughan \etal 2001, Xu \etal 2003) 
and none which were observed by \xte with sufficient time to
obtain a 3$\sigma$ detection at 100 keV.
The final sample of 23 AGN contains a variety of AGN subclasses, including 
17 Seyferts, two broad-line radio galaxies (BLRGs), two 
flat-spectrum radio quasars (FSRQs), and two BL~Lac objects.  Some objects 
have PCA + HEXTE spectra already published, but in most cases
we update these spectra by including additional data.

Table \ref{tabsample} lists the source name, AGN type, redshift ($z$), and the 
Galactic column density \NHgal, as well as
the total exposure time for each of the \xte instruments.
Lightcurve references are also included. 

For the purposes of analysis, we have most-often broken our sample down into
four categories: Seyfert 1's, Compton-thick Seyfert 2's, Compton-thin Seyfert 2's,
and blazars.  Note that since Seyfert 1's and 2's are classified optically, there is not always a 
firm correspondence between the levels of X-ray absorption along the line of sight and the classification.
Seyfert 1's include Seyfert 1--1.5's while Seyfert 2's include Seyfert 1.8--2's.
The distinction between Compton-thick and Compton-thin Seyfert 2's
is important to us since the Compton-thick sources tend to have much more complicated
X-ray spectra potentially causing model degeneracy in fitting.  
More details can be found in the notes on each Compton-thick source. 
Since 3C~273 has X-ray spectral properties more closely resembling a BLRG than a typical blazar 
(displays an Fe line and a very weak Compton hump), it is included with these solely for the purpose 
of spectral analysis in this paper.

%--------------------------------------------------------------------------------------------------%

\subsection{Proportional Counter Array (PCA) Data Reduction}

For all PCA and HEXTE data extraction and analysis we used HEASOFT version 6.7 software.
The PCA consists of five large-area, collimated proportional counter 
units (PCUs).  Reduction of the PCA data followed standard extraction 
and screening procedures. 
PCA STANDARD-2 data were collected from PCU's 0, 1 and 2 prior to 1998 December 23; PCU's
0 and 2 from 1998 December 23 until 2000 May 12; and PCU 2 only after 2000 May 12.
PCU 0 lost its propane veto layer starting on 2000 May 12, 
PCUs 1, 3, and 4 have been known to suffer from repeated breakdown during on-source time,
and PCU 1 lost its propane layer on 2006 December 25.
In addition, PCU 2 is the best calibrated of the PCU's and has maintained consistent functionality
for the duration of \xte's mission (Jahoda et al.\ 2006).  

To maximize signal-to-noise ratios we extract events
from the top Xe layer of the PCA only.
Data were rejected if they were 
gathered less than 10$\degr$ from Earth's limb, within 20 minutes of 
satellite passage through the South Atlantic Anomaly (SAA), if the satellite's 
pointing offset was greater than 0$\fdg$01, or if the ELECTRON2 value (a measure of particle flux) was $>$ 0.1. 
As the PCA is a non-imaging instrument, the background was estimated using 
the ``L7-240'' background models, appropriate for faint sources when the 
total count rate was below 40 counts\,s$^{-1}$\,PCU$^{-1}$ ; see e.g., Edelson \& Nandra (1999) 
for details on PCA background subtraction, the dominant source of systematic 
uncertainty (e.g., in total broadband count rate) in these data.
Some observations of Cen\,A and Mkn\,421 
fell above this flux threshold, and we used
the Sky-VLE background model on those observations.

Observations of 3C\,454.3 were obtained using a pointing position   
0$\fdg$52 to the southeast of 3C\,454.3 to completely eliminate
contributions to the spectrum from the cataclysmic variable, IM Peg, located about 0$\fdg$72 
to the northwest of 3C\,454.3 (see Jorstad et al.\ 2010).
At this offset position, the PCA and HEXTE collimator efficiencies
are both 45$\%$ (e.g., Jahoda et al.\ 2006); we corrected for this when 
generating the response matrices (below) and during spectral fitting.

Source and background spectra were extracted for each on-source ID,
and added using \textsc{sumpha} to create time-average source and background spectra.
Source spectra were binned to a minimum of 50 counts bin$^{-1}$ to ensure use of the 
$\chi^2$ statistic, although for nearly all energy bins, the systematic uncertainties 
dominate over the statistical uncertainties. A systematic uncertainty of 0.5\% was added to 
the source PCA spectra.  As the response of the PCA slowly hardens slightly
over time due to the gradual leak of xenon gas into the propane layer in each 
PCU, response files were generated for each separate observation ID (ObsID) using 
\textsc{pcarmf} version 11.7 and the latest calibration files available as of 2009 October.  
A time-averaged response file was generated by summing the individual response 
files and weighting them by their good exposure time and number of active PCU's.

%--------------------------------------------------------------------------------------------------%

\begin{deluxetable*}{lcccccc}
%   \tablewidth{0pc}
   \tablecaption{20--100 keV band \label{tab20}}
   \tablecolumns{6}
\startdata
\hline
\hline
Source    &     Photon Index    & HEXTE-A Flux  &  HEXTE-B Flux   &    $\%$ of Background  &  $\chi^{2}$/dof\\
&      & ($10^{-10}$\,erg\,s$^{-1}$) & ($10^{-10}$\,erg\,s$^{-1}$)  &    A, ~~~~~B&\\[1mm]
\hline
NGC~4151      &  1.84$\pm 0.02$ &   4.45    &   4.58    &  6.2,  ~~~6.4  &   536/53\\[1mm]
IC~4329A      &  1.89$\pm 0.04$ &   1.91    &   1.92    &  4.0,  ~~~4.5  &   140/49\\[1mm]
NGC~3783      &  1.82$\pm 0.05$ &   1.19    &   1.29    &  1.9,  ~~~2.2  &   90/43\\[1mm]
NGC~5548      &  1.79$\pm 0.08$ &   0.78    &   0.79    &  1.1,  ~~~1.1  &   53/33\\[1mm]
Mkn~509       &  1.87$\pm 0.09$ &   0.80    &   0.73    &  1.4,  ~~~1.5  &   51/37\\[1mm]
MR~2251--178   &  1.8$\pm 0.2$   &   0.70    &   0.71    &  0.9,  ~~~1.0  &   20/23\\[1mm]
NGC~3516      &  1.8$\pm 0.1$   &   0.89    &   0.91    &  1.1,  ~~~1.2  &   51/29\\[1mm]
NGC~3227      &  1.77$\pm 0.08$ &   0.76    &   0.76    &  1.2,  ~~~1.4  &   53/39\\[1mm]
NGC~4593      &  1.76$\pm 0.09$ &   0.70    &   0.66    &  1.4,  ~~~1.4  &   52/37\\[1mm]
NGC~7469      &  1.8$\pm 0.2$   &   0.47    &   0.51    &  0.7,  ~~~0.8  &   29/31\\[1mm]
3C~111        &  1.8$\pm 0.1$   &   0.82    &   0.81    &  1.6,  ~~~1.6  &   48/47\\[1mm]
3C~120        &  1.78$\pm 0.06$ &   0.80    &   0.75    &  1.3,  ~~~1.4  &   62/47\\[1mm]
Cen~A         &  1.78$\pm 0.01$ &   6.02    &   7.38    &  9.9,  ~~~13.7 &   62/53\\[1mm]
NGC~5506      &  1.87$\pm 0.04$ &   1.75    &   1.74    &  2.2,  ~~~2.3  &   194/37\\[1mm]
MCG--5-23-16  &  1.9$\pm 0.1$   &   1.48    &   1.52    &  1.9,  ~~~1.9  &   56/35\\[1mm]
NGC~4507      &  1.6$\pm 0.1$   &   1.33    &   1.73    &  1.7,  ~~~1.7  &   40/35\\[1mm]
Circinus      &  1.72$\pm 0.07$ &   2.02    &   2.16    &  2.6,  ~~~2.9  &   259/35\\[1mm]
NGC~7582      &  1.8$\pm 0.3$   &   0.58    &   0.60    &  1.5,  ~~~1.5  &   30/13\\[1mm]
NGC~4945      &  1.32$\pm 0.03$ &   1.77    &   1.80    &  3.4,  ~~~3.9  &   894/53\\[1mm]
3C~273        &  1.74$\pm 0.02$ &   1.80    &   1.98    &  2.9,  ~~~3.5  &   79/53\\[1mm]
3C~454.3      &  1.5$\pm 0.3$   &   0.64    &   0.46    &  1.7,  ~~~1.4  &   9/20\\[1mm]
Mkn~421       &  2.70$\pm 0.04$ &   1.11    &   1.04    &  2.3,  ~~~2.2  &   47/41\\[1mm]
1ES~1959+650   &  2.32$\pm 0.08$ &   0.78    &   0.81    &  1.9,  ~~~2.2  &   47/45\\[1mm]
\enddata
\tablecomments{These are results for simple power-law fits to the 20--100 keV band only using data from
both HEXTE clusters.  $\Gamma$ is tied between HEXTE-A and -B for all objects.}
\end{deluxetable*}

%--------------------------------------------------------------------------------------------------%
%------------------------------------------------------------------%

\subsection{High Energy X-ray Timing Experiment (HEXTE) Data Reduction}

HEXTE consists of two independent clusters (A and B), each containing four 
NaI(Tl)/CsI(Na) phoswich scintillation counters (Rothschild et al.\ 1998) 
which share a common 1$\degr$ FWHM field of view. Each of the eight detectors 
has a net open area of about 200 cm$^{2}$.  Source and background spectra were 
extracted from each individual \xte\ visit using Science Event data and standard 
extraction procedures. Data were rejected if they were gathered less than $10\degr$ from 
Earth's limb, within 30 minutes of satellite passage through the SAA, or if the 
satellite's pointing offset was $> 0\fdg$01.

To measure real-time background measurements, the two HEXTE 
clusters each undergo two-sided rocking to offset positions 1$\fdg$5 off-source, 
with default switching every 16 s before 1998 January 8 and every 32 s thereafter.
This allows for background subtraction for any timescale from 32 s up to the lifetime
of the mission.

We used the online resources HEXTErock\footnote{http://mamacass.ucsd.edu/cgi-bin/HEXTErock.html} 
and HEASARC's skyview tool\footnote{http://skyview.gsfc.nasa.gov/} (looking at \swiftbat and \rxte all-sky slew 
survey images) to ensure that there were no possible contaminating bright sources above 20 keV within 
2\degr of each source.  

Cluster A data taken during the following times were excluded, as the cluster did 
not rock on/off-source: 2004 December 13 -- 2005 January 14, 2005 December 12 -- 2006 January 4, 
during 2006 January 25, and after 2006 March 14.  Detector 2 in cluster B lost spectral 
capabilities in 1996; spectra were derived from the other three detectors only. Deadtime corrections were applied to 
account for cluster rocking, pulse analyzer electronics, and the recovery time following 
scintillation pulses caused by high-energy charged particles; typically, the HEXTE 
deadtime is around $30\%-40\%$.

We did not combine cluster A and B data; for observations after spring of 2006
we used only cluster B data and many of our sources have data from this period.
We used standard response matrices, hexte\_00may26\_pwa.arf, hexte\_97mar20c\_pwa.rmf,
hexte\_00may26\_pwb013.arf, and hexte\_97mar20c\_pwb013.rmf.

%------------------------------------------------------------------%

\subsection{HEXTE Background Considerations}

The HEXTE background is sufficiently bright that it dominates over the source spectra
for typical AGN fluxes.
The HEXTE background subtraction is well-understood for source fluxes down to approximately 1\% of the 
background (Gruber \etal 1996).  Sources with 20--100 keV fluxes fainter than approximately 1\% of the background
and/or with insufficient exposure time therefore had very large error bars in their net spectra
near 30 and 70 keV, the energies of the strongest activation lines that comprise the HEXTE background,
or were not detected out to 100 keV.
For sources whose net fluxes were in the range 0.7$\%$--1.2$\%$, 1.2$\%$--1.5$\%$, and 1.5$\%$--2.0$\%$ of the background, 
the good exposure times per HEXTE cluster after screening required to yield acceptable-quality spectra
were approximately 250, 100, and 50 ks, respectively.

The average HEXTE background rate has dropped by $\sim$50\% 
(from $\sim 12$ to $6 \times 10^{-9}$\,erg\,cm$^{-2}$\,s$^{-1}$) 
over a decade as the altitude of the \xte spacecraft has decreased, thus 
intercepting smaller regions of the SAA (F\"urst et al.\ 2009), resulting in a smaller 
particle flux through the detectors.
The reader is referred to Appendix A for further details of the composition and long-term
variability of the HEXTE background.

We combined 16 s and 32 s rocking within each cluster, however this proved to be
non-trivial because \textsc{sumpha} does not take into account the difference in 
background exposure time relative to on-source time which is 3/4 for 16 s and 7/8 
for 32 s.  To eliminate this problem we performed background subtraction on the 16 s 
and 32 s separately before summing the spectra.  We also created summed background 
files to use as correction files for fitting (see Section 3 below).

%==================================================================%

\section{Methods and Analysis}

All spectral fitting below was done with \textsc{xspec} version 12.5.1k utilizing
the abundances of Wilms et al.\ (2000) and cross-sections of Verner et al.\ (1996). 
Unless otherwise stated, uncertainties given are 90\% confidence ($\Delta \chi^2$ = 2.71 
for one interesting parameter).  

We aimed to truncate each source spectrum at an energy where our signal-to-noise ratio dropped 
below 3$\sigma$.  This energy is different for each source.
We also made use of the ``\textsc{recorn}'' component in \textsc{xspec} to correct for
imperfections in the background subtraction (PCA) and dead time estimation as a function of count rate (PCA and HEXTE).
Typical values were around 1\% of the background for the PCA and 0.16\% for the HEXTE clusters. 
Additionally, free cross-normalization factors for HEXTE-A and -B with respect to the PCA had average values
of 0.79 and 0.81 respectively.
%standard deviations for constants are 0.06 and 0.07.  cornorm stddevs = 1\%, and 0.10\% 
%--------------------------------------------------------------------------------------------------%

\begin{deluxetable*}{lccccccccc}
%   \tablewidth{0pc}
   \tablecaption{Baseline Model Fits \label{tabbase}}
   \tablecolumns{10}
\startdata
\hline
\hline
Source     &     $\Gamma$   &  $A$\tablenotemark{1} & \NH  & Partial Covering & Covering & Fe Line   &   $\sigma$   &   $I_\textrm{Fe}$ ($10^{-4}$)  & $\chi^{2}$/dof\\[1mm]
&       &                   ($10^{-2}$)   &   (10$^{22}$\,cm$^{-2}$)               &  \NH (10$^{22}$\,cm$^{-2}$)    &  Frac. ($f_\textrm{cov}$)      &  Energy (keV)  &  (keV)   &   (ph\,cm$^{-2}$s$^{-1}$)  & \\[1mm]
\hline

NGC~4151   &  1.84$\pm 0.02$   &   13.3 $\pm 0.1$   & 3.4(fixed) & 97$\pm 2$ & 0.66$\pm0.04$ &  6.06$\pm 0.05$  &  0.9$\pm 0.1$    &  12.0$\pm 0.6$  &  682/84\\[1mm]
IC~4329A   &  1.75$\pm 0.01$   &   3.84 $\pm 0.05$   &                 &&&  6.25$\pm 0.09$  &  $< 0.5$         &  1.8$\pm 0.2$   &  381/81\\[1mm]
NGC~3783   &  1.66$\pm 0.03$   &   2.0 $\pm 0.1$     &                 &&&  6.3$\pm 0.2$    &  0.3$\pm 0.1$    &  1.7$\pm 0.9$   &  242/71\\[1mm]
NGC~5548   &  1.67$\pm 0.01$   &   1.40 $\pm 0.04$   &                 &&&  6.1$\pm 0.2$    &  0.5$\pm 0.4$    &  1.0$\pm 0.3$   &  102/59\\[1mm]
Mkn~509    &  1.70$\pm 0.01$   &   1.40 $\pm 0.03$   &                 &&&  6.2$\pm 0.8$    &  $< 0.8$         &  0.7$\pm 0.1$   &  92.8/63\\[1mm]
MR~2251--178  & 1.63$\pm 0.02$ &   0.97 $\pm 0.05$   &                 &&&  6.3$\pm 0.4$    &  0.9$\pm 0.6$    &  0.9$\pm 0.3$   &  75.9/49\\[1mm]
NGC~3516   &  1.72$\pm 0.03$   &   1.5 $\pm 0.2$   &   &&0.55\tablenotemark{2}$\pm0.10$    &  6.23$\pm 0.06$  &  $< 0.3$         &  0.9$\pm 0.1$   &  122/61\\[1mm]
NGC~3227   &  1.60$\pm 0.01$   &   0.89 $\pm 0.03$   &   0.3$\pm 0.3$  &&&  6.2$\pm 0.2$    &  0.4$\pm 0.3$     &  0.9$\pm 0.2$   &  117/70\\[1mm]
NGC~4593   &  1.77$\pm 0.01$   &   1.12 $\pm 0.03$   &                 &&&  6.36$\pm 0.06$  &  $< 0.4$         &  1.0$\pm 0.1$   &  118/60\\[1mm]
NGC~7469   &  1.78$\pm 0.02$   &   0.89 $\pm 0.03$   &                 &&&  6.27$\pm 0.09$  &  $< 0.5$         &  0.6$\pm 0.2$   &  96.7/54\\[1mm]
\textbf{3C~111}     &  1.67$\pm 0.02$   &   1.24 $\pm 0.05$   &                 &&&  6.2$\pm 0.2$    &  0.7$\pm 0.3$    &  1.2$\pm 0.4$   &  102/73\\[1mm]
3C~120     &  1.68$\pm 0.03$   &   1.18 $\pm 0.07$   &                 &&&  6.3$\pm 0.3$    &  $< 0.8$         &  1.2$\pm 0.5$   &  97.5/77\\[1mm]
3C~273     &  1.66$\pm 0.01$   &   2.71 $\pm 0.02$   &                 &&&  7.0\err{*}{-0.8}  &  $< 1.0$       &  0.4$\pm 0.2$   &  158/83\\[1mm]
\textbf{Cen~A}      &  1.83$\pm 0.01$   &   16.0 $\pm 0.2$    &   16.9$\pm 0.3$ &&&  6.38$\pm 0.09$  &  $< 0.4$         &  4.9$\pm 0.7$   &  143/84\\[1mm]
NGC~5506   &  1.62$\pm 0.02$   &   2.2 $\pm 0.1$     &   $<$ 1.0       &&&  5.8$\pm 0.1$    &  0.9$\pm 0.1$    &  6.6$\pm 1.0$   &  527/70\\[1mm]
MCG--5-23-16 & 1.67$\pm 0.02$ &   2.5 $\pm 0.1$     &   1.6$\pm 0.4$  &&&  6.26$\pm 0.06$  &  0.4$\pm 0.1$    &  2.5$\pm 0.2$   &  135/68\\[1mm]
NGC~4507   &  1.62$\pm 0.06$   &   1.4 $\pm 0.2$     &   85$\pm 2$     &&&  6.4$\pm 0.5$    &  $< 0.8$         &  1.7$\pm 0.2$   &  148/71\\[1mm]
NGC~7582   &  1.12$\pm 0.08$   &   0.13 $\pm 0.02$   &   11$\pm 2$     &&&  6.2$\pm 0.8$    &  $< 0.8$         &  0.8$\pm 0.1$   &  179/34\\[1mm]
\textbf{NGC~7582-\textsc{PC}}\tablenotemark{3} &  1.8$\pm 0.1$   &  1.8\err{1.2}{-0.7}  &   24$\pm 4$     & 360$\pm 60$   & 0.6\err{0.6}{0.3}  &  6.2$\pm 0.1$    &  $< 0.5$         &  0.7$\pm 0.1$   &  40/32\\[1mm]
\hline
             &       &    &    &  $\Gamma_\textrm{SXPL}$\tablenotemark{4} & $A_\textrm{SXPL}$\tablenotemark{1}($10^{-2}$) &     &    &  &   \\[0.8mm]
\hline
Circinus  &  1.62$\pm 0.03$   &   2.8 $\pm 0.3$     & 395$\pm 9$ & 1.62$\pm 0.03$  & 0.41$\pm 0.02$ &  6.43$\pm 0.01$  &  0.25$\pm 0.03$  &  5.8$\pm 0.1$   &  338/64\\[1mm]
NGC~4945   &  1.73$\pm 0.04$  &   3.53 $\pm 0.03$   & 600$\pm 20$& 1.8$\pm 0.1$ & 0.15 $\pm 0.03$&  6.45$\pm 0.07$  &  $< 0.4$         &  0.65$\pm 0.07$ &  196/82\\
\enddata

\tablecomments{Best fit model parameters for our baseline model fit (no Compton hump or high-energy rollover modeled).  \NH~ is the full-covering cold
column in excess of \NHgal.  Source name in bold indicates that we conclude this model is the best description of the spectrum.}

\tablenotetext{1}{Power law normalization (ph\,keV$^{-1}$\,cm$^{-2}$\,s$^{-1}$ at 1 keV)}
\tablenotetext{2}{Partial covering warm absorber with log($\xi$) fixed at 2.19 and column density fixed at $2 \times 10^{23}$ cm$^{-2}$ but covering fraction left as a free parameter.}
\tablenotetext{3}{Alternative model with partial covering Compton-thick absorber as suggested by Turner \etal (2000).  See Appendix A.19 for details.}
\tablenotetext{4}{SXPL parameters are for the soft X-ray power law due to leaked or scattered emission.}

\end{deluxetable*}

%--------------------------------------------------------------------------------------------------%

%------------------------------------------------------------------%
\subsection{The 20--100 keV Bandpass}

To explore target properties in the 20--100 keV bandpass we fit a simple power law 
to HEXTE-only data. We fit HEXTE-A and B spectra with the power-law normalizations untied
due to the fact that for several sources, the HEXTE-B data covered a longer time span than HEXTE-A.
We kept the photon indices tied between A and B since untying the photon indices
did not yield a significant improvement in the fit for most sources.
The only exception to this was 3C~273, for which untying the photon indices and power law normalizations
led to an improvement in fit of $\Delta \chi^{2}$/dof = --8/2 with photon indices of $1.79 \pm 0.02$ 
and $1.71 \pm 0.02$ for cluster A and B respectively.
Tied photon indices, absorbed fluxes, and goodness of fits are listed in Table \ref{tab20}.
Brightness level as a percent of background in the 20--100 keV band is also listed.
Note that there are about a half-dozen sources where a simple power law is a bad fit, due to extra
curvature possibly indicating a Compton reflection hump or a high-energy rollover.

%--------------------------------------------------------------------------------------------------%

\begin{deluxetable*}{lcccccccccc}
%   \tablewidth{0pc}
   \tablecaption{Reflection Models \label{tabpexrav}}
   \tablecolumns{11}
\startdata
\hline
\hline
Source     &     $\Gamma$   &  $A$\tablenotemark{1} & \NH   & Partial Cov. & Covering &   Fe Line   &   $\sigma$   &   $I_\textrm{Fe}$ ($10^{-4}$) & $R$ &  $\chi^{2}$/dof \\[1mm]
    &       &    ($10^{-2}$)  & (10$^{22}$\,cm$^{-2}$)    &  \NH (10$^{22}$\,cm$^{-2}$)    &  Frac. ($f_\textrm{cov}$)      &  Energy (keV)  &  (keV)   &   (ph\,cm$^{-2}$s$^{-1}$) & & \\[1mm]
\hline
\textbf{NGC~4151}     &  1.90$\pm 0.02$   &   9.9 $\pm 0.5$    & 3.4(fixed) & 37$\pm 3$& 0.66$\pm0.06$  &  6.4$\pm 0.1$  &  $<$0.5           &  2.9 $\pm 0.4$  &  0.7$\pm 0.1$       &  165/83 \\[1mm]
\textbf{IC~4329A}     &  1.88$\pm 0.02$   &   4.5 $\pm 0.1$     &                 &&&  6.5$\pm 0.1$    &  0.5$\pm 0.3$  	 &  2.1 $\pm 0.4$  &  0.39$\pm 0.05$  	 &  110/79 \\[1mm]
\textbf{NGC~3783}     &  1.86$\pm 0.03$   &   2.5 $\pm 0.1$     &     	        &&&  6.4$\pm 0.7$    &  $<$0.4  	 &  1.5 $\pm 0.3$  &  0.41$\pm 0.08$  	 &  109/70 \\[1mm]
\textbf{NGC~5548 }    &  1.72$\pm 0.02$   &   1.51 $\pm 0.04$   &     	        &&&  6.2$\pm 0.1$    &  0.6$\pm 0.2$  	 &  1.1 $\pm 0.2$  &  0.13$\pm 0.04$  	 &  69.3/58 \\[1mm]
\textbf{Mkn~509}      &  1.75$\pm 0.02$   &   1.48 $\pm 0.05$   &     	        &&&  6.3$\pm 0.2$    &  $<$0.6  	 &  0.7 $\pm 0.2$  &  0.15$\pm 0.05$  	 &  70.9/62 \\[1mm]
MR~2251-178  &  1.63$\pm 0.02$   &   0.97$\pm 0.08$    &     	        &&&  6.3$\pm 0.3$    &  0.9$\pm 0.5$  	 &  0.9 $\pm 0.3$  &  $<$0.01            &  76.0/48 \\[1mm]
\textbf{NGC~3516 }    &  1.82$\pm 0.04$   &   1.7$\pm 0.2$     &&      & 0.55\tablenotemark{2}$\pm0.10$ &  6.2$\pm 0.1$  &  $<$0.3 &  0.9 $\pm 0.2$  &  0.31$\pm 0.09$     	 &  58.2/60 \\[1mm]
\textbf{NGC~3227 }    &  1.79$\pm 0.04$   &   1.3 $\pm 0.1$     &   2.7$\pm 0.9$  &&&  6.4$\pm 0.2$    &  0.6$\pm 0.5$ &  0.6 $\pm 0.2$  &  0.41$\pm 0.09$  	 &  62.4/69 \\[1mm]
\textbf{NGC~4593}     &  1.85$\pm 0.03$   &   1.25 $\pm 0.05$   &     	        &&&  6.4$\pm 0.1$  &  0.4$\pm 0.1$  	 &  1.0 $\pm 0.1$  &  0.34$\pm 0.09$ 	 &  77.8/59 \\[1mm]
\textbf{NGC~7469 }    &  1.88$\pm 0.04$   &   1.02 $\pm 0.06$   &     	        &&&  6.3$\pm 0.1$    &  $<$0.5           &  0.5 $\pm 0.1$  &  0.4$\pm 0.1$   	 &  65.4/53 \\[1mm]
3C~111  	    &  1.67$\pm 0.01$   &   1.2\err{0.2}{0.8} &     	        &&&  6.2$\pm 0.2$    &  0.7$\pm 0.2$  	 &  1.2 $\pm 0.3$  &  $<$0.02            &  102/72 \\[1mm]
\textbf{3C~120 } 	    &  1.82$\pm 0.03$   &   1.50 $\pm 0.01$   &     	        &&&  6.5$\pm 0.2$    &  $<$0.7    	 &  0.6 $\pm 0.4$  &  0.24$\pm 0.06$  	 &  69.8/76 \\[1mm]
\textbf{3C~273}  	    &  1.69$\pm 0.01$   &   2.82 $\pm 0.04$   &     	        &&&  7.0\err{0*}{2.0}   &  0.9\err{1.0}{0.6} & 0.7$\pm 0.3$   &  0.07$\pm 0.03$     &  126/82 \\[1mm]
Cen~A  	    &  1.83$\pm 0.01$   &   16.0 $\pm 0.3$    &   16.9$\pm 0.3$ &&&  6.38$\pm 0.09$  &  $< 0.5$         &  4.9$\pm 0.7$   &  <0.005 &  143/83\\[1mm]

\textbf{NGC~5506 }    &  1.93$\pm 0.03$   &   3.6 $\pm 0.2$     &   1.9$\pm 0.5$  &&&  6.1$\pm 0.2$    &  1.0$\pm 0.2$  	 &  5.4 $\pm 0.9$  &  1.2$\pm 0.1$   	 &  118/69 \\[1mm]
\textbf{MCG--5-23-16} &  1.84$\pm 0.03$   &   3.3 $\pm 0.2$     &   3.5$\pm 0.5$  &&&  6.4$\pm 0.6$    &  0.2$\pm 0.1$  	 &  1.8 $\pm 0.3$  &  0.4$\pm 0.1$  	 &  81.5/67 \\[1mm]
\textbf{NGC~4507}     &  1.72$\pm 0.04$   &   1.8 $\pm 0.3$     &   87$\pm 2$     &&&  6.45$\pm 0.05$  &  $<$0.4     	 &  1.4 $\pm 0.2$  &  0.4$\pm 0.1$   	 &  126/70 \\[1mm]
\textbf{NGC~7582-\textsc{REFL}}     &  1.79$\pm 0.10$   &   0.34$\pm 0.06$    &   14$\pm 3$     &&&  6.2$\pm 0.1$    &  $<$0.5    	 &  0.7 $\pm 0.2$  &  3.3\err{1.6}{0.9}   &  36.3/33 \\[1mm]
%\textbf{NGC~7582pc} &  1.82$\pm 0.$   &  0.88  &   18$\pm 3$     & 360$\pm $   & 0.49  &  6.2$\pm 0.1$    &  $< 0.7$   &  0.7$\pm 0.$ & 1.7$\pm 0.$  &  37/31\\[1mm]
\hline
             &       &    &    & $\Gamma_\textrm{SXPL}$\tablenotemark{3} & $A_\textrm{SXPL}$\tablenotemark{1}($10^{-2}$)   &     &      &    &  &   \\[0.8mm]
\hline
Circinus    &  2.01$\pm 0.06$   &   10 $\pm 2$  & 690$\pm 50$ &  2.01$\pm 0.06$  & 0.44$\pm 0.02$ &  6.44$\pm 0.01$  &  $<$0.15  &  4.7 $\pm 0.2$  &  0.53$\pm 0.04$  	 &  138/62 \\[1mm]
%Circinus-tied    &  2.01$\pm 0.06$   &   10 $\pm 2$  & 690$\pm 50$ &  tied  & 0.44$\pm 0.02$ &  6.44$\pm 0.01$  &  $<$0.15  &  4.7 $\pm 0.2$  &  0.53$\pm 0.04$  	 &  138/63 \\[1mm]
NGC~4945     &  1.71$\pm 0.03$   &   3.0 $\pm 0.4$  &   670$\pm 35$   & 4.0$\pm 0.5$ & 2\err{2}{1} &  6.4$\pm 0.1$    &  0.4$\pm 0.2$  	 &  0.7 $\pm 0.2$  &  0.24$\pm 0.05$  	 &  175/81 \\
\enddata

\tablecomments{Best fit model parameters when a Compton reflection hump is added to the baseline model.  Symbols are the same as in Table 3.  
Source name in bold indicates that we conclude this model is the best description of the spectrum. An asterisk (*) indicates parameter pegged at hard limit.}
\tablenotetext{1}{Power law normalization (ph\,keV$^{-1}$\,cm$^{-2}$\,s$^{-1}$ at 1 keV)}
\tablenotetext{2}{Partial covering warm absorber with log($\xi$) fixed at 2.19 and column density fixed at $2 \times 10^{23}$ cm$^{-2}$ but covering fraction left as a free parameter.}
\tablenotetext{3}{SXPL parameters are for the soft X-ray power law due to leaked or scattered emission.}

\end{deluxetable*}

%--------------------------------------------------------------------------------------------------%
%--------------------------------------------------------------------------------------------------%

\begin{deluxetable*}{lccccccccccc}
%   \tablewidth{0pc}
   \tablecaption{Rollover Models \label{tabcutoff}}
   \tablecolumns{12}
\startdata
\hline
\hline
Source   &     $\Gamma$   &   $A$\tablenotemark{1} &   \NH  & $\Gamma_\textrm{SXPL}$\tablenotemark{2} & $A_\textrm{SXPL}$\tablenotemark{2} &   Fe Line   &   $\sigma$   &   $I_\textrm{Fe}$($10^{-4}$) &   $R$  & \eroll \tablenotemark{3} &  $\chi^{2}$/dof\\[1mm]
       &     &    ($10^{-2}$)        & (10$^{22}$\,cm$^{-2}$) &   &  ($10^{-2}$)  &  Energy (keV)  &  (keV)   &   (ph\,cm$^{-2}$s$^{-1}$) &         &  (keV) &\\
\hline
\textbf{MR~2251--178}  &  1.56$\pm 0.03$   & 0.91\err{0.03}{0.07}  &&&                                           & 6.2\err{0.3}{0.6} & 0.8$\pm 0.6$ & 0.8$\pm 0.5$ &  $<$0.06      &  100\err{40}{30} &  65.8/47 \\[1mm]
\textbf{Circinus}     &  1.2$\pm 0.2$     & 1.2\err{0.9}{0.6}     & 920\err{120}{150} & 2.5$\pm0.4$       & 1.1$\pm0.4$  & 6.42$\pm 0.01$ & 0.20$\pm 0.04$ & 5.2$\pm 0.1$ & 1.1$\pm$ 0.3  &  41\err{6}{10}   &  47.1/62 \\[1mm]
\textbf{NGC~4945}     &  1.0$\pm 0.1$     & 0.24\err{0.27}{0.05}  & 468\err{30}{14}   & 2.1\err{0.6}{0.1} & 0.5$\pm0.2$  & 6.5$\pm 0.1$ & 0.4$\pm 0.2$ & 0.8$\pm 0.2$ & $<$0.03       &  70\err{10}{10}  &  90.6/80 \\[1mm]

\tablecomments{Symbols are the same as in Table 3.  The Fe line parameters for these sources do not vary greatly when a rollover is added.  
See Tables 3 and 4 for these values.  Source name in bold indicates that we conclude this model is the best description of the spectrum.}  
%In the cases of Circinus and NGC~4945 this does not correspond to the lowest $\chi^{2}$ value since model degeneracies lead to unrealistic 
%values for $\Gamma$ and unexpectedly low values of $R$, see Discussion?}
\tablenotetext{1}{Power law normalization (ph\,keV$^{-1}$\,cm$^{-2}$\,s$^{-1}$ at 1 keV)}
\tablenotetext{2}{SXPL parameters are for the soft X-ray power law due to leaked or scattered emission.}
\tablenotetext{3}{\eroll is the energy of the high-energy rollover.}
%\tablenotetext{*}{Reached hard lower limit.}
\end{deluxetable*}

%--------------------------------------------------------------------------------------------------%
%--------------------------------------------------------------------------------------------------%

\begin{deluxetable}{lcccccc}
   \tablewidth{0pc}
   \tablecaption{Blazar Best-fit Parameters \label{tabblaz}}
   \tablecolumns{7}
\startdata
\hline
\hline
Source         &  $\Gamma_1$   &   Break  &  $\Gamma_2$   &  A\tablenotemark{1}           &  $\chi^{2}$/dof\\[1mm]
&&  Energy \\[1mm]
&&  (keV)  \\[1mm]
\hline
\textbf{3C~454.3}    &   1.67$\pm 0.05$  &                 &       &  1.8$\pm 0.2$\tablenotemark{2}       & 18.2/38\\[1mm]
\textbf{Mkn~421}     &   2.49$\pm 0.08$  &   7.1$\pm 0.8$  &  2.79$\pm 0.01$   &  31$\pm 4$        & 94.6/79\\[1mm]
\textbf{1ES~1959+650} &   2.01$\pm 0.02$    &   5.7$\pm0.8$     &  2.12$\pm 0.01$    &  6.2$\pm 0.3$\tablenotemark{3}      &  40.7/62\\[1mm]
\enddata
\tablecomments{Best fit model parameters for the jet-dominated sources in our sample.  Power law normalization refers to
the normalization at 1 keV for the simple power law fit and the normalization at the break energy for the broken power
law fits.  For best fit parameters of 3C~273 please see Table \ref{tabpexrav}.} 
\tablenotetext{1}{Power law normalization ($10^{-2}$ ph\,keV$^{-1}$\,cm$^{-2}$\,s$^{-1}$) at 1 keV for a power law and at the break energy for a broken power law.}
\tablenotetext{2}{Off-source pointing taken into account, see Section 2.2 for details.}
\tablenotetext{3}{Uncertainty determined with the break energy held fixed.}
\end{deluxetable}

%--------------------------------------------------------------------------------------------------%

\subsection{Broadband Fitting of PCA and HEXTE Combined Data}

Since Seyfert galaxies share many common spectral attributes we applied
similar models to all the sources that were Seyferts or Seyfert-like, including
the BLRGs and the FSRQ 3C\,273 in which we find evidence for an Fe line and possibly
even a weak Compton reflection hump, both of which are atypical of blazars.
The remaining three blazars were treated separately (see Section 3.3) since they do
not show many of the spectral features seen in the rest of our sample. 

With the latest calibration it was not necessary to add instrumental features to the models,
such as a Gaussian at 8.05 keV for Cu emission or an Xe L edge near 4--5 keV,
as was done by Rothschild \etal (2006) for \xte observations of Cen~A.

The Fe K line is virtually ubiquitous in Seyfert spectra. 
The PCA cannot deconvolve broad and narrow Fe lines, nor Fe K$\alpha$ and K$\beta$ lines.
In addition, the PCA cannot resolve an Fe line except in cases of extreme relativistic broadening.
We therefore use a single Gaussian to fit the Fe K emission complex.
The reader is referred to results from instruments with superior energy resolution at 6 keV,
namely \xmm, \chandra-HETGS, or \suzaku, to determine if the Fe line we model is dominated 
by a broad or a narrow Fe component, or a blend of both in each object.
For example, the presence of both broad and narrow lines has been confirmed with \textit{XMM-Newton} or 
\textit{Suzaku} for MCG--5-23-16 (Reeves et al.\ 2007), NGC 3516 (Markowitz et al.\ 2008), and 3C 120 
(Kataoka et al.\ 2007), while broad lines have not been confirmed in Cen~A (Markowitz et al.\ 2007) and  
NGC~5548 (Liu et al.\ 2010).  In other objects, the Fe line is moderately broadened (e.g., NGC 4593, 
$v_\textrm{FWHM}$ $\sim$ 10000 km s$^{-1}$; Brenneman et al.\ 2007).

One of our primary goals was to make use of the 3--100 keV spectra to test for the 
presence of Compton humps and high-energy cut-offs.  To begin, we created a baseline 
model consisting of a primary power law with Galactic absorption plus an Fe line but with
no Compton reflection or high-energy cutoff of the continuum. 
This alone was a suitable baseline for most of the Seyfert 1's.  In the case of both 
Compton-thin and Compton-thick Seyfert 2s, the baseline model additionally included a ``\textsc{zphabs}'' 
component to model cold absorption in excess of \NHgal.  In the case of the Compton-thick 
Seyfert 2s, Circinus and NGC\,4945, we also included a soft power-law
absorbed only by \NHgal, to model e.g., scattered nuclear X-ray continuum emission, 
thermal emission from diffuse plasma, and/or ``leaked'' continuum emission in the case of a 
partial-covering absorber.  Best-fit results are listed in Table \ref{tabbase} for all objects.

For each object, we included warm absorber components as needed by searching through publications 
relying on \chandra-HETGS and/or \xmm-RGS data. The PCA is not highly sensitive to discrete 
narrow absorption lines. However, if a warm absorber phase's spectral signature was to induce continuum 
curvature below 5 keV of more than 1$\%$--2$\%$, then we modeled it with an XSTAR table component, 
keeping the parameters frozen at the column density and ionization parameter specified in the 
literature. See notes on individual targets for details.

There were three exceptions to the above scheme: NGC\,4151 was modeled to have a cold partial 
coverer in the line of sight (see Appendix A.1 for details).  NGC\,3227 was modeled to have full-covering 
cold absorption in excess of the Galactic column (see Appendix A.8 for details). 
For NGC\,7582 we obtained two good fits, a ``reflection-dominated'' model and a ``partial-covering'' 
model. We report both model fits in Table \ref{tabbase} (see Appendix A.19 for details).  

Next we added a Compton reflection component using \pexrav (Magdziarz \& Zdziarski 1995) with the photon 
index tied to that of the continuum, all abundances set to solar, and no high-energy rollover.  
The inclination was set to 30$\degr$ for Seyfert 1-1.5's and 45$\degr$ for Seyfert 2's.
Results are listed in Table \ref{tabpexrav}.  
The value $R$ given by the model is the fraction of light reflected assuming an isotropic
X-ray source above a semi-infinite slab such as a disk covering 50\% of the sky as viewed by the
illuminating source.
Significant improvement in fit was found for all objects except \mr, 3C\,111, and Cen\,A.

Next we tested for the presence of high-energy continuum cutoffs.  
We tested two forms for the cut-off, ``\textsc{CutoffPL}'' and ``\textsc{HighEcut}''.
\textsc{CutoffPL} (which is a bit of a misnomer) has a slow rollover where the continuum at \eroll is $1/e$ times the initial value.
\textsc{HighEcut} has a much more abrupt cut-off governed by the following equations:
$A(E)$ = $e^{(E_\mathrm{cut}-E)/E_\mathrm{fold}}$ for $E > E_\mathrm{cut}$;
$A(E)$ = 1 for $E < E_\mathrm{cut}$. Since we did not see evidence for such abrupt cut-offs in the data, 
results found using this model are not presented in this paper.  
E$_{roll}$ was limited to a hard lower bound of 50 keV in all cases with the exception of
Circinus which shows evidence for a rollover at slightly lower energies ($\sim$40 keV).
We included the \pexrav component in these fits with $\Gamma$ and $E_{roll}$ tied to those of  
\textsc{CutoffPL}.

Best-fit \textsc{CutoffPL} results are listed in Table \ref{tabcutoff}. We found significant 
improvement in fit ($\Delta\chi^2 > 50$) for two objects, Circinus and NGC\,4945 and
marginal evidence ($\Delta\chi^2 \sim 10$) for a rollover in \mr ~(see notes on individual sources).
In all other cases, we found only lower limits for rollovers as $\chi^2$ did not improve,
nor was there improvement in data/model residuals at high energies.  
Unfortunately, for the two cases (both Compton-thick Seyfert 2's) where the rollover substantially improved the fit,
systematic degeneracies present between $\Gamma$, \eroll and $R$ 
lead to unrealistically low values of the photon index when a rollover is modeled.
For this reason we present analysis on both models.

\begin{figure}
\plotone{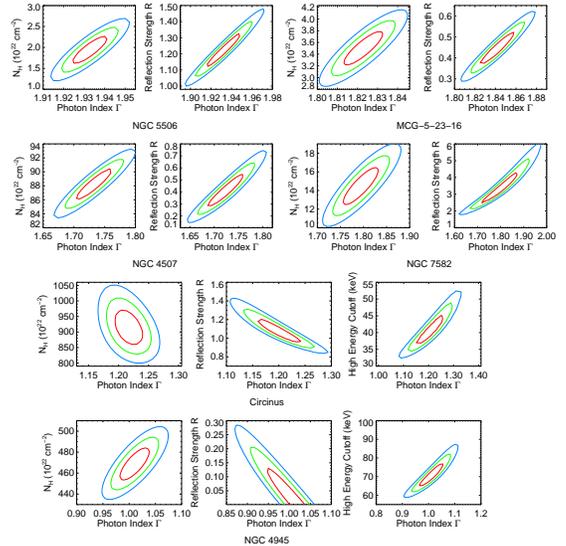}
\caption{Contour plots of various parameters for Seyfert 2's, illustrating the degree of statistical degeneracy seen 
in these objects.}
   \label{contours}
\end{figure}

\begin{figure}[H]
\plotone{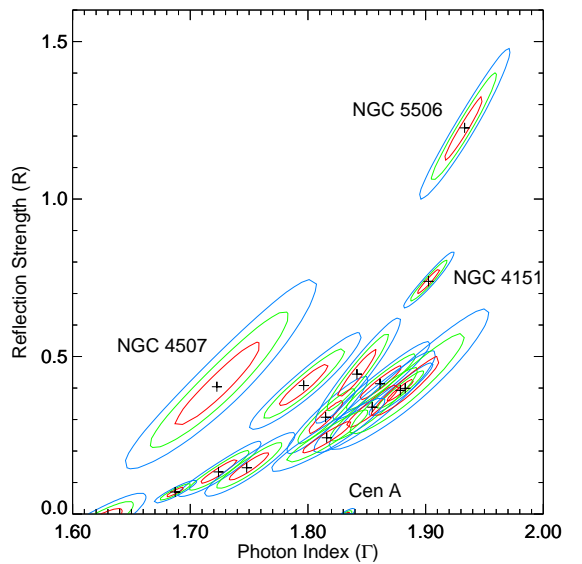}
\caption{Contour plot of $R$ versus $\Gamma$ for all objects with well-constrained values, ie.\ excluding Compton-thick sources.
The scatter in this plot seems to indicate that there is no strong correlation between $R$ and $\Gamma$ and that our fits are
robust in determining these parameters.}
   \label{contall}
\end{figure}

We include contour plots of \NH ~versus $\Gamma$, $R$ versus $\Gamma$ and 
\eroll versus $\Gamma$ (in instances where a rollover was detected)
in Figure \ref{contours} for Seyfert 2's to identify potential model degeneracies.
In nearly all cases statistical degeneracies are minimal, the one exception being NGC~7582, for which
the reflection strength is very poorly constrained.
Additionally, Figure \ref{contall} shows $R$ versus $\Gamma$ for all objects with well-constrained/realistic 
values, ie.\ excluding the Compton-thick sources NGC~7582, NGC~4945 and Circinus.

Spectra are shown in Figures \ref{NGC4151spec}--\ref{1ES1959+650spec} along with the best-descriptor model in panel (a),
$\chi$ values for the base model in panel (b), and any models that show improvement past the base in subsequent panels.
Data to model ratios for the best-fit reflection and rollover models that demonstrate the extra curvature in Circinus and NGC~4945 
that indicates the presence of a high-energy rollover are shown in Figures \ref{CIRCrat} and \ref{N4945rat}. 

%------------------------------------------------------------------%
\subsection{Blazars}

Blazars are jet-dominated sources whose continua are very slowly bending
through the X-ray bandpass, and whose X-ray spectra are typically fit by power laws, 
broken power laws, or other slowly bending continuum models.  Fit results are listed in Table \ref{tabblaz}
for all but 3C\,273 which we tested for evidence of an Fe line and Compton hump based on previous
detections of these parameters (see Appendix A.20 for details).  Best fit parameters 
for 3C\,273 are listed in Table \ref{tabpexrav}.

For 3C\,454.3, a power law absorbed only by \NHgal\ yielded a good fit.
For Mkn\,421 and 1ES~1959+650, we found that a broken power law fit, absorbed only by \NHgal,
yielded a significantly better fit compared to an unbroken power law.

Broadband SED fitting to constrain synchrotron/inverse Compton emission parameters
is beyond the scope of this paper. Thus we do not discuss this class of objects further,
however it is our hope that future work may be done incorporating time-averaged spectral
properties in other bands.

%------------------------------------------------------------------%
\subsection{Further Analysis}

Additional information for the ``best descriptor'' models is given in Tables \ref{tabeqw} and
\ref{tabflux} which list Fe line equivalent widths (EW's) and flux/luminosity information
respectively.
Luminosities were calculated following, e.g.,
Alexander et al.\ (2003, their Equation (1)).
Luminosity distances were taken from the NASA Extragalactic Database.
For each object, the mean of the available
redshift-independent estimates was used if available,
otherwise the luminosity distance based on redshift
(using the 3K cosmic microwave background radiation as a reference
frame) was used.
%==================================================================%
%Figures

%==================================================================%

%==================================================================%
% This is Section 4

\section{Discussion}

Compared to X-ray data below $\sim$10 keV, 
higher energies have been relatively unexplored 
because of the difficulty in obtaining simultaneous high-quality spectra above 20 keV.  
By using \textit{RXTE} PCA + HEXTE monitoring data for X-ray bright sources,
we have been able to construct spectra
featuring broadband sensitivity that extends up to $\geq$100 keV, 
and long total exposure times.

HEXTE offers several advantages compared to other
instruments which operated or are operating above 20 keV.
HEXTE has realtime background measurement, as opposed to relying on 
modeling as with the \suzaku HXD. \sax PSD had a collecting area equivalent to only one HEXTE cluster
but with significantly less background due to its equatorial orbit,
while OSSE, \textit{INTEGRAL}-IBIS, and \swiftbat have much higher background due to larger fields of view.

Because the \xte monitoring has spanned a very long baseline (more than a decade 
for several sources), the time-averaged spectral properties derived
can be taken to be true averages over time and over a wide range of fluxes.
Judging by the 2--10 keV light curves, for most sources, we have sampled
the full range typically displayed by a given source in the recent past.
\swiftbat, with its large field of view, has also been able to provide
long-term monitoring in the 14--195 keV energy range since 2005,
however the spectra currently have only eight channels which adds to the difficulty
of fitting multi-component spectra.  

One of our main goals was to obtain constraints on the strength of the 
Compton reflection component, which we detected in the majority of Seyferts 
(16--17 out of 20 Seyferts, depending on the form of the model used).
As detailed below, we found an average reflection strength $\langle R \rangle$ of
0.35$\pm 0.16$ for Seyfert 1's and 0.67$\pm 0.46$ for Compton-thin Seyfert 2's.  
The implications for the geometry of the Compton-thick gas are discussed
below in Section 4.2.
Another goal was to search for high-energy rollovers in the 
power-law continuum, expected if the power-law is produced by thermal
Comptonization in an X-ray corona.
We found evidence for high-energy rollovers in only
three sources in the sample, and we caution that the evidence is somewhat tentative
in all three cases. Lower limits to \eroll were obtained for the
remaining Seyferts and the implications for the
X-ray corona are discussed in Section 4.3.

%==================================================================%

%==================================================================%
%==================================================================%

\begin{figure}[H]
  %\epsscale{0.7}
  \plotone{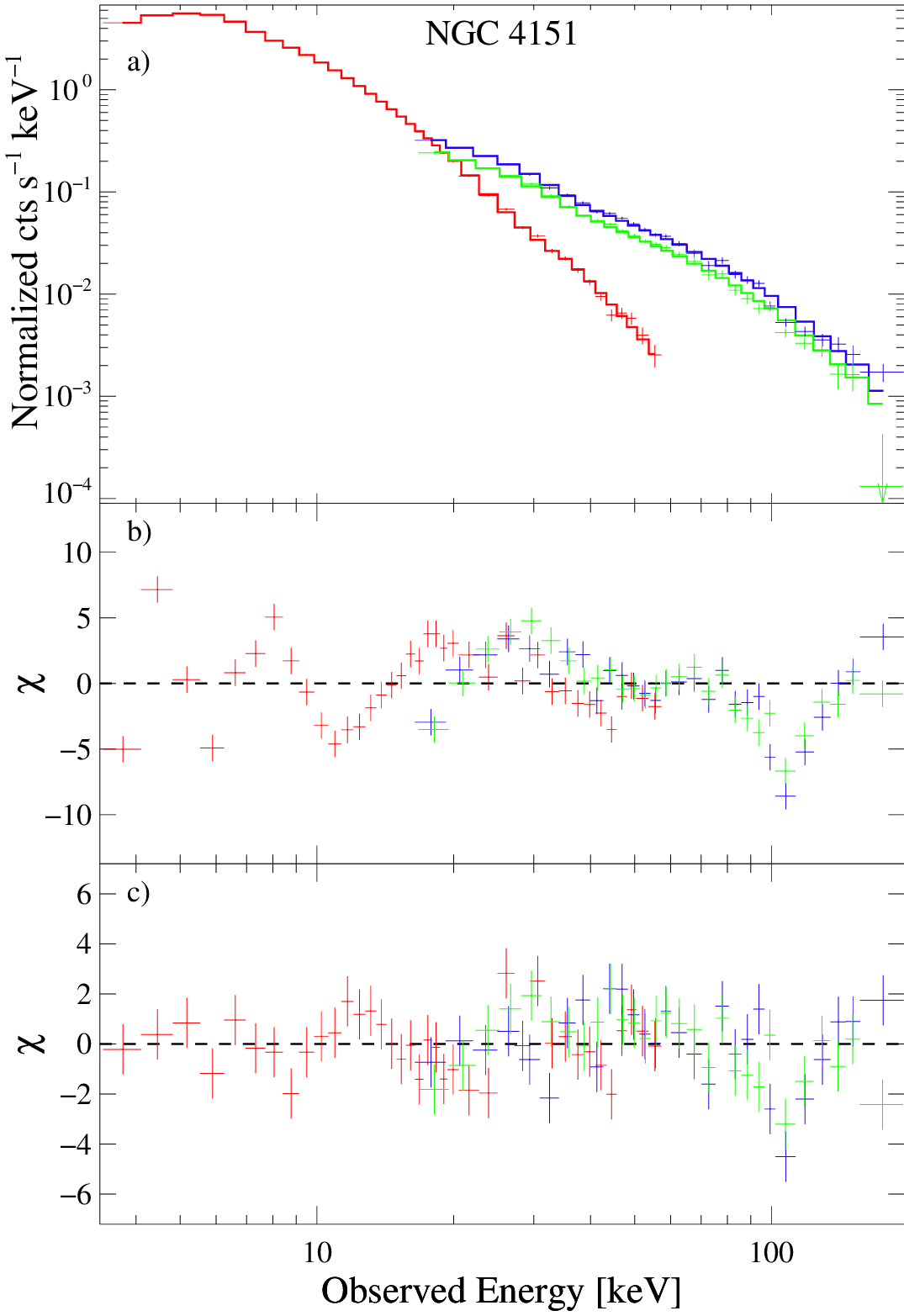}
  \caption{Data, model and data--model residuals for NGC~4151.  Panel (a) shows the PCA and HEXTE data along with the 
           best-fit model (solid line); panel (b) shows residuals for the baseline model; and panel (c) shows residuals
           for the best-fit model (parameters for the best-fit model are listed in Table \ref{tabpexrav}).}
  \label{NGC4151spec}
\end{figure}

\begin{figure}[H]
%\epsscale{0.7}
  \plotone{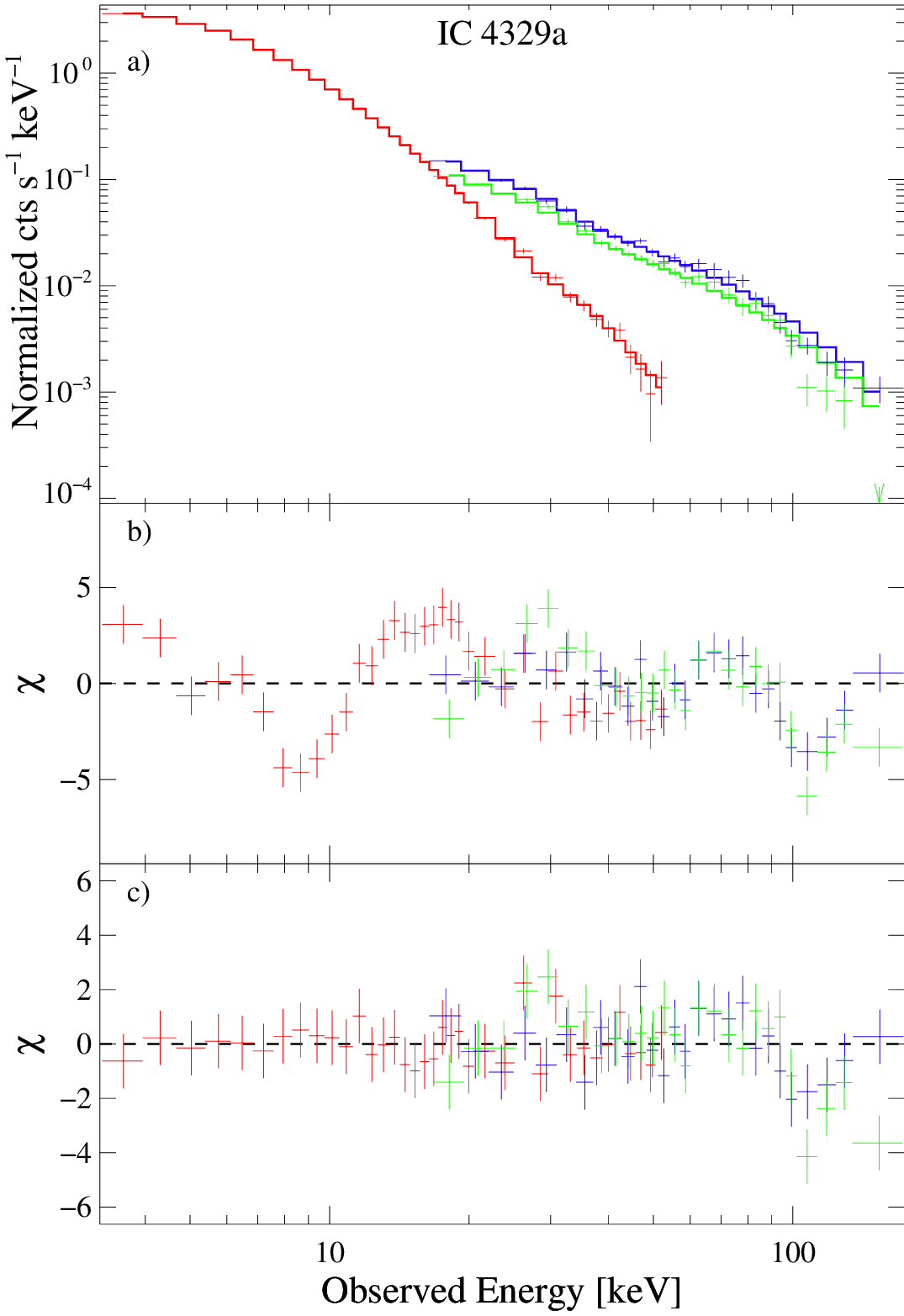}
  \caption{Data, model and data--model residuals for IC~4329a.  Panel (a) shows the PCA and HEXTE data along with the 
           best-fit model (solid line); panel (b) shows residuals for the baseline model; and panel (c) shows residuals
           for the best-fit model (parameters for the best-fit model are listed in Table \ref{tabpexrav}).}
  \label{IC4329Aspec}
\end{figure}

\begin{figure}[H]
 %\epsscale{0.7}
  \plotone{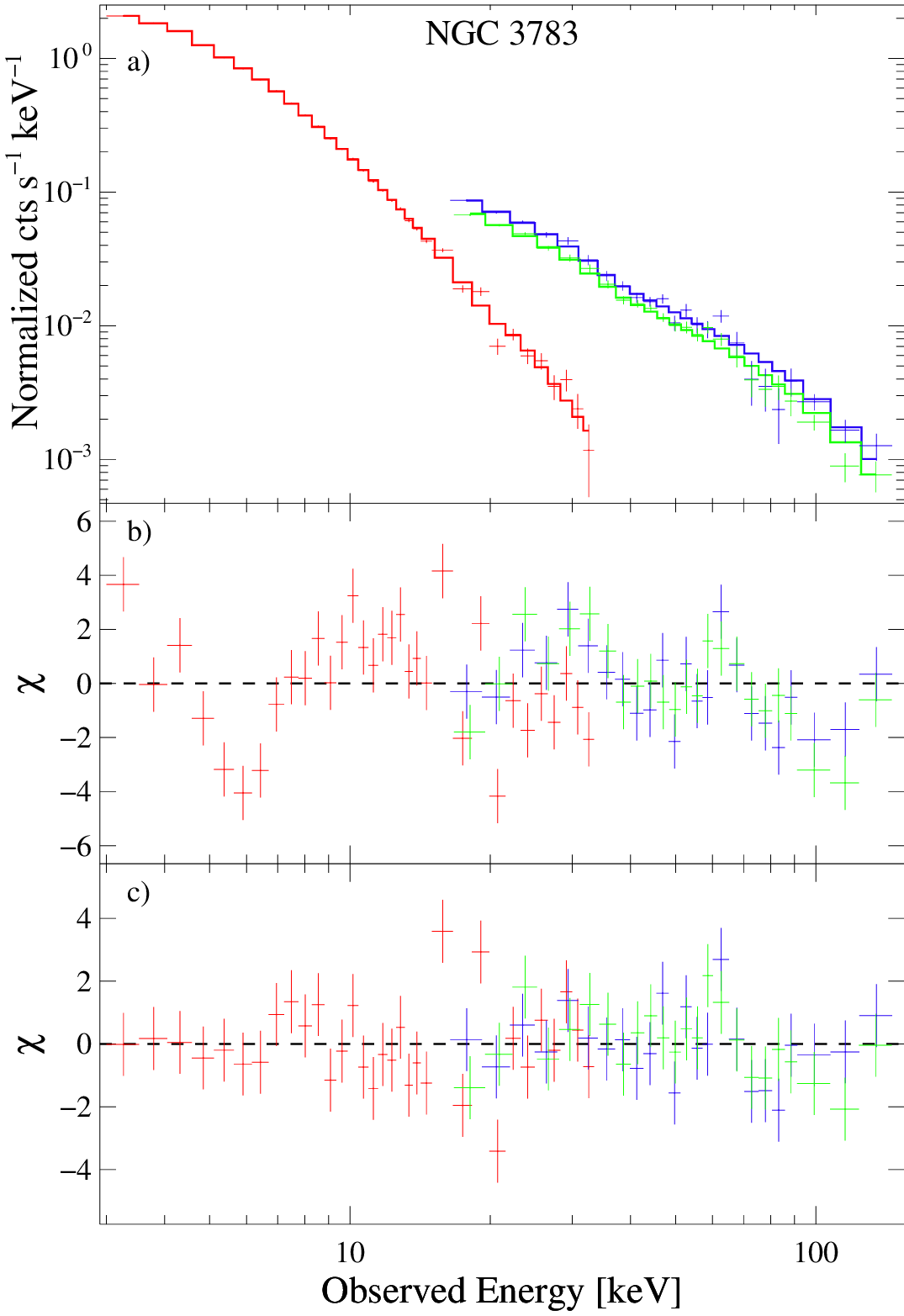}
  \caption{Data, model and data--model residuals for NGC~3783.  Panel (a) shows the PCA and HEXTE data along with the 
           best-fit model (solid line); panel (b) shows residuals for the baseline model; and panel (c) shows residuals
           for the best-fit model (parameters for the best-fit model are listed in Table \ref{tabpexrav}).}
  \label{NGC3783spec}
\end{figure}

\begin{figure}[H]
%\epsscale{0.7}
  \plotone{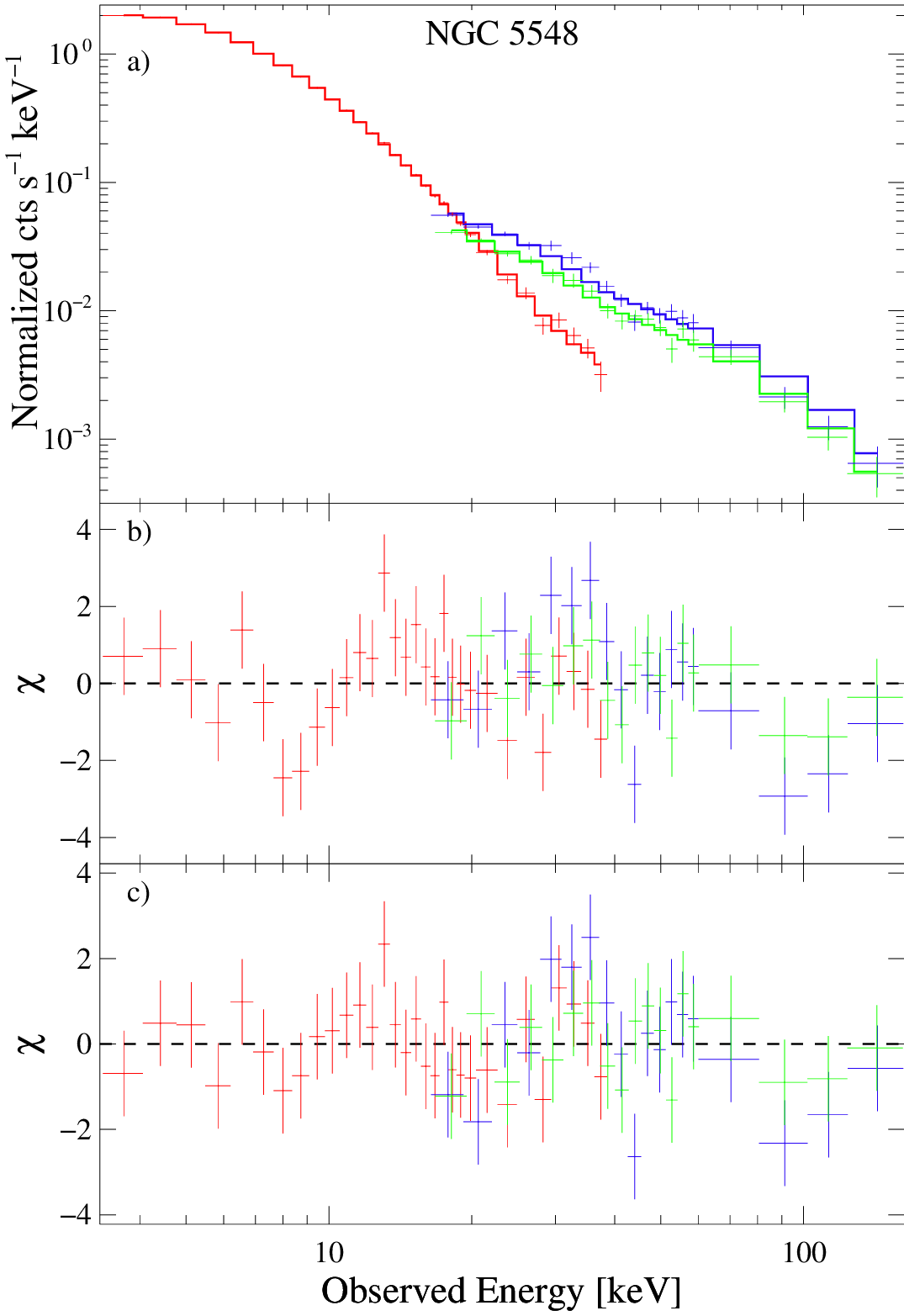}
  \caption{Data, model and data--model residuals for NGC~5548.  Panel (a) shows the PCA and HEXTE data along with the 
           best-fit model (solid line); panel (b) shows residuals for the baseline model; and panel (c) shows residuals
           for the best-fit model (parameters for the best-fit model are listed in Table \ref{tabpexrav}).}
  \label{NGC5548spec}
 \end{figure}

\begin{figure}[H]
 %\epsscale{0.7}
 \plotone{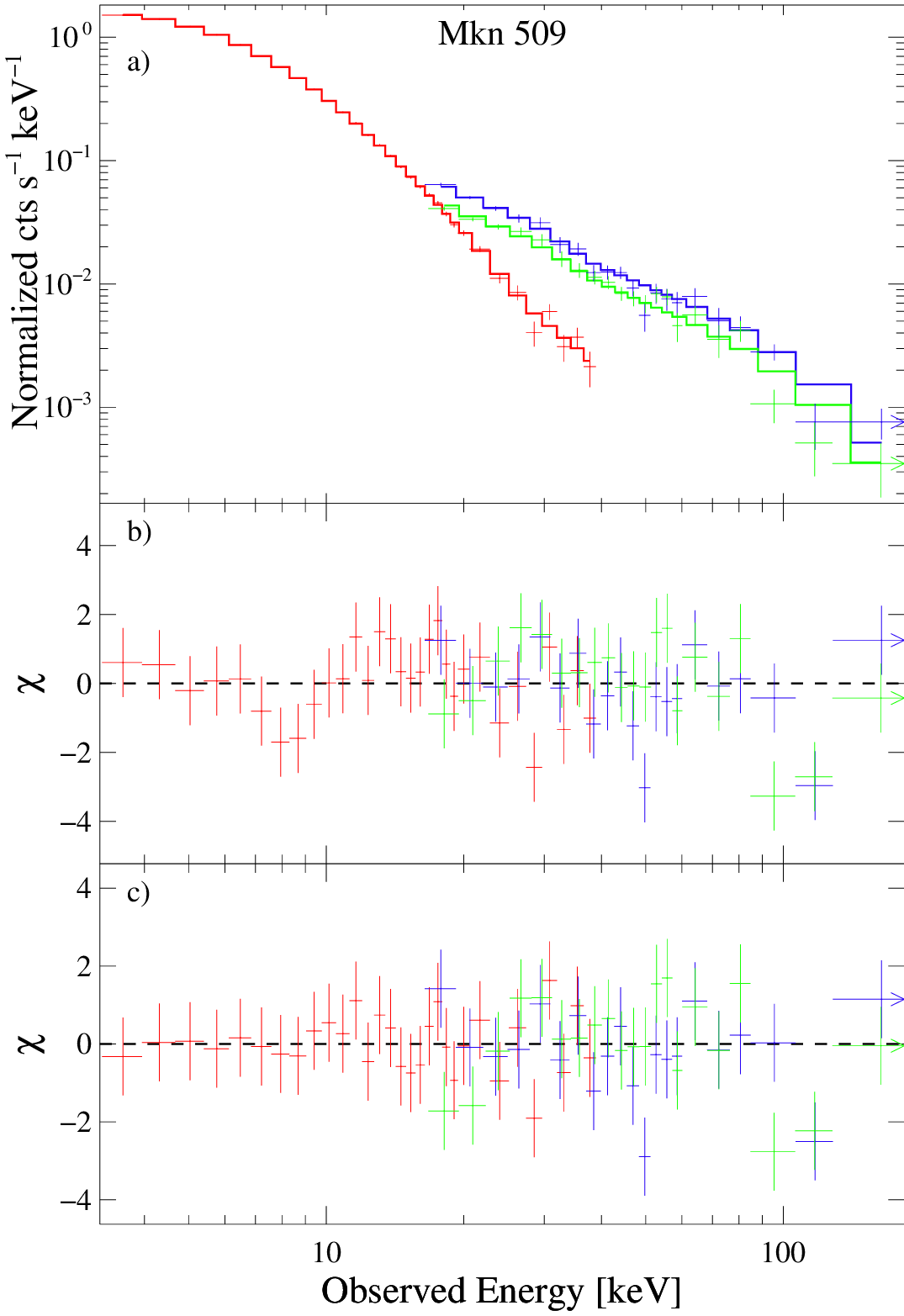}
  \caption{Data, model and data--model residuals for Mkn~509.  Panel (a) shows the PCA and HEXTE data along with the 
           best-fit model (solid line); panel (b) shows residuals for the baseline model; and panel (c) shows residuals
           for the best-fit model (parameters for the best-fit model are listed in Table \ref{tabpexrav}).}
 \label{MKN509spec}
 \end{figure}

\begin{figure}[H]
 \epsscale{0.9}
 \plotone{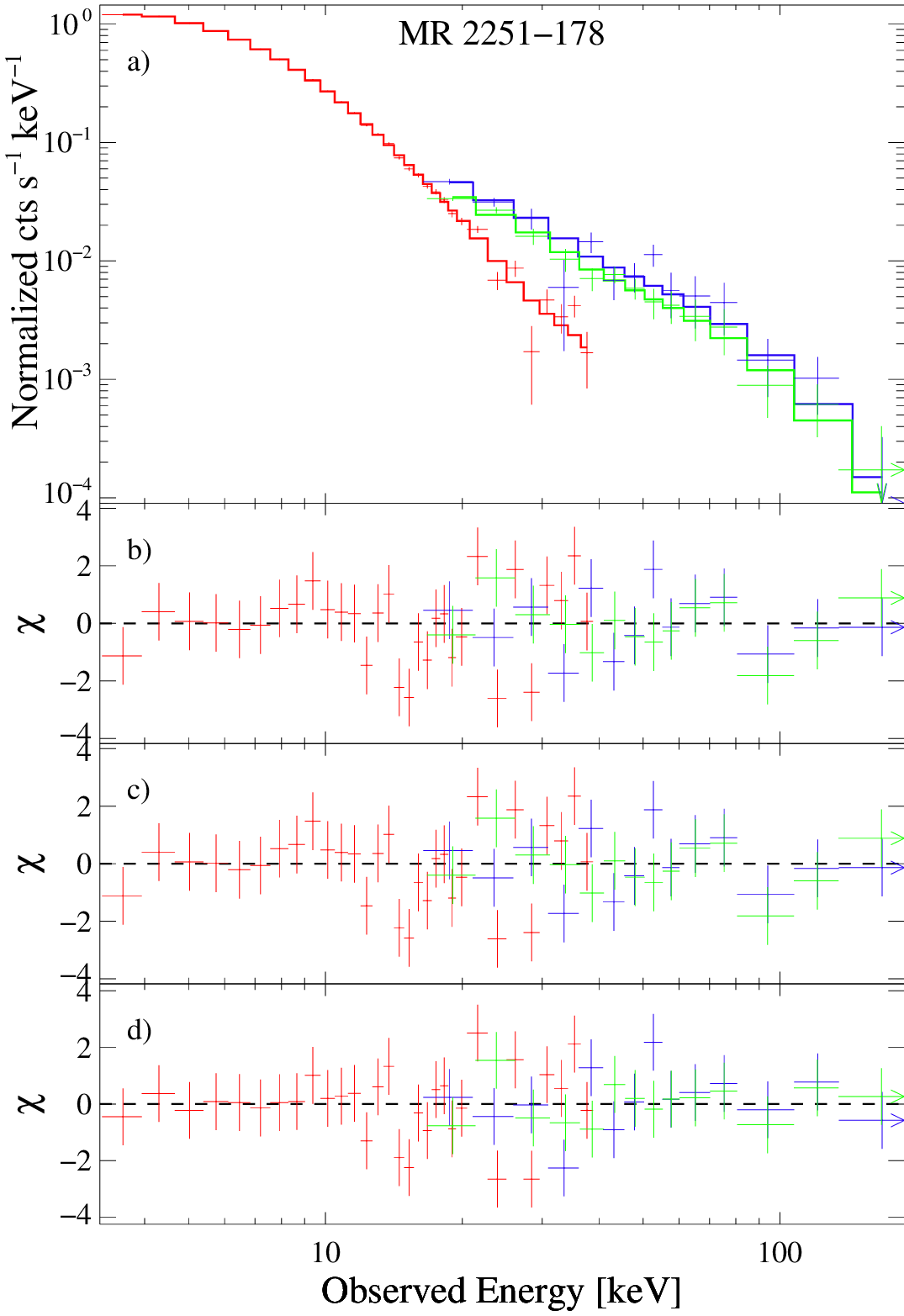}
  \caption{Data, model and data--model residuals for MR2251--178.  Panel (a) shows the PCA and HEXTE data along with the 
           best-fit model (solid line); panel (b) shows residuals for the baseline model; panel (c) showss residuals for the
           \pexrav model; and panel d) shows residuals for the best-fit model with \pexrav and a high-energy rollover 
           (parameters for the best-fit model are listed in Table \ref{tabcutoff}).}
 \label{MR2251-178spec}
\end{figure}

\begin{figure}[H] 
%\epsscale{0.7}
\plotone{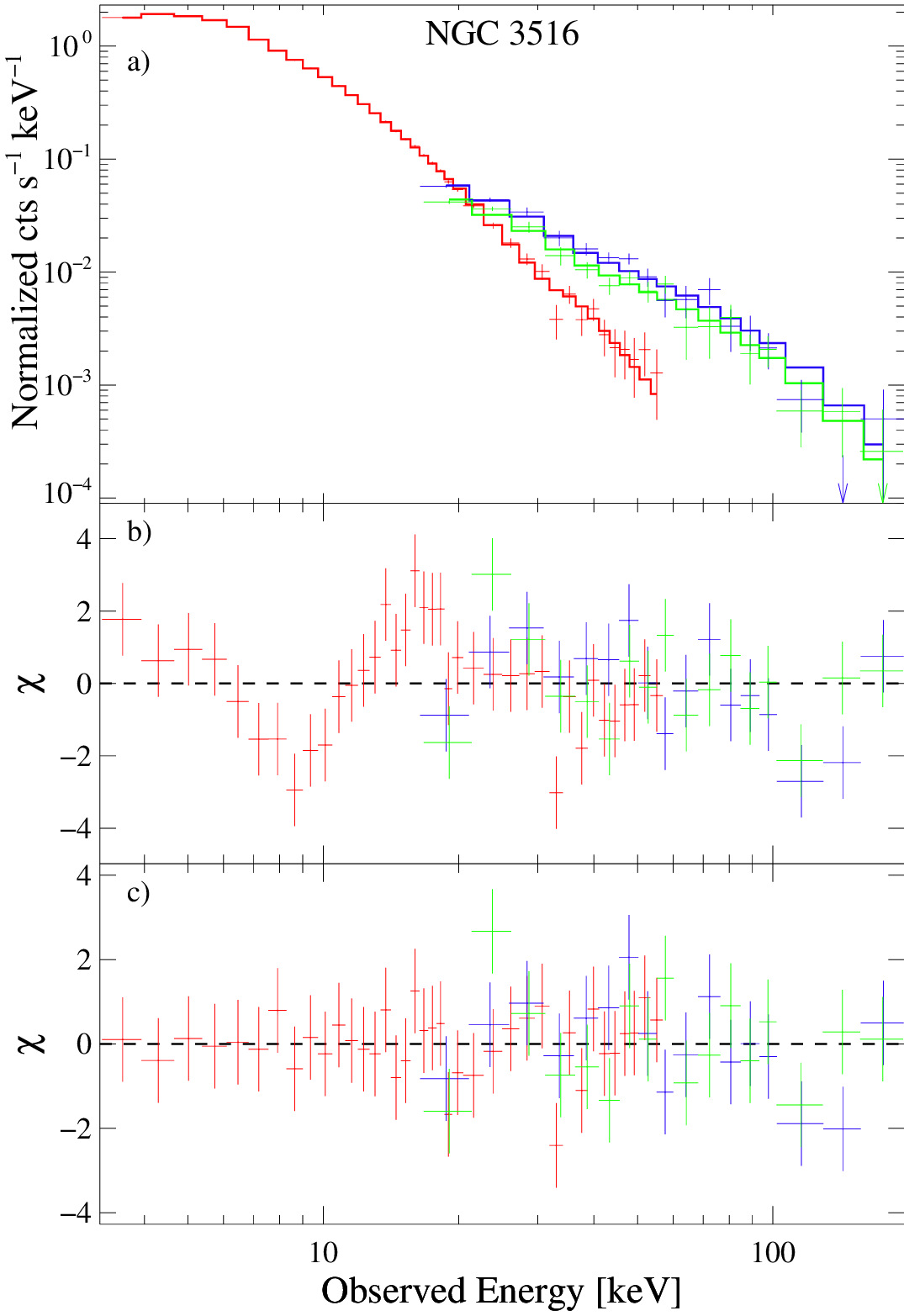}
  \caption{Data, model and data--model residuals for NGC~3516.  Panel (a) shows the PCA and HEXTE data along with the 
           best-fit model (solid line); panel (b) shows residuals for the baseline model; and panel (c) shows residuals
           for the best-fit model (parameters for the best-fit model are listed in Table \ref{tabpexrav}).}
  \label{NGC3516spec}
 \end{figure}

\begin{figure}[H] 
%\epsscale{0.7}
\plotone{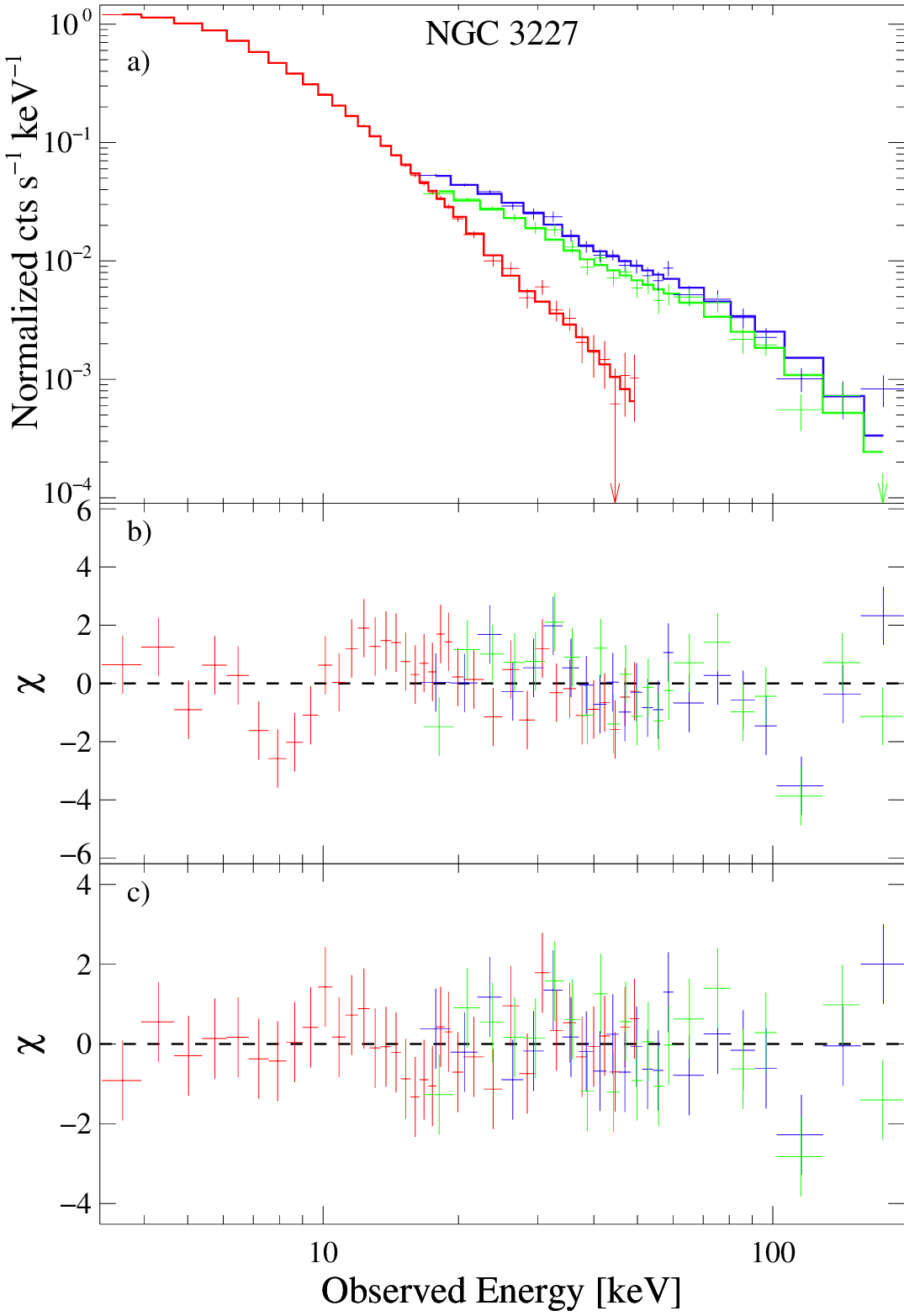}
  \caption{Data, model and data--model residuals for NGC~3227.  Panel (a) shows the PCA and HEXTE data along with the 
           best-fit model (solid line); panel (b) shows residuals for the baseline model; and panel (c) shows residuals
           for the best-fit model (parameters for the best-fit model are listed in Table \ref{tabpexrav}).}
  \label{NGC3227spec}
\end{figure}

\begin{figure}[H]
 %\epsscale{0.7}
 \plotone{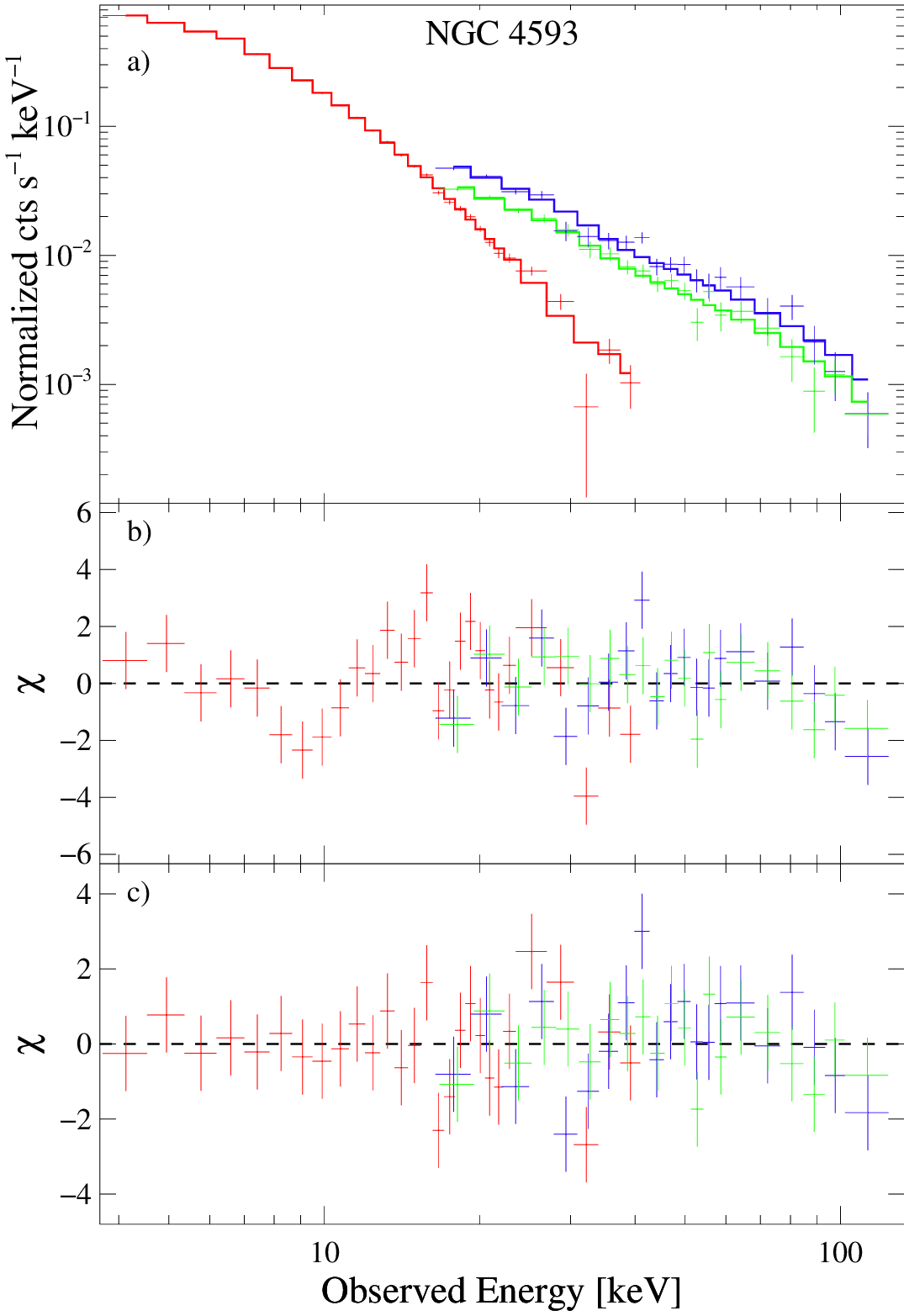}
\caption{Data, model and data--model residuals for NGC~4593.  Panel (a) shows the PCA and HEXTE data along with the 
           best-fit model (solid line); panel (b) shows residuals for the baseline model; and panel (c) shows residuals
           for the best-fit model (parameters for the best-fit model are listed in Table \ref{tabpexrav}).}
   \label{NGC4593spec}
\end{figure}

\begin{figure}[H] 
%\epsscale{0.7}
\plotone{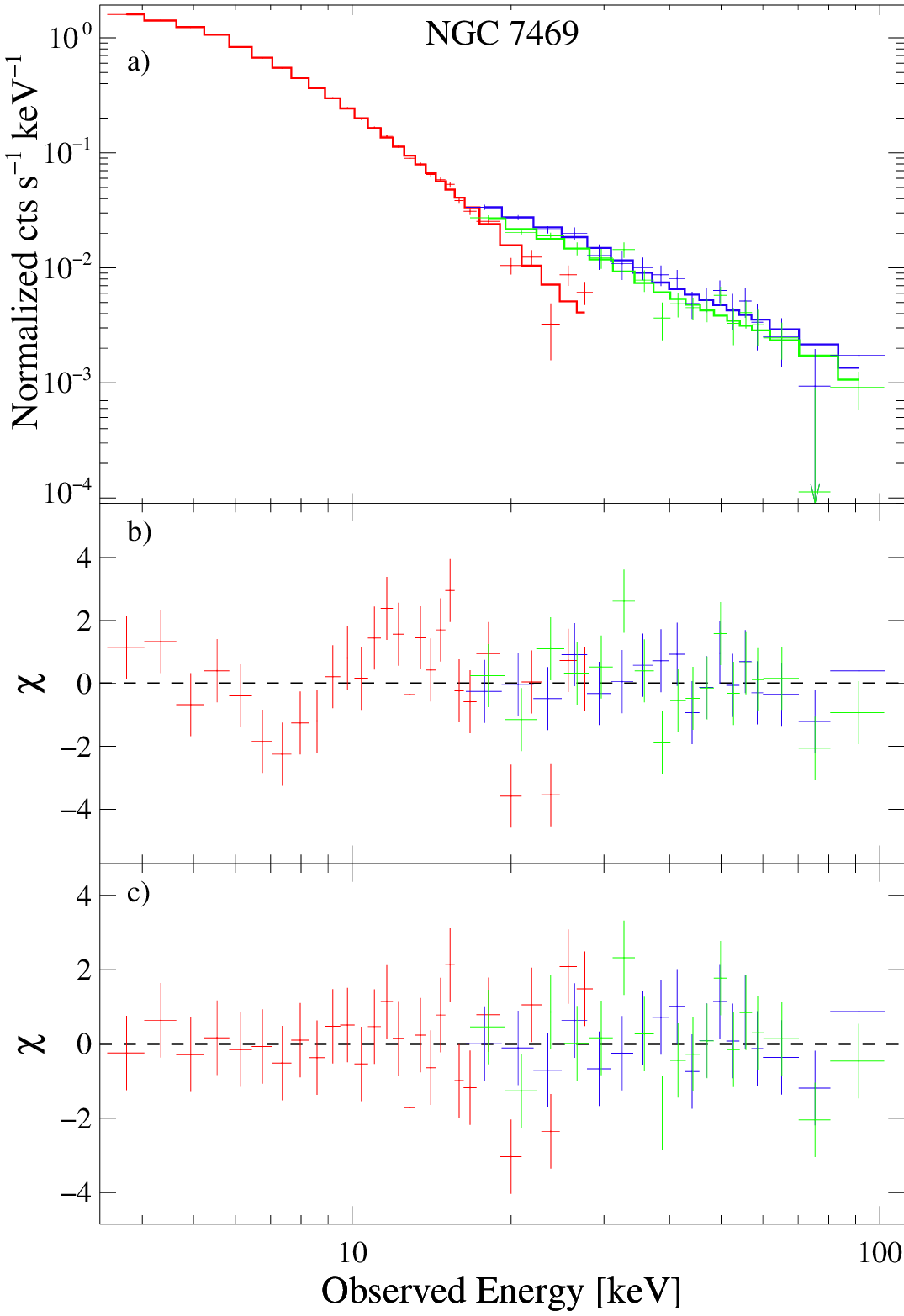}
\caption{Data, model and data--model residuals for NGC~7469.  Panel (a) shows the PCA and HEXTE data along with the 
           best-fit model (solid line); panel (b) shows residuals for the baseline model; and panel (c) shows residuals
           for the best-fit model (parameters for the best-fit model are listed in Table \ref{tabpexrav}).}
   \label{NGC7469spec}
\end{figure}

\begin{figure}[H]
 %\epsscale{0.7}
 \plotone{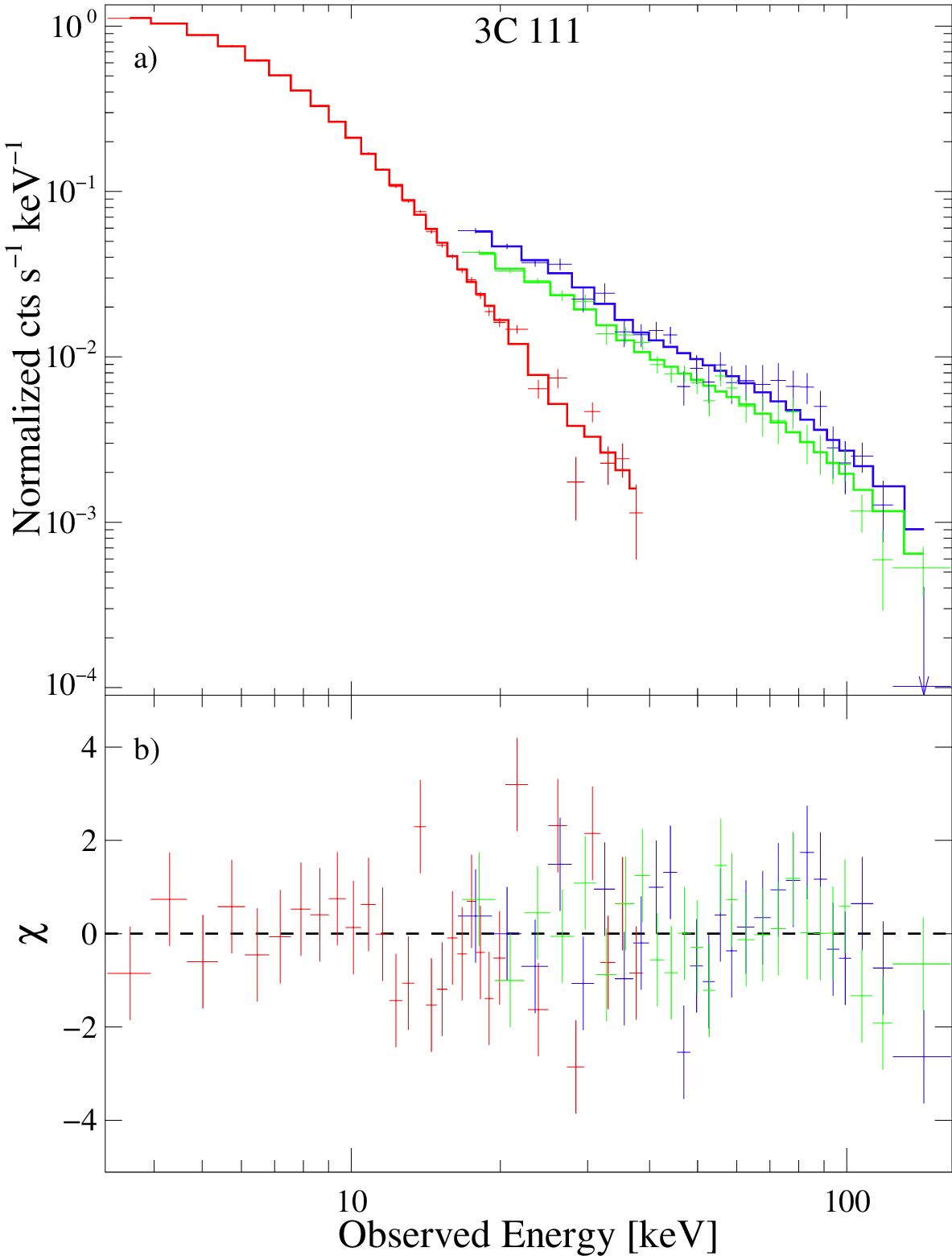}
 \caption{Data, model and data--model residuals for 3C~111.  Panel (a) shows the PCA and HEXTE data along with the 
           best-fit model (solid line); panel (b) shows residuals for the baseline model which is also the best-fit
           model for this source (parameters are listed in Table \ref{tabbase}).}
 \label{3C111spec}
\end{figure}

\begin{figure}[H] 
%\epsscale{0.7}
\plotone{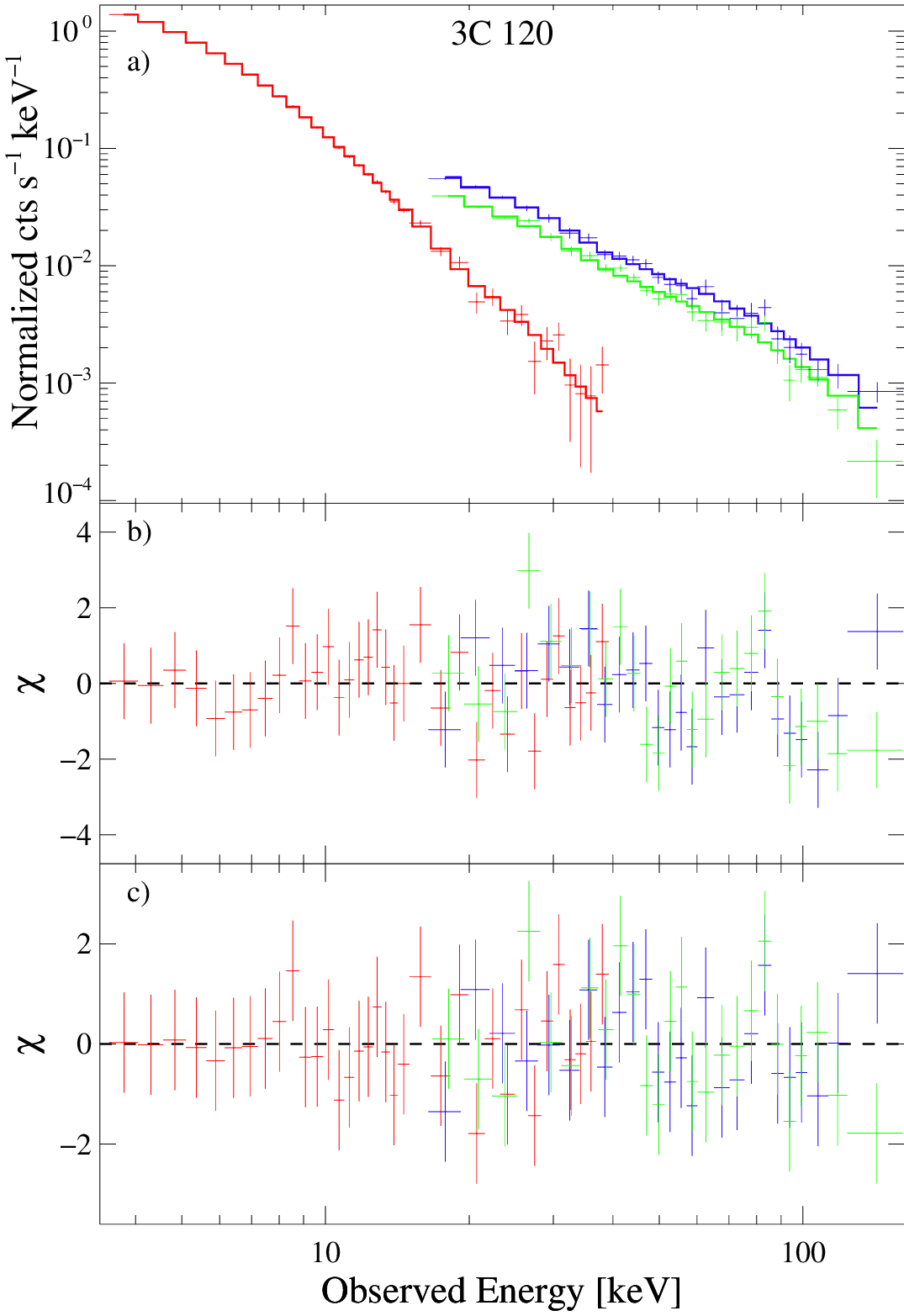}
\caption{Data, model and data--model residuals for 3C~120.  Panel (a) shows the PCA and HEXTE data along with the 
           best-fit model (solid line); panel (b) shows residuals for the baseline model; and panel (c) shows residuals
           for the best-fit model (parameters for the best-fit model are listed in Table \ref{tabpexrav}).}
   \label{3C120spec}
\end{figure}

\begin{figure}[H]
 %\epsscale{0.7}
 \plotone{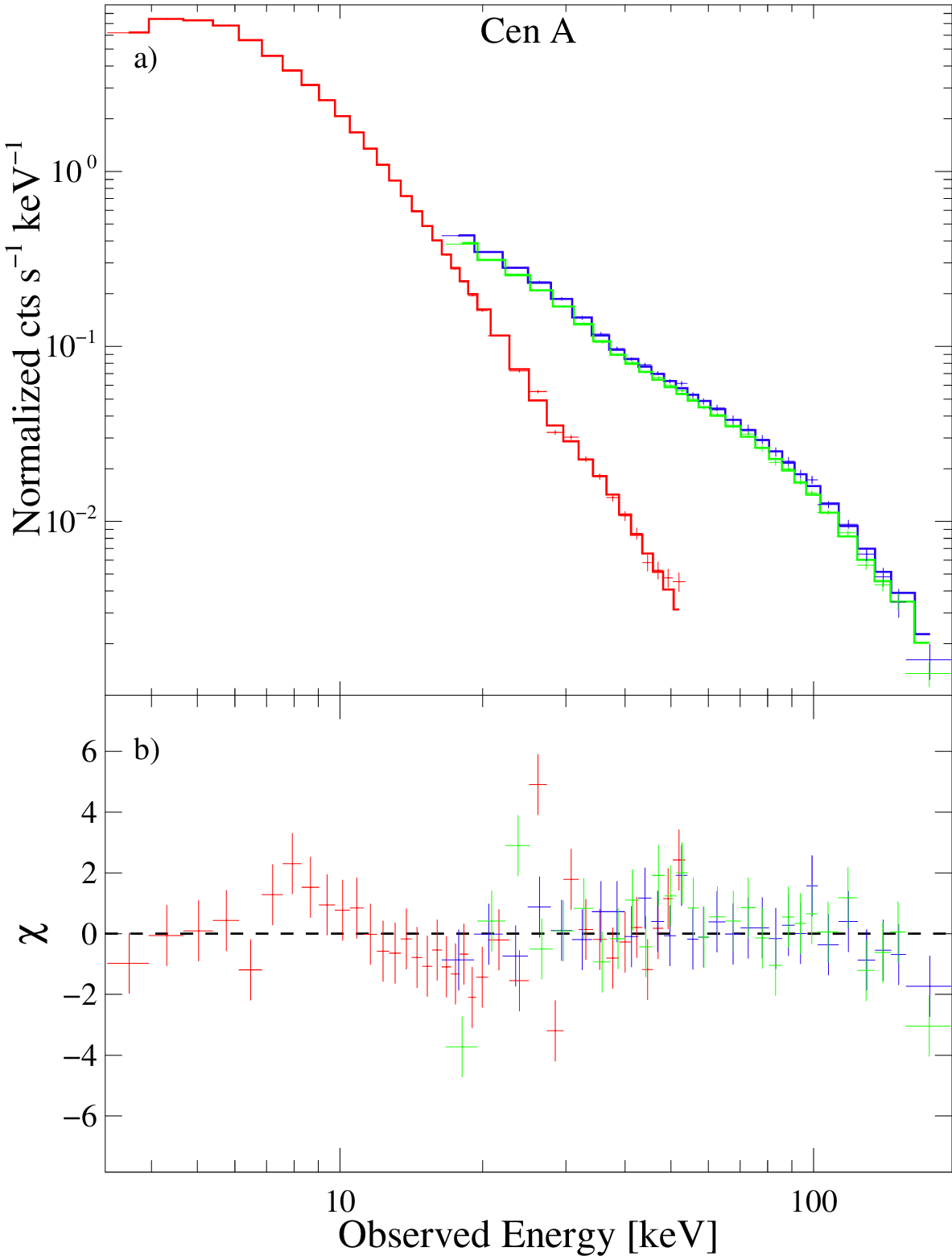}
 \caption{Data, model and data--model residuals for Cen~A.  Panel (a) shows the PCA and HEXTE data along with the 
           best-fit model (solid line); panel (b) shows residuals for the baseline model which is also the best-fit
           model for this source (parameters are listed in Table \ref{tabbase}).}
 \label{CENAspec}

\end{figure}

\begin{figure}[H]
 %\epsscale{0.7}
 \plotone{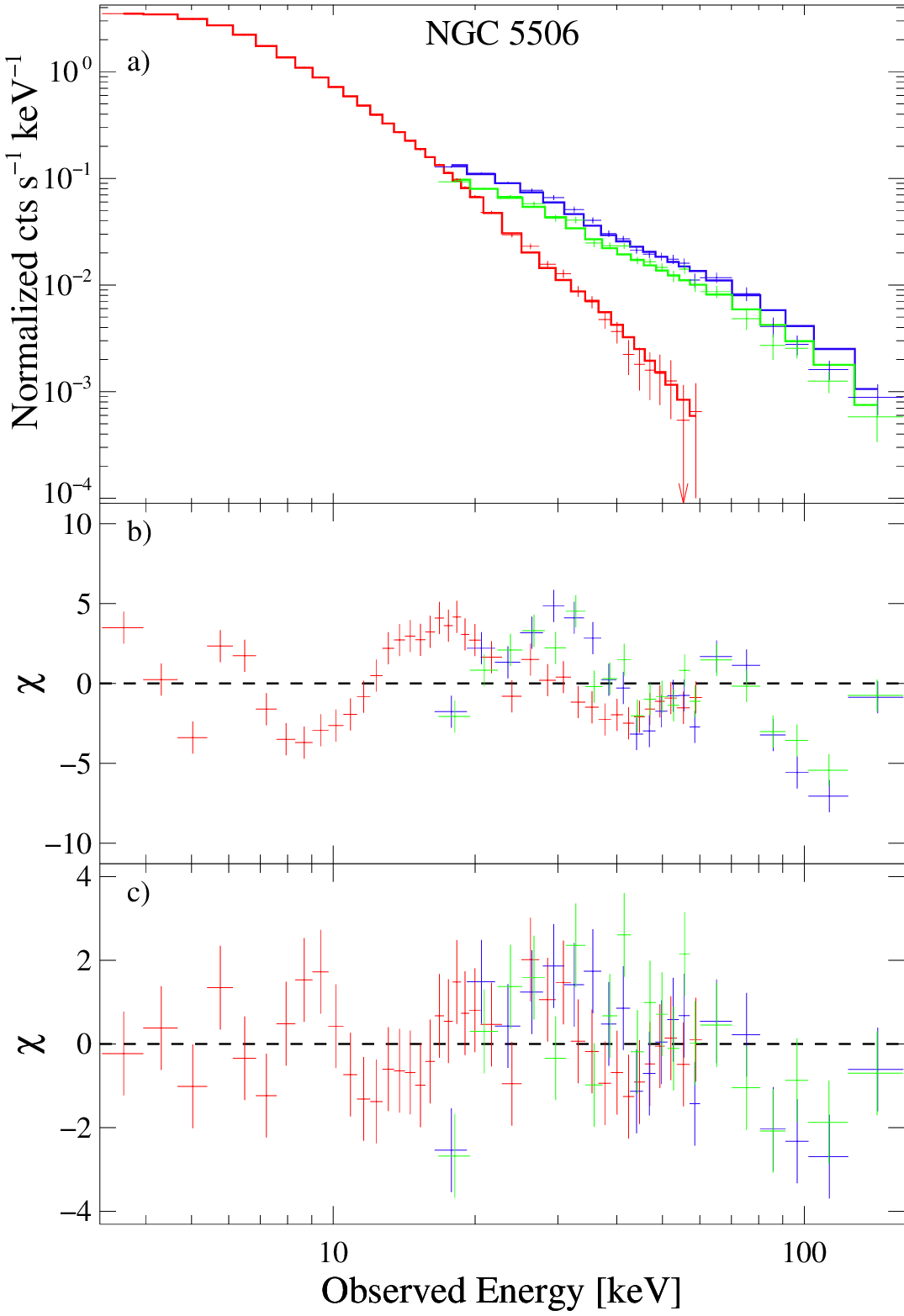}
 \caption{Data, model and data--model residuals for NGC~5506.  Panel (a) shows the PCA and HEXTE data along with the 
           best-fit model (solid line); panel (b) shows residuals for the baseline model; and panel (c) shows residuals
           for the best-fit model (parameters for the best-fit model are listed in Table \ref{tabpexrav}).}
 \label{NGC5506spec}
\end{figure}

\begin{figure}[H]
 %\epsscale{0.7}
 \plotone{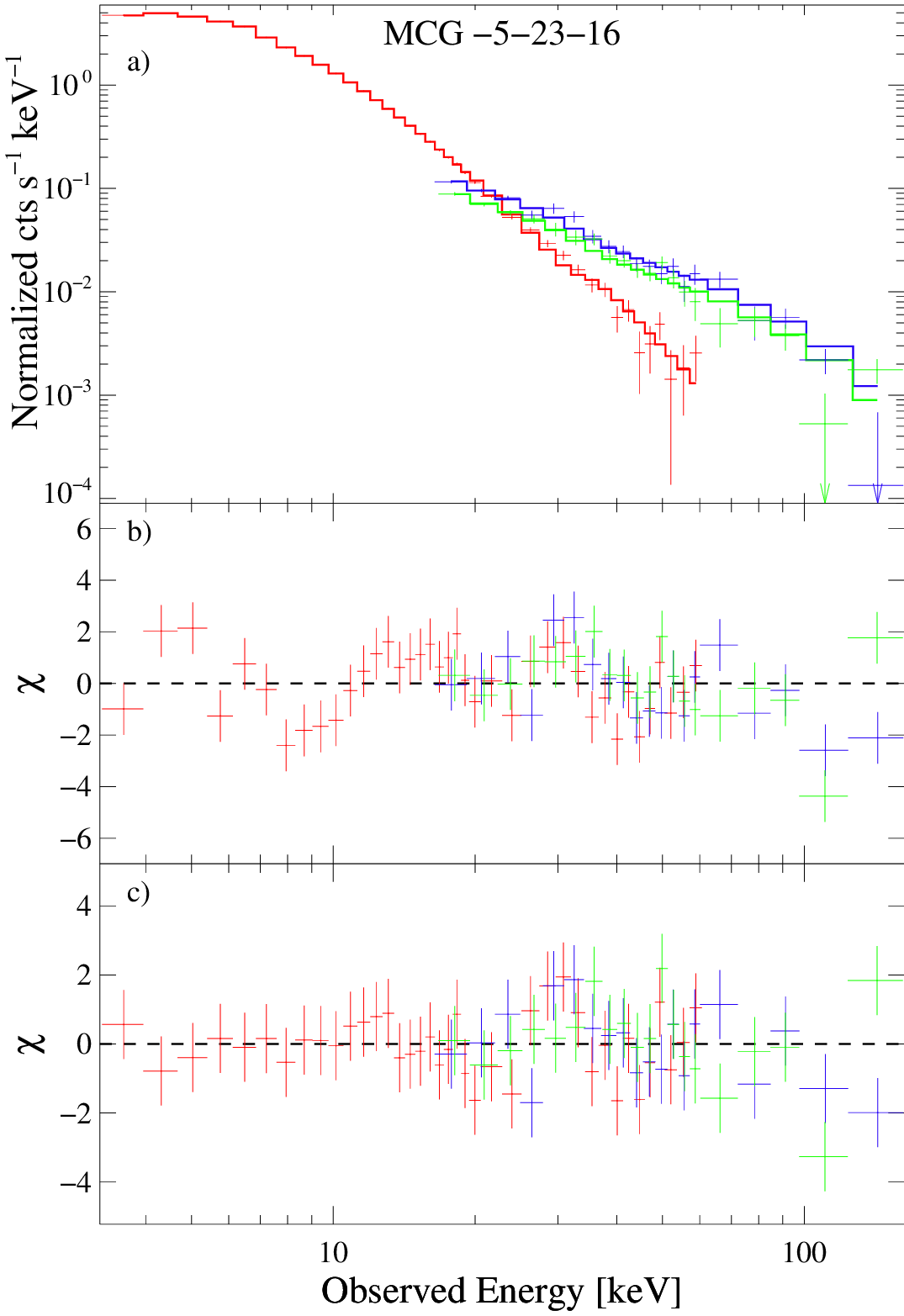}
 \caption{Data, model and data--model residuals for MCG--5-23-16.  Panel (a) shows the PCA and HEXTE data along with the 
           best-fit model (solid line); panel (b) shows residuals for the baseline model; and panel (c) shows residuals
           for the best-fit model (parameters for the best-fit model are listed in Table \ref{tabpexrav}).}
 \label{MCG-5-23-16spec}
\end{figure}

\begin{figure}[H]
 %\epsscale{0.7}
 \plotone{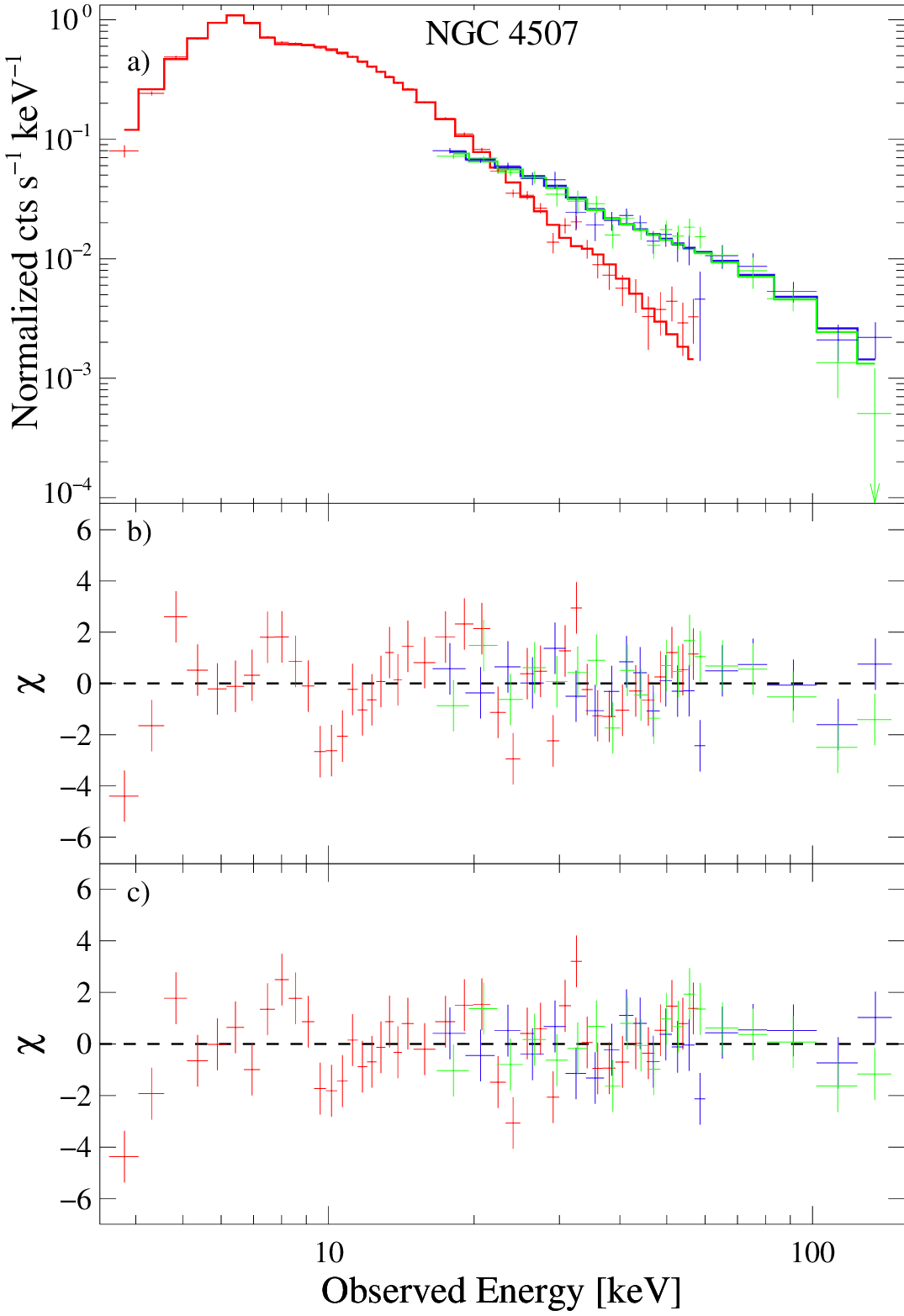}
 \caption{Data, model and data--model residuals for NGC~4507.  Panel (a) shows the PCA and HEXTE data along with the 
           best-fit model (solid line); panel (b) shows residuals for the baseline model; and panel (c) shows residuals
           for the best-fit model (parameters for the best-fit model are listed in Table \ref{tabpexrav}).}
 \label{NGC4507spec}
 \end{figure}

\begin{figure}[H]
 \epsscale{0.9}
 \plotone{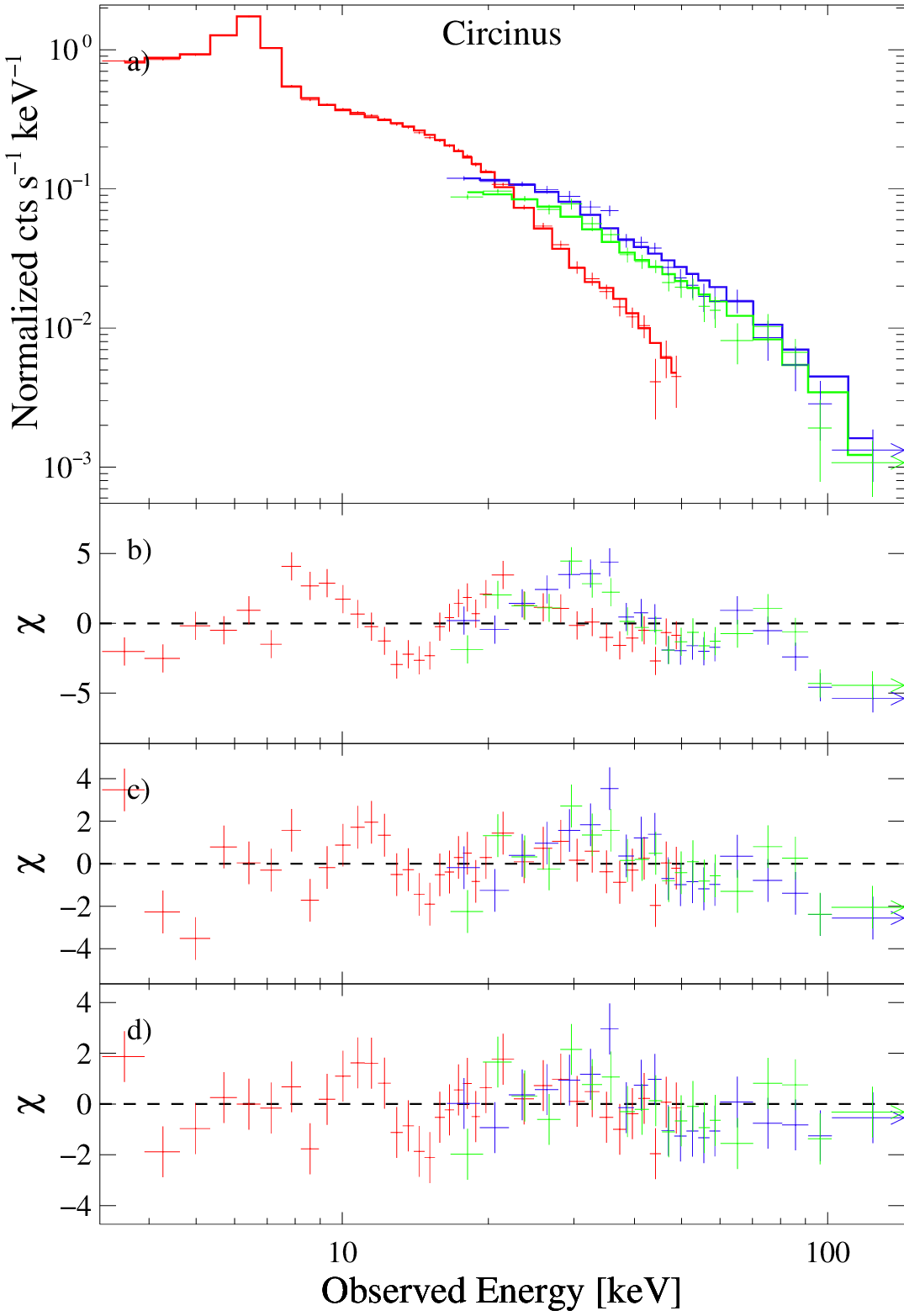}
 \caption{Data, model and data--model residuals for Circinus.  Panel (a) shows the PCA and HEXTE data along with the 
           best-fit model (solid line); panel (b) shows residuals for the baseline model; panel (c) showss residuals for the
           \pexrav model; and panel d) shows residuals for the best-fit model with \pexrav and a high-energy rollover 
           (parameters for the best-fit model are listed in Table \ref{tabcutoff}).}
 \label{CIRCINUSspec}
\end{figure}

\begin{figure}[H]
 %\epsscale{0.7}
 \plotone{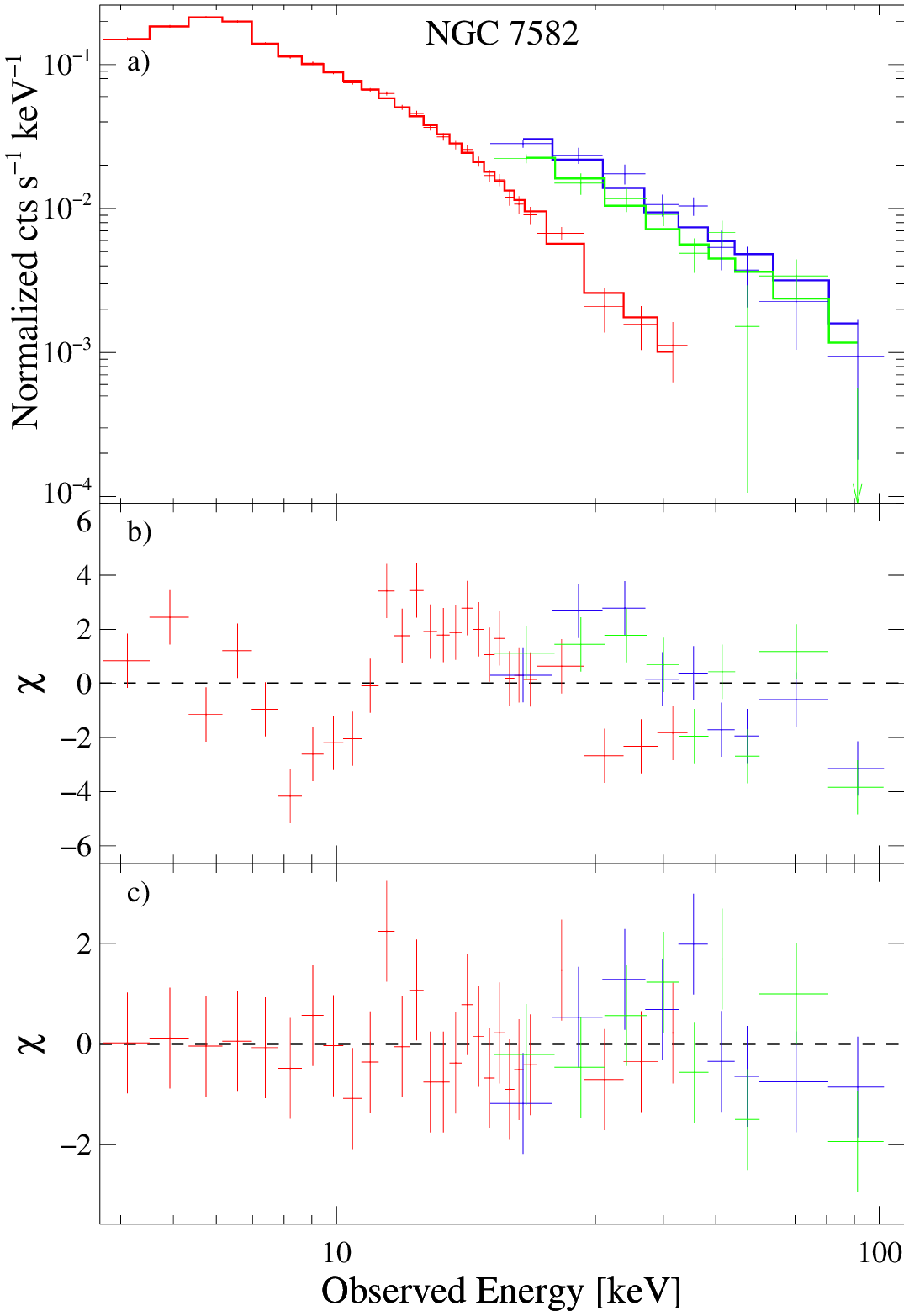}
 \caption{Data, model and data--model residuals for NGC~7582.  Panel (a) shows the PCA and HEXTE data along with the 
           best-fit model (solid line); panel (b) shows residuals for the baseline model; and panel (c) shows residuals
           for the best-fit model (parameters for the best-fit model are listed in Table \ref{tabpexrav}).}
 \label{NGC7582spec}
\end{figure}

\begin{figure}[H]
 \epsscale{0.9}
 \plotone{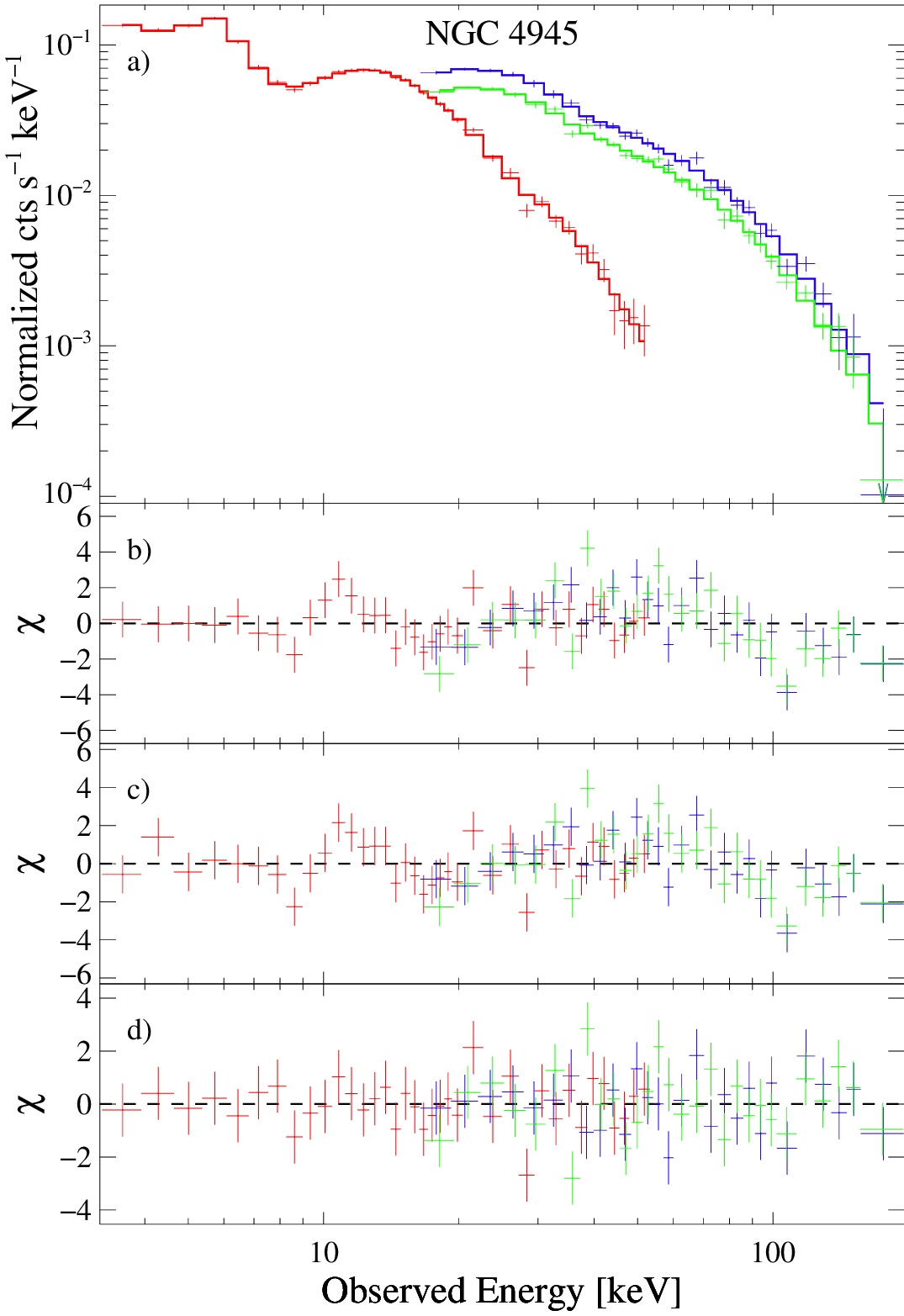}
  \caption{Data, model and data--model residuals for NGC~4945.  Panel (a) shows the PCA and HEXTE data along with the 
           best-fit model (solid line); panel (b) shows residuals for the baseline model; panel (c) showss residuals for the
           \pexrav model; and panel d) shows residuals for the best-fit model with \pexrav and a high-energy rollover 
           (parameters for the best-fit model are listed in Table \ref{tabcutoff}).}
 \label{NGC4945spec}
\end{figure}

\begin{figure}[H]
 %\epsscale{0.9}
 \plotone{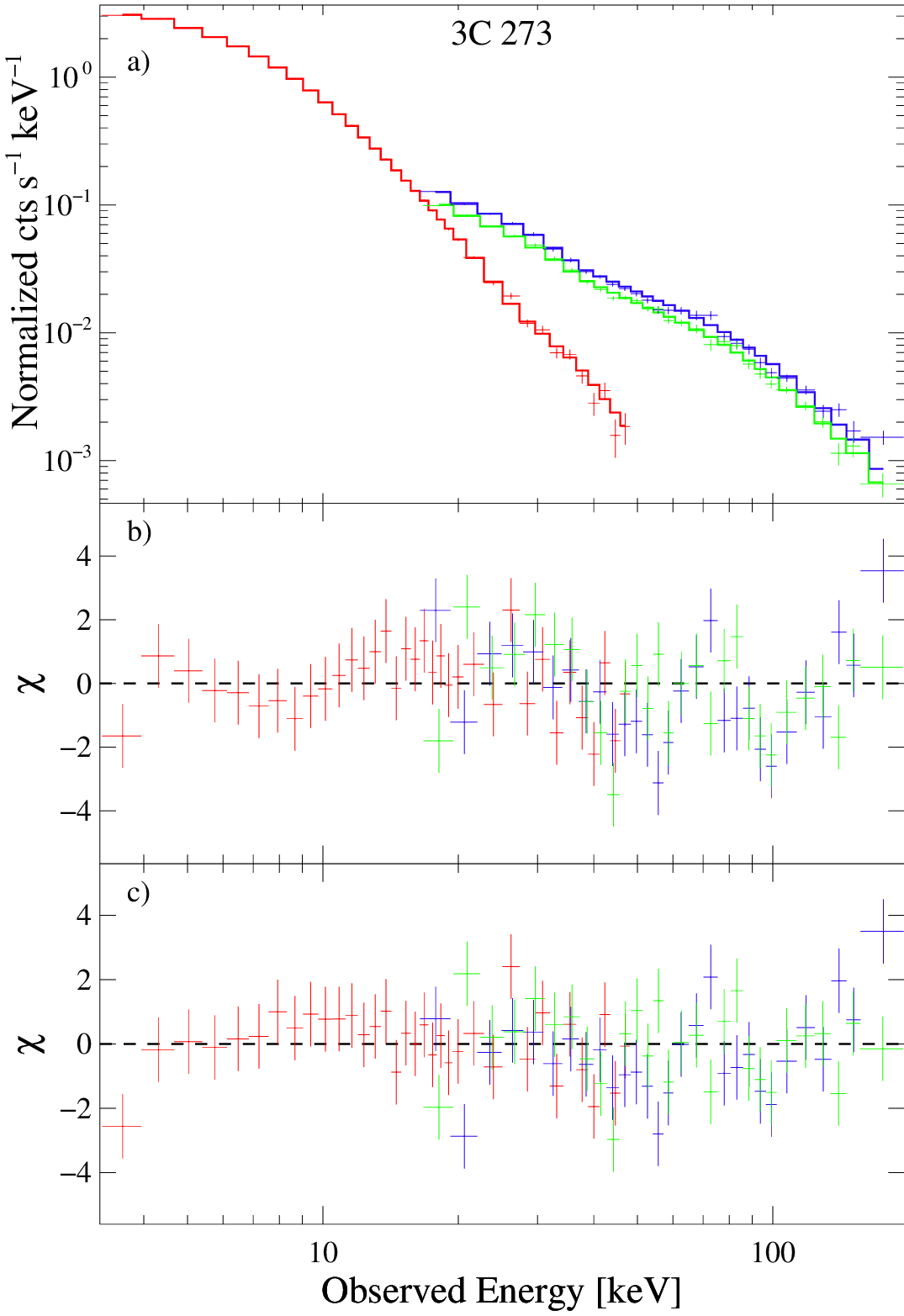}
\caption{Data, model and data--model residuals for 3C~273.  Panel (a) shows the PCA and HEXTE data along with the 
           best-fit model (solid line); panel (b) shows residuals for the baseline model; and panel (c) shows residuals
           for the best-fit model (parameters for the best-fit model are listed in Table \ref{tabpexrav}).}
 \label{3C273spec}
\end{figure}

\begin{figure}[H]
 %\epsscale{0.7}
 \plotone{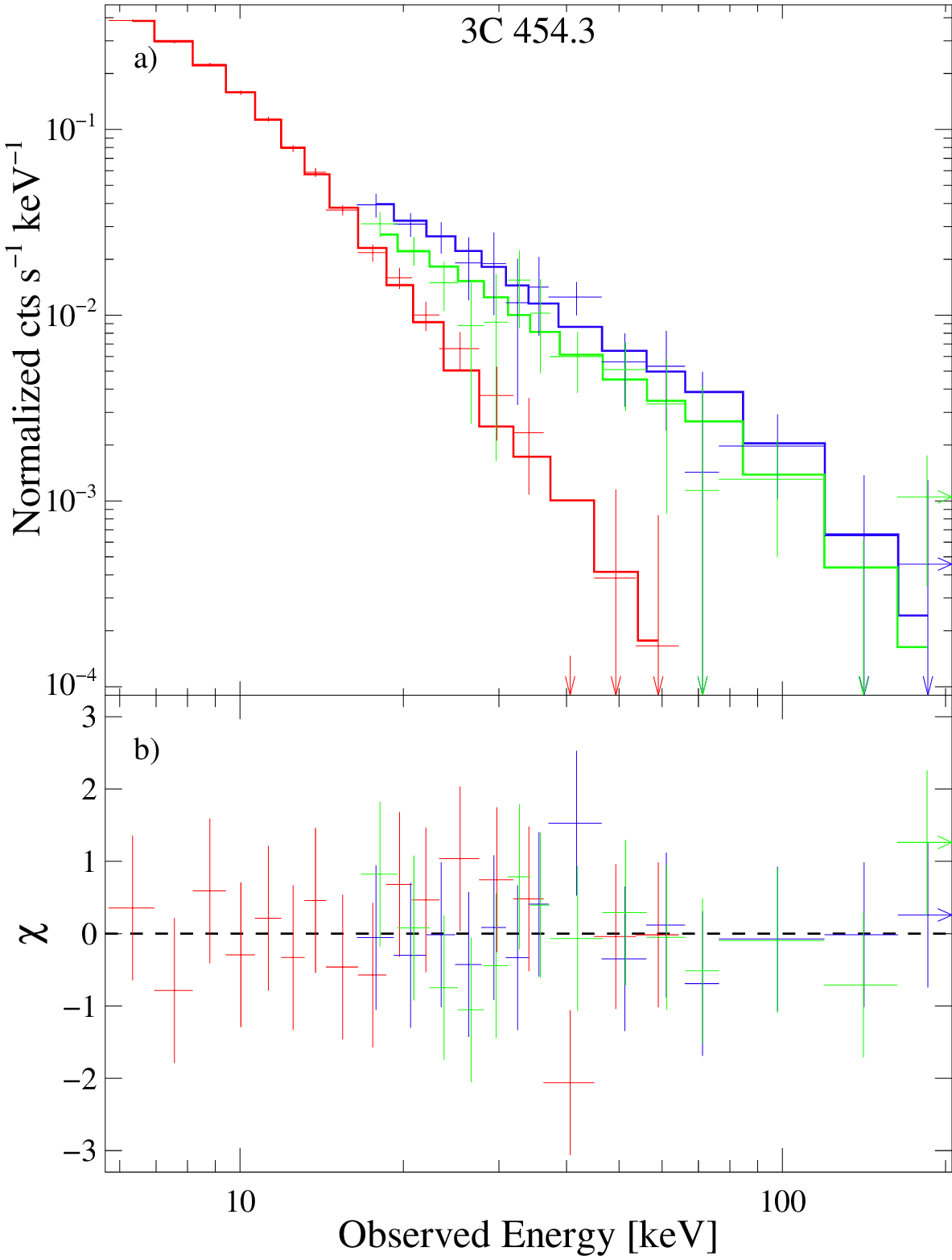}
\caption{Data, model and data--model residuals for 3C~454.3.  Panel (a) shows the PCA and HEXTE data along with the 
           best-fit model (solid line); panel (b) shows residuals for the best-fit model (parameters are listed in 
           Table \ref{tabblaz}).}
   \label{3C454.3spec}
\end{figure}

\begin{figure}[H]
 %\epsscale{0.7}
 \plotone{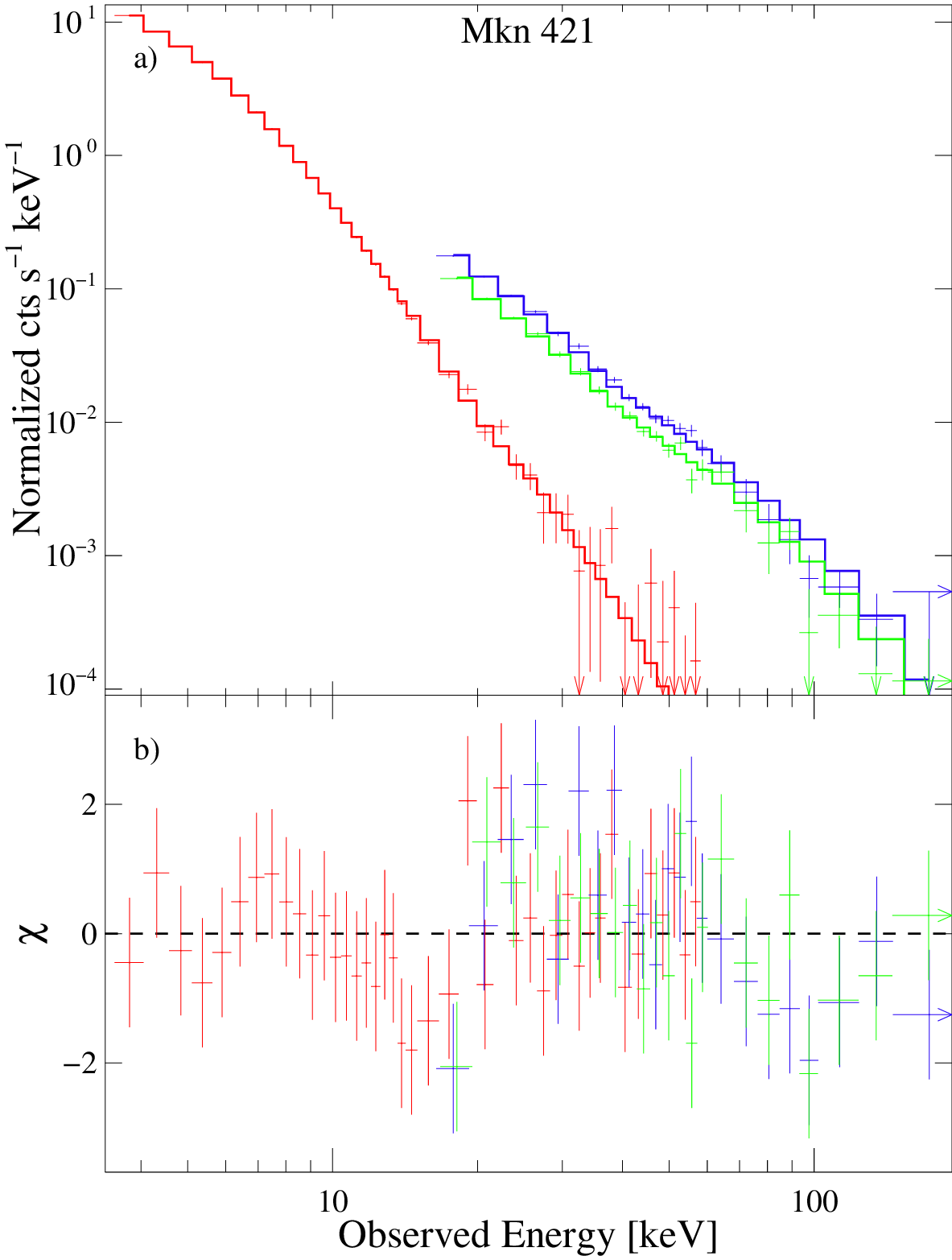}
 \caption{Data, model and data--model residuals for Mkn~421.  Panel (a) shows the PCA and HEXTE data along with the 
           best-fit model (solid line); panel (b) shows residuals for the best-fit model (parameters are listed in 
           Table \ref{tabblaz}).}
  \label{MKN421spec}
\end{figure}

\begin{figure}[H]
 %\epsscale{0.7}
 \plotone{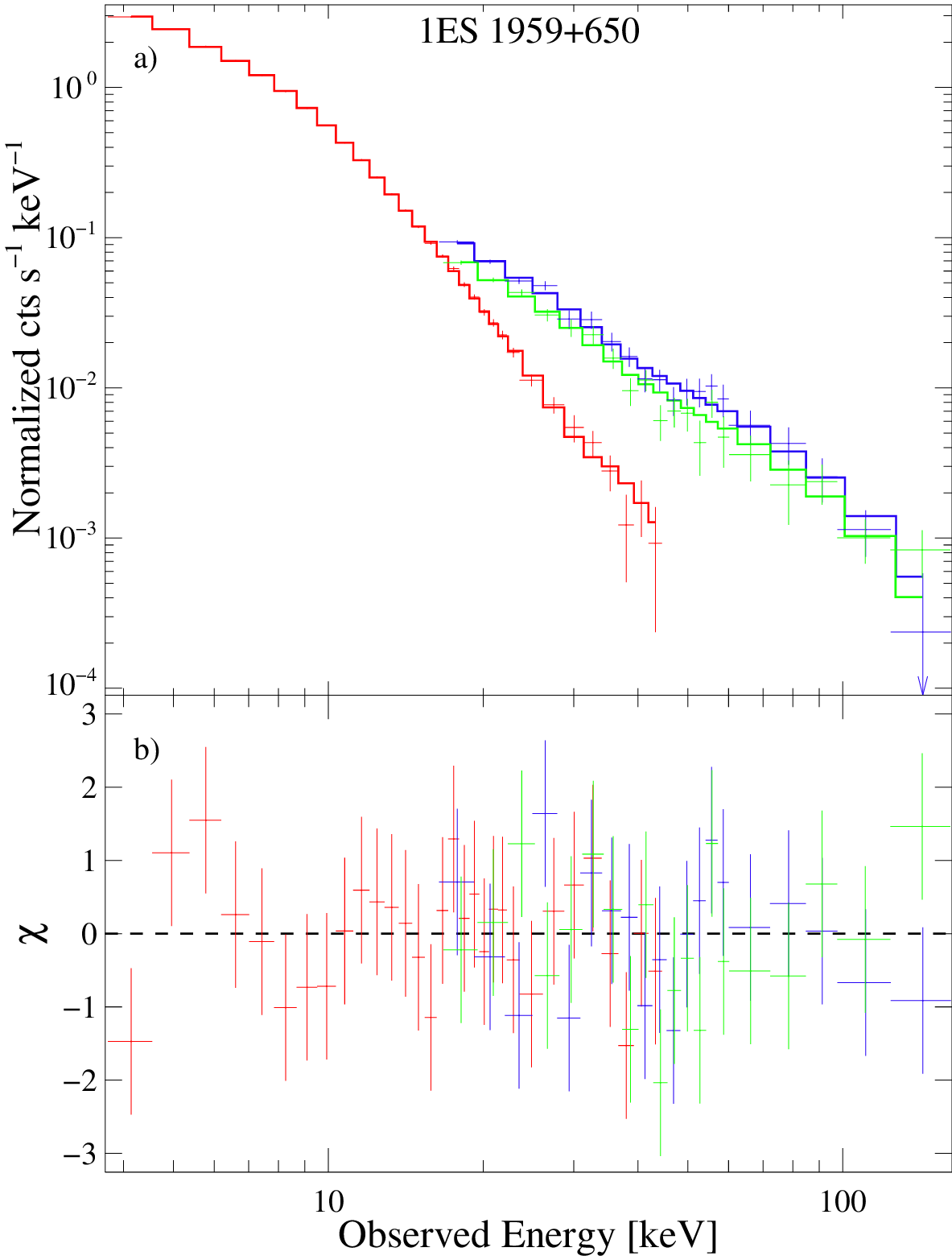}
 \caption{Data, model and data--model residuals for 1ES~1959+650.  Panel (a) shows the PCA and HEXTE data along with the 
           best-fit model (solid line); panel (b) shows residuals for the best-fit model (parameters are listed in 
           Table \ref{tabblaz}).}
  \label{1ES1959+650spec} 
\end{figure}

%==================================================================%

\begin{figure}[H]
\plotone{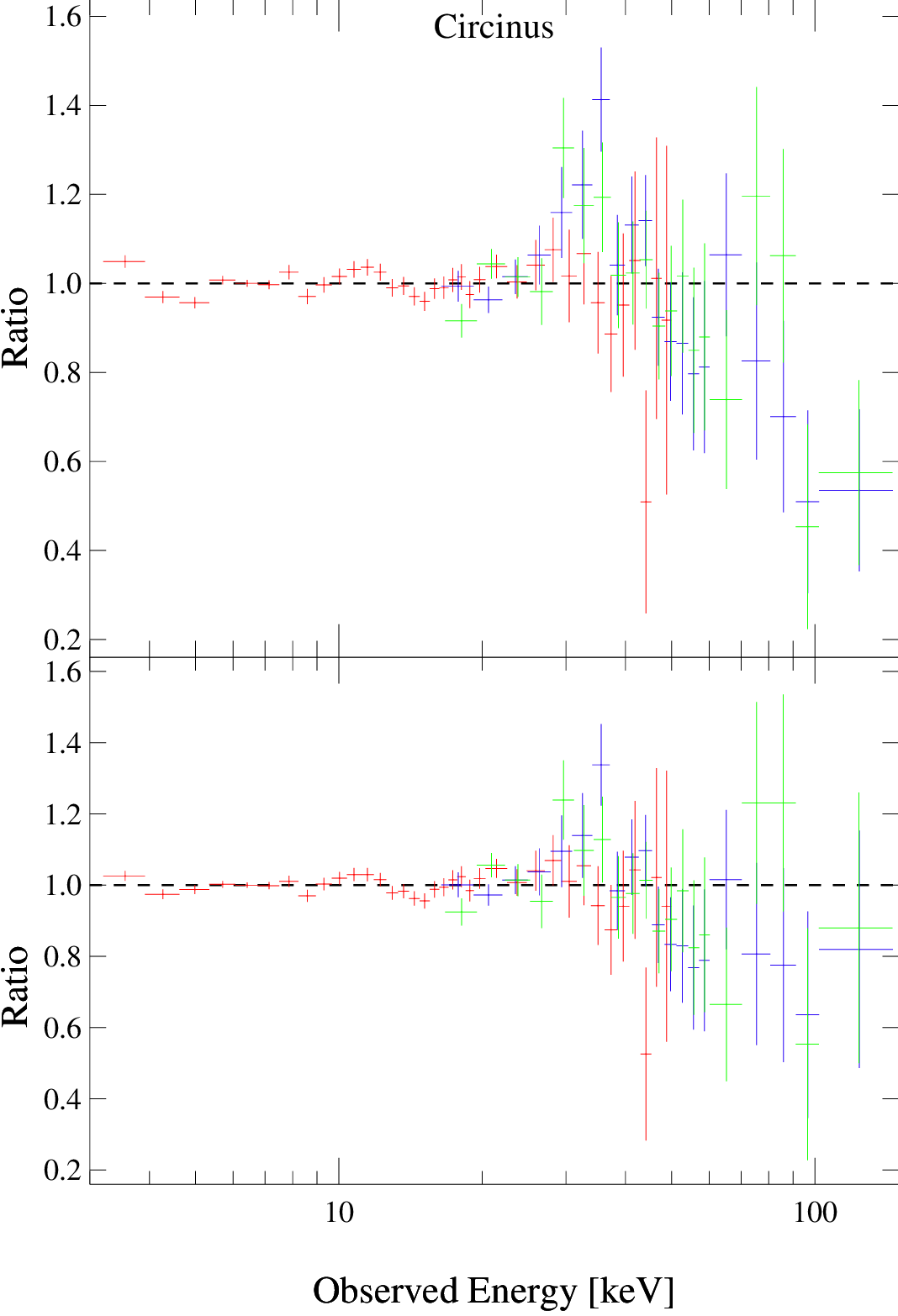}
\caption{Data to model ratios for the best-fit reflection model ({\it top}) and best-fit rollover model ({\it bottom}). 
Parameters are given in Tables \ref{tabpexrav} and \ref{tabcutoff} respectively. }
   \label{CIRCrat}
\end{figure}

\begin{figure}[H]
\plotone{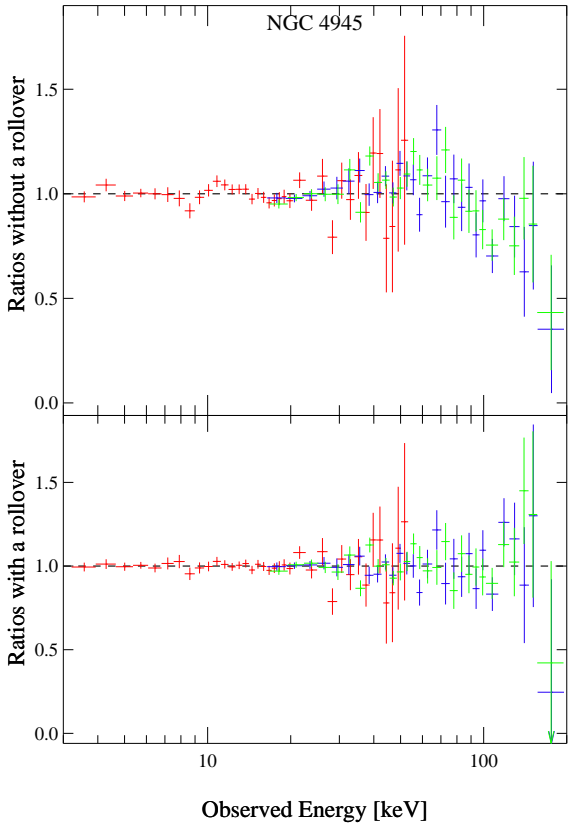}
\caption{Data to model ratios for the best-fit reflection model ({\it top}) and best-fit rollover model ({\it bottom}).
Parameters are given in Tables \ref{tabpexrav} and \ref{tabcutoff} respectively. }
   \label{N4945rat}
\end{figure}

%==================================================================%

%---------------------------------------%  This is Section 4.1
\subsection{Comparisons to Previous Surveys}

Most hard X-ray surveys of the extragalactic sky have concentrated on flux measurements and generation of luminosity functions,
e.g.\ Piccinotti \etal (1982; \textit{HEAO-1}-A2), Tueller \etal (2008; \swiftbat), and Bassani \etal (2006; \integral-IBIS).
Fits to spectra using non-simultaneous observations by various missions, e.g., Gondek \etal (1996; \textit{Ginga}, \textsl{CGRO}-OSSE, \textit{EXOSAT}, 
\textit{HEAO-1}), Winter \etal (2008; \swiftbat, \xmm), and Molina \etal (2009; \integral-IBIS, \xmm, \chandra, \textsl{ASCA}), 
have yielded interesting insights, but the lack of simultaneity and
systematic uncertainties in cross-instrument normalizations has precluded definitive conclusions.

The first X-ray spectral surveys of Seyfert AGN were compiled by Mushotzky \etal (1980), Rothschild \etal (1983), and Mushotzky (1984)  
using broadband data from \textit{HEAO-1}-A2 and A4. Using mainly simple power-law or broken power-law models, 
these surveys established that the dominating X-ray continuum displayed a moderately narrow range of $\Gamma$.
\textit{EXOSAT} and \textit{Ginga} observations established the necessity to model the Fe K$\alpha$ line 
(e.g., Nandra et al.\ 1989, Pounds et al.\ 1989).  Nandra \& Pounds (1994) provided the next major spectral compilation, 
of 27 Seyferts observed by \textsl{Ginga} in the 1.5--37 keV band.  They found the need to model the Compton reflection hump 
and found a mean reflection strength of $\langle R \rangle =1.60\pm 0.06$.  They found a mean photon index of 
$\Gamma$=1.95$\pm0.05$ with a dispersion of $\sigma$=0.15$\pm 0.04$ when reflection was included.  They did not, however, 
have the advantage of sensitivity above 20 keV with which to test for curvature beyond that of the reflection component.  

A large spectral survey of Seyfert spectra above 50 keV performed by Zdziarski \etal (2000) 
using \textsl{CGRO}-OSSE, determined photon indices in the 50--150 keV band, although uncertainties were quite high (15\%--30\% 
for 1$\sigma$ uncertainties) despite exposure times of 100s--1000s of ks.  Of the 27 Seyfert galaxies observed by OSSE, 
19 have measured power-law photon indices with mean values of $\Gamma$=2.37$\pm 0.11$ for Seyfert 1's and 2.06$\pm 0.15$
for Seyfert 2's.  When a cut-off power law was employed, the average values of $\Gamma$ dropped to 1.69\err{0.57}{0.81}
and 1.33\err{0.56}{0.52}, with exponential cut-off energies of 120\err{220}{60} keV and 130\err{220}{50} keV respectively
for Seyfert 1's and 2's.  For comparison, our results from the 20--100 keV band only modeled with a simple power law 
gave an average values of $\Gamma$=1.81 for Seyfert 1's and $\Gamma$=1.71 for Seyfert 2's. %errs?
%--------------------------------------------------------------------------------------------------%

\begin{deluxetable}{lc}
   \tablecaption{Fe Line Equivalent Widths for Best-fit Models \label{tabeqw}}
   \tablecolumns{2}
   \tablewidth{0pc}
%   \tablewidth{3in}
\startdata
\hline
\hline
Source & EW (eV)\\[1mm]
\hline
NGC~4151 &         ~110 $\pm ~  20$ \\[1mm]
IC~4329a &         ~130 $\pm ~  30$ \\[1mm]
NGC~3783 &         ~170 $\pm ~  30$ \\[1mm]
NGC~5548 &         ~150 $\pm ~  30$ \\[1mm]
Mkn~509 &          ~110 $\pm ~  30$ \\[1mm]
MR~2251--178 &      ~130 $\pm ~  40$ \\[1mm]
NGC~3516 &         ~150 $\pm ~  30$ \\[1mm]
NGC~3227 &         ~110 $\pm ~  40$ \\[1mm]
NGC~4593 &         ~230 $\pm ~  20$ \\[1mm]
NGC~7469 &         ~150 $\pm ~  30$ \\[1mm]
3C~111 &           ~180 $\pm ~  60$ \\[1mm]
3C~120 &           ~110 $\pm ~  70$ \\[1mm]
3C~273 &           ~~40 $\pm ~  20$ \\[1mm]
Cen~A &            ~~90 $\pm ~  10$ \\[1mm]
NGC~5506 &         ~410 $\pm ~  70$ \\[1mm]
MCG--5-23-16 &     ~140 $\pm ~  20$ \\[1mm]
NGC~4507 &         ~180 $\pm  ~ 30$ \\[1mm]
Circinus &        2400 $\pm  ~100$ \\[1mm]
NGC~7582 &         ~330 $\pm  ~ 90$ \\[1mm]
NGC~4945 &         1600 $\pm  ~500$ \\[1mm]
\enddata
\tablecomments{Fe line equivalent widths (EW) for all objects using the best fit model for each.}
\end{deluxetable}
%--------------------------------------------------------------------------------------------------%

%---------------------------------------% 
\begin{figure}
  %\epsscale{1.0}
  \plotone{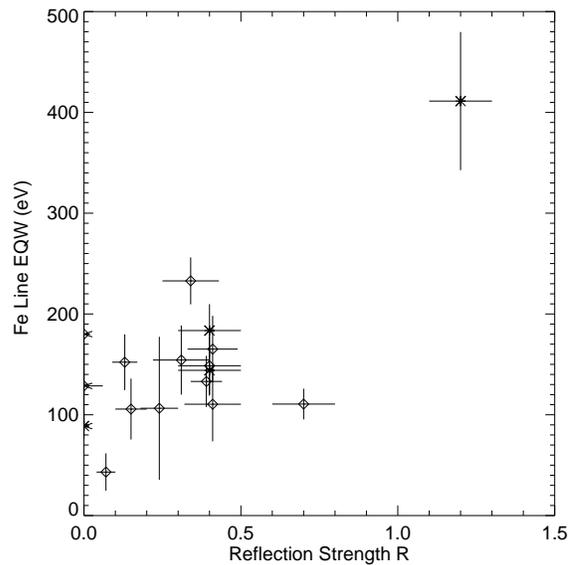}
  \caption{Fe line equivalent width versus $R$ for all Seyferts excluding NGC~7582 
   which has a poorly constrained $R$ value, and also Circinus and NGC~4945 which have
   unusually high Fe line equivalent widths due to extreme absorption of the power laws. 
   Diamonds indicate Seyfert 1's, X's indicate Seyfert 2's.}
  \label{WvsR}
\end{figure}
%---------------------------------------% 

Unification schemes imply that Seyfert 1's and 2's should share a common range of photon indices, since according
to this model only the amount of absorption distinguishes the different types of Seyferts while the intrinsic central engine
is the same.  Gondek \etal (1996), who created an average radio-quiet 
Seyfert 1 spectrum with \textsl{Ginga}+OSSE and \textsl{EXOSAT}+OSSE samples, found $\Gamma = 1.90 \pm 0.05$,
a rollover energy between 400 and 2700 keV, $R = 0.76 \pm 0.15$, and an Fe line EW = 120 $\pm 40$ eV.
Smith \& Done (1996) analyzed a \textsl{Ginga} sample of Seyfert 2's, finding mean $\Gamma$ = 2.0, very close to that of Seyfert 
1s, and with similar dispersion, all of which is consistent with unification schemes.
Our results for the power-law photon indices gave an average $\Gamma = 1.79$ with a standard deviation of 0.09 for all Seyferts,
$\Gamma = 1.79$ with a standard deviation of 0.09 for Seyfert 1's and $\Gamma = 1.81$ with a 
standard deviation of 0.08 for Seyfert 2's, consistent with previous results and with unification schemes.  

Zdziarski \etal (1995) created an average Seyfert spectrum also from \textsl{Ginga}+OSSE data with rollovers 
near a few hundred keV.  Perola \etal (2002) studied nine Seyfert 1's with \sax, measuring rollovers between 
80 and 300 keV with large error bars (up to 100's of keV). 
In contrast, our results for the high-energy rollover yielded mainly lower limits, with tentative
detections for only 3/20 Seyferts, as discussed in Section 4.3.

The recent 22-month \swiftbat survey (Tueller \etal 2010) contained all 23 of 
our selected sources covering an energy range 14--195 keV, comparable to our 
HEXTE-only data.  Photon indices derived from their hardness ratios were 
slightly steeper than those derived from a simple power-law fit to HEXTE 
data (see Table \ref{tab20}) by about 0.1--0.2, probably due in part to 
differences in calibration for the two instruments (BAT calibration to the 
Crab used $\Gamma$=2.15 while HEXTE used a value of 2.062 ; Tueller \etal 
2010 and Rothschild \etal 1998 respectively; Jourdain \& Roques 2009 used a 
broken power-law with $\Gamma = \sim$ 2.07 below 100 keV to model the Crab 
using the extensively calibrated SPI instrument aboard \textsl{INTEGRAL}-SPI).  
BAT spectral analysis is ongoing.

%Additionally, using simulated BAT data and the formula for converting HR to \Gamma given by Tueller etal, 
%we typically derived a value of \Gamma ~0.2 steeper than that obtained through spectral fitting.

%---------------------------------------------------------------------------------------------------------%

\begin{deluxetable*}{llllll}
   \tablewidth{0pc}
   \tablecaption{Fluxes and Luminosities }
   \tablecolumns{6}
\startdata
\hline
&        $F_{2-10}$   &       $F_{20-100}$  &   $L_{2-10}$ (Rest Frame)  &  $L_{20-100}$ (Rest Frame) &  $L_\textrm{CR}/L_\textrm{(PL+CR)}$  \\[1mm]
Source  &  Observed     &        Observed     &   Power-law only    &     Power Law Only    &    in 20--100 keV \\[1mm]
Name    & (10$^{-11}$ erg cm$^{-2}$ s$^{-1}$)& (10$^{-11}$ erg cm$^{-2}$ s$^{-1}$)& (erg s$^{-1}$)    &   (erg s$^{-1}$)  &   (Rest Frame)\\[1mm]
\hline
NGC~4151 & 16.92        &   43.48    &    5.14$\times 10^{43}$    &       4.64$\times 10^{42}$          &           39\% \\[1mm]
IC 4329a & 13.97        &   19.37    &    7.42$\times 10^{43}$    &       7.65$\times 10^{43}$          &           26\% \\[1mm]
NGC~3783 &  7.34        &   12.12    &    1.39$\times 10^{43}$    &       1.57$\times 10^{43}$          &           27\% \\[1mm]
NGC~5548 &  5.97        &    8.57    &    3.88$\times 10^{43}$    &       4.94$\times 10^{43}$          &           12\% \\[1mm]
Mkn 509  &  5.77        &    8.75    &    1.32$\times 10^{44}$    &       1.81$\times 10^{44}$          &           12\% \\[1mm]
MR 2251--178 & 4.46     &    6.25    &    3.78$\times 10^{44}$    &       5.32$\times 10^{44}$          &        < 4.6\% \\[1mm]
NGC~3516 &  4.45        &    8.80    &    1.02$\times 10^{43}$    &       1.24$\times 10^{43}$          &           22\% \\[1mm]
NGC~3227 &  4.11        &    7.49    &    2.34$\times 10^{42}$    &       2.84$\times 10^{42}$          &           27\% \\[1mm]
NGC~4593 &  4.19        &    6.29    &    9.27$\times 10^{42}$    &       1.16$\times 10^{43}$          &           23\% \\[1mm]
NGC~7469 &  3.33        &    4.53    &    1.47$\times 10^{43}$    &       1.57$\times 10^{43}$          &           26\% \\[1mm]
3C~111   &  5.21        &    8.66    &    2.73$\times 10^{44}$    &       4.27$\times 10^{44}$          &        < 0.8\% \\[1mm]
3C~120   &  5.30        &    7.52    &    1.16$\times 10^{44}$    &       1.40$\times 10^{44}$          &           18\% \\[1mm]
Cen A   &  28.42        &   66.01    &    8.59$\times 10^{41}$    &       1.08$\times 10^{42}$          &       < 0.02\% \\[1mm]
NGC~5506 & 10.65        &   16.88    &    1.01$\times 10^{43}$    &       9.70$\times 10^{42}$          &           42\% \\[1mm]
MCG--5-23-16 & 9.60     &   16.00    &    2.03$\times 10^{43}$    &       2.35$\times 10^{43}$          &           21\% \\[1mm]
NGC~4507 &  1.43        &   14.94    &    2.41$\times 10^{43}$    &       4.05$\times 10^{43}$          &           19\% \\[1mm]
Circinus &  2.63        &   20.62    &    2.22$\times 10^{41}$    &       3.13$\times 10^{41}$          &           64\% \\[1mm]
NGC~4945 &  0.61        &   19.89    &    1.15$\times 10^{41}$    &       4.02$\times 10^{41}$          &        < 0.4\% \\[1mm]
NGC~7582REFL &  1.04    &    5.62    &    6.03$\times 10^{41}$    &       7.24$\times 10^{41}$          &           75\% \\[1mm]
NGC~7582PC   &  1.01    &    6.54    &    3.09$\times 10^{42}$    &       3.47$\times 10^{42}$          &           --   \\[1mm]
3C~273   &  11.90       &   19.92    &    7.23$\times 10^{45}$    &       1.15$\times 10^{46}$         &            6\%  \\[1mm]
3C~454.3 &  7.41        &   13.05    &    2.06$\times 10^{47}$    &       3.60$\times 10^{47}$          &           --   \\[1mm]
Mkn 421  & 40.00        &   10.14    &    8.63$\times 10^{44}$    &       2.18$\times 10^{44}$          &           --   \\[1mm]
1ES 1959+650 & 15.48    &    9.37    &    1.17$\times 10^{45}$    &       7.02$\times 10^{44}$          &           --   \\[-1mm]
\enddata

\tablecomments{Columns (2) and (3) list the observed frame, model-dependent 2--10 and 20--100 keV fluxes, not corrected for absorption, 
from each source's best-fit model. %%%% (e.g., rollovers were included in the power-law continuum for MR 2251--178, NGC~4945 and Circinus).
The 20--100 keV fluxes listed in col.\ (3) were determined from HEXTE responses, averaged between the
two HEXTE clusters and weighted by exposure time and effective area.
%%For NGC~7582, parameters from the best-fit model with reflection ("REFL," as opposed to
%%the Compton-thick partial-covering model, "PC") are listed.
Column (4) lists the absorption corrected, rest-frame 2--10 keV luminosity of the power-law continuum component only
i.e., without the Fe line, Compton reflection hump, and after correcting for all 
cold (including Galactic) and warm absorbers.
Isotropic emission was assumed (i.e., there is no correction for beaming effects in blazars).
Similarly, column (5) lists the absorption corrected, rest-frame 20--100 keV luminosity determined using HEXTE responses,
weighted by exposure time and effective area between the two clusters.
Column (6) lists the ratio of the Compton reflection component emission to the total (power-law plus
Compton reflection hump) emission in the rest-frame 20--100 keV band 
for each source's best-fit model, again using HEXTE responses and
weighted between HEXTE A and B by exposure time and effective area.}
\label{tabflux}
\end{deluxetable*} 

%---------------------------------------------------------------------------------------------------------%
%---------------------------------------% This is Section 4.2

\subsection{The Compton Reflection Component}

X-ray reflection of the illuminating power-law continuum
is expected to occur at the surface of the optically thick, radiatively efficient, geometrically thin
accretion disk whose thermal emission comprises the ``big blue bump'' in the optical/UV.
Compton-thick material can also be present in the form of the putative homogeneous torus
commonly invoked in classical Seyfert 1/2 unification schemes and thought to be located
$\sim$1 pc from the black hole (e.g., Antonucci 1993); more recent models incorporate sub-pc scale Compton-thick clumps
(e.g., Nenkova et al.\ 2008; Elitzur \& Shlosman 2006).

Constraining the level of reflection relative to that of the illuminating power-law
can, in principle, yield information on the sky-covering fraction of the reflecting material (as seen from the X-ray continuum source,
assumed to be the central X-ray corona), thereby providing constraints on how
much Compton-thick material is in the form of a torus, disk, etc.
In practice, however, this is not straightforward.
The variability of the illuminating X-ray continuum (in both the photon index and normalization)
and the fact that there is a non-zero
light-travel time from the X-ray corona to the reflecting material can result in a lag between
the properties of the observed power-law continuum and
the properties of the power-law continuum illuminating the reflecting material. 

A spectral survey such as ours, wherein we have used data spanning as long a baseline as possible, thus has
an advantage compared to measurements obtained via single-epoch spectral fitting: there is a much higher likelihood
that the long-term average normalizations and photon indices of the observed and illuminating power-laws
will be equal, provided that the lag time between the variations in the 
continuum and the responses in the strength of the Compton reflection component
(assumed to be dominated by the light travel time) is substantially less than the duration of the monitoring. 
For most of the objects in the current sample, this condition is satisfied
provided that the bulk of the reflecting material in Seyferts lies within a few light-years of the
black hole and corona. If the reflecting material is spatially extended, then rapid variations will be smoothed out anyway.

Moreover, an additional complication is the geometry of the reflecting material and its orientation relative to the observer.
In this work, for simplicity, we used \textsc{pexrav}, which assumes a semi-infinite
slab, most applicable to a ``lamppost'' geometry wherein the illuminating X-ray 
source lies above an accretion disk. However, as discussed by Murphy \& Yaqoob (2009),
\textsc{pexrav} does not account for transmission through the slab. 
In addition, as Compton scattering in the Thompson regime is an isotropic process, 
photons are preferentially scattered forward and backward (parallel and anti-parallel to the path of the incident
photons). For a torus-shaped structure, whose inner face is preferentially illuminated by a central continuum source,
photons will be preferentially scattered parallel to the equatorial plane instead of 
in a direction parallel to the rotation axis.
For a torus with its equatorial plane oriented perpendicular to the observer, the amount of 
reflected emission reaching the observer will be relatively less.
Meanwhile, in a face-on lamppost/disk geometry, the bulk of scattered photons will
be parallel/anti-parallel to the line of sight.
Consequently, for a given incident flux, the amount of reflected flux reaching the observer
will be far less in the case of a torus than for a disk; this factor is $\sim$6 for the
reflected continuum near 6 keV and is $\sim$3 near 20 keV
(Murphy \& Yaqoob 2009; their Figure 7). 
Models for reflected emission from a torus geometry are still 
under development and can be applied to the current sample in future work.

In the current work, where we have used \textsc{pexrav},
we find a value of $\langle R \rangle \sim 0.35$ with a standard deviation of 0.16 for Seyfert 1's as a whole (excluding the FSRQ 3C\,273
and objects that do not show evidence for a reflection hump) 
and $\langle R \rangle \sim 0.67$ with a standard deviation of 0.46 for the three Compton-thin Seyfert 2's (excluding Cen A), but
we must rely on several assumptions in order to infer
physically meaningful constraints on the geometry of the reflecting material for the cases of a lamppost/disk geometry
or a centrally-illuminated torus.
(For the remainder of this section, we omit the Compton-thick sources due to 
possible model degeneracy problems resulting from the spectral complexity
and large systematic uncertainties on $R$).
Angles of inclination have also been assumed for classes of objects  
for simplicity (30$\degr$ and 60$\degr$ for Seyfert 1's and Compton-thin Seyfert 2's, respectively)
as opposed to relying on model-dependent
methods such as relativistic Fe K$\alpha$ line fitting, (e.g., Reeves et al.\ 2007),
and using the inclination of the host galaxy as a proxy is not feasible since there is no guarantee that
the planes of the accretion disk and galaxy are aligned.

Assuming that the lamppost/disk geometry and the \textsc{pexrav} modeling are valid,
the covering fractions (as seen from the corona) implied by the above values of $\langle R \rangle$ are
(0.35)2$\pi$ sr for Seyfert 1's and (0.67)2$\pi$ sr for Compton-thin Seyfert 2's. These values
suggest several possibilities: the radiatively efficient disk may not extend down to the innermost stable orbit,
may be too highly ionized at the inner radii to emit substantial amounts of reflected emission, and/or
the entire disk may not be well illuminated due to self-shielding or the X-ray continuum emission being preferentially 
beamed away from the disk.
If the inner disk is indeed very highly-ionized, then we would expect some line emission from highly-ionized Fe; 
however given the PCA resolution; that line flux may be overwhelmed by emission from neutral Fe.
The majority of our energy centroid values for the Fe K lines (16 out of the 20 sources in which an Fe line is 
detected) are consistent with 6.4 keV, i.e., with the bulk of the Fe line photons originating in cold/neutral gas.

If instead all the reflected emission originates in a torus, the 
covering factors may be 3--6 times larger, suggesting that a very large fraction of the sky
may be covered; one possibility is that the illuminated inner face of the torus may be substantially vertically extended. 
A more complex geometry (such as a distribution of Compton-thick clumps) of course cannot be excluded and
indeed could explain how it is possible to obtain values of $R$ > 1 as in the case of NGC~5506.

Finally, combining knowledge of the Compton hump strength and the EW of the Fe K emission
line can yield estimates of how much of the line emission originates in Compton-thick material as opposed to Compton-thin.
A relativistically broadened Fe K$\alpha$ component indicates a radiatively-efficient disk
(as opposed to a radiatively inefficient flow, e.g., Narayan \etal 1998); whereas if only a narrow Fe line component is confirmed, then the sites of
Compton reflection are more likely in the distant torus.  

For the lamppost/disk geometry if the disk is inclined at 30$\degr$ (60$\degr$) to the line of sight and 
illuminated by a power-law with $\Gamma = 1.8$, a value of $R=1$ corresponds to EW$ \sim 125$ (90) eV  
(George \& Fabian 1991; their Figure 14 with correction factor corresponding to the abundances of Wilms et al.\ 2000).
We assume solar Fe abundance for purpose of this discussion.
For our average values of $R$ for the Seyfert 1's, we therefore predict EW$_{\rm disk} \sim 40$ eV.  For the Compton-thin Seyfert 2's, 
EW$_{\rm disk} \sim 60$ eV.
Alternatively, for a torus geometry, assuming a continuum with $\Gamma = 1.9$ illuminating gas with a 
column density of $2 \times 10^{24}$ cm$^{-2}$, with the torus inclined by 30$\degr$; Murphy \& Yaqoob (2009; their Figure 9)
gives similar values: EW$_{\rm torus }\sim$50 eV for the Seyfert 1's and $\sim$40 eV for the Compton-thin Seyfert 2's. 

The average observed equivalent width $\langle$ EW$_{\rm obs} \rangle$ for just the Seyfert 1's 
(again, not including 3C~273, whose observed EW may be diluted due to contamination from the jet) 
is 145$\pm 35$ eV, so we estimate that, on average, $\sim$25$\%$--35$\%$ of the Fe line originates in Compton-thick 
material in Seyfert 1s. For the Compton-thin Seyfert 2's, we find $\langle$ EW$_{\rm obs} \rangle$ = 205 eV, 
suggesting that, on average, $\sim$20$\%$--30$\%$ of the Fe line originates in Compton-thick material in these objects.

Risaliti (2002) analyzed \sax data of Compton-thin Seyfert 2's, finding that on average the Fe line EW increased slightly when
\NH was above $\sim$ 3$\times 10^{23}$.  He concluded that this must mean there is contribution to the Fe line flux
that does not come from the accretion disk (otherwise the Fe line would be just as absorbed as the continuum).
We find a similar and far more dramatic increase in EW for our two heavily absorbed sources, 
Circinus and NGC 4945 (see Table \ref{tabeqw}).
This confirms the findings of Risaliti (2002) and is consistent with the idea that a substantial amount of reflection/emission
may come from within the torus.

Figure \ref{WvsR} shows the measured values of EW plotted against the Compton reflection strength
In the reflection models discussed so far, one may expect a correlation between
these parameters if the bulk of the Fe line originates in Compton-thick material.
In this plot, we have omitted the Compton-thick sources.
%%%XXX include ones with R = zero: if the notion of having all the line originate in C-thick gas is correct, then R=0 $\rightarrow$ EW = 0, too XXX}.  
The Pearson correlation coefficient $r_{\rm P}$ = 0.79, with a null hypothesis
probability (probability of obtaining the correlation by chance) $P_r$ = 7$\times 10^{-4}$. 
NGC 5506 stands out as an outlier as it has the largest values of both $R$ and EW$_{\rm obs}$
for the non-Compton-thick sources. When omitted from the 
correlation, $r_{\rm P}$ falls to 0.28 ($P_r$ = 0.35). 
The lack of a strong correlation between $R$ and EW is consistent with the
notion above that it is very unlikely that 
the bulk of the Fe line emission originates in Compton-thick material.
This is consistent with the findings of Risaliti (2002) who found no correlation between EW and $R$ (his Figure 5).

%---------------------------------------% This is Section 4.3
\subsection{High-Energy Rollovers}

High-energy rollovers in the power-law continua of Seyferts are
expected if the power-law is generated via thermal Comptonization of
soft seed photons, most likely blackbody emission from the accretion disk,
(e.g., Haardt et al.\ 1994, Titarchuk 1994, Poutanen \& Svensson 1996, and references therein).
In the simplest models, if all the electrons in the corona
are at the same temperature $T_{\rm e}$, a high-energy continuum rollover 
can be expected in the spectrum near an energy $\sim$3 times $k_{\rm B}T_{\rm e}$ (assuming
an optical depth near unity).

Such rollovers have been detected in the X-ray spectra for many accreting Galactic
black hole (GBH) systems (e.g., Takahashi et al.\ 2008; Wilms \etal 2006; Kalemci \etal 2005; 
Yamaoka et al. 2005; Frontera et al. 2001). 
As Seyferts are generally X-ray fainter than GBH systems, constraining rollovers in AGN
has been observationally challenging, however for sources with $\Gamma < 2$, rollovers 
are expected in order to prevent divergence of the total energy.
In general, the rollovers modeled in \textit{BeppoSAX} spectra of Seyferts reported in the 
literature span the range $\sim 50--400$ keV  with uncertainties anywhere from tens to a 
few hundred keV (e.g., Petrucci \etal 2001; Perola et al.\ 2002; Quadrelli et al.\ 2003).
In our \xte sample, adding high-energy rollovers to our spectral models 
yielded mostly only lower limits to \eroll of $\sim$200--300 keV.
This is similar to the results of Risaliti (2002) who concluded that high-energy cutoffs 
must not be an ubiquitous property of Seyfert galaxies.

We detect a rollover in Circinus around 40~keV, and NGC~4945 around 70~keV, and, tentatively, in \mr ~around 100~keV.  
Extra curvature in the data/model residuals when no rollover is modeled suggest that the rollovers detected in Circinus and 
NGC~4945 are not spurious.  However, problems encountered with model degeneracies during fitting led to unexpectedly low 
(and likely unrealistic) values for the photon index $\Gamma_{\rm HX}$ of the absorbed hard X-ray power law in each case.
Given the extreme amount of absorption and the requirement to model an additional power-law component below $\sim$10 keV
in each object, Circinus and NGC~4945 have two of the most complex spectra in our
sample.  Successfully deconvolving all components, and accurately constraining rollovers and Compton reflection
strengths in these objects may be possible in the future via observations spanning 
a broad energy range and having sufficient energy resolution in the Fe K bandpass.
For the purposes of this paper, we have simply presented the results of model fits with and without rollovers,
with the understanding for the reader that for the models with rollovers, best-fit values may have large systematic uncertainties.

For objects in which a rollover was not detected,
we can be reasonably certain that the quality of our data precludes rollover energies below $\sim$ 100 keV.  
Excluding the blazars, 3C~273, and Cen A (each of which has X-ray emission
dominated by or likely contaminated by X-ray emission from a relativistic jet),
the lower limits on \eroll for our objects range from 120 to 350 keV, with an 
average lower limit of 225 keV, which is very close to the upper limit of the HEXTE energy range. 
This would suggest that, on average, the electron temperatures in our Seyfert coronae 
must be at least $\gtrsim$75~keV.

The potential bias produced by our selection criteria may have had an adverse
effect on the likelihood of finding rollovers below 100 keV in our sample since our goal was to
have a 3$\sigma$ detection at 100 keV.
In addition, as thermal Comptonization is thought to be at work in both Seyfert and GBH coronae, then 
if there exists a relation between $\Gamma$ and \eroll in Seyferts as seen in GBHs (Yamaoka et al.\ 2005),
then our tendency to select flat-spectrum objects may be another source of bias against finding relatively low values of \eroll
\footnote{See also Petrucci et al.\ (2001) for investigations into a possible correlation between $\Gamma$ and corona 
temperature in Seyferts.}.

%---------------------------------------%
\section{Conclusion}

Thanks to the combination of the longevity of the \xte mission,
the sustained monitoring campaigns on many AGN, and 
the simultaneous operation of the PCA and HEXTE instruments,
we have been able to extract broadband (3 to $\geq$100 keV) X-ray spectra 
with long integration times and moderately good spectral resolution. 
Although the HEXTE instrument is sometimes overlooked due to its low sensitivity on short timescales, 
we have demonstrated that HEXTE can indeed yield spectra out to at least 100 keV
for 23 AGN with a sufficient combination of brightness and exposure time.
The fact that the spectra presented here are long-term averaged spectra means the
ambiguity inherent in single-epoch spectral fitting caused by variability of the source
is eliminated. In summary, we have presented the first high-quality broadband
spectra sample with which to study the long-term average properties of X-ray bright Seyferts
in the $>$10 keV sky.

Observed fluxes and unabsorbed power-law luminosities for the 2--10 and the 
20--100~keV range are given in Table \ref{tabflux},
along with the modeled
fraction of reflected flux in the 20--100 keV band.
As these quantities are long-term averages in nearly all cases, 
they can be used by the community,
e.g., for purposes of planning future observations of these objects, or
for constraining models which seek to 
determine the contributions of low-$z$ Seyferts to the observed
cosmic X-ray background (CXB) and to the accretion history of the universe.

Our long-term averaged values for $\Gamma$ were generally consistent with previous results for individual 
objects; our averages by AGN classification were also consistent with previous surveys.
Unsurprisingly, we found that the Fe line complex was necessary to model in all Seyfert spectra. 
Using a single Gaussian to model this component we found typical equivalent widths around 
100--200~eV, roughly consistent with previous results.  
In contrast to the Fe line, Compton reflection components have not been well studied
as, historically, the $>$20 sky has not been as well observed as the 2--10 keV band.  
For our sample, we found that the Compton reflection component was required in the model for 
a vast majority of the Seyferts studied, again consistent with previous results such as Nandra \& Pounds (1994).
Using the \textsc{pexrav} model in \textsc{xspec}, we find a average reflection fraction 
of $\sim$0.3/0.7 in Seyfert 1/2's coming from either a truncated disk or a torus.  
Assuming solar Fe abundance, only 25\%--35\% of the Fe line emission comes from Compton-thick material  
with the remainder arising in Compton-thin gas.   
We found tentative evidence for high-energy rollovers in the spectra of only three objects,
Circinus, NGC~4945, and MR~2251--178; we obtained lower limits of very roughly 100 keV
for the remaining objects implying kT $\gtrsim$75 keV in thermal Comptonization models.

The time-averaged model fits in this paper may be used in future
time- or flux-resolved analyses of \xte monitoring data
to test for variations in the absorbing or reflecting components
to constrain their locations and/or distributions.
We conclude by commenting that future in-depth study of targets whose spectra require
highly complex model fits, such as
Circinus and NGC~4945, can yield fewer model degeneracies and better 
constraints on $\Gamma$, \eroll, etc., provided that
observations are made with instruments featuring a combination of 
CCD-like (or better) energy resolution and broad energy coverage up
to at least hundreds of keV. Such observations are also required for further
progress in better constraining the high-energy rollovers in Seyferts and understanding
the properties of the X-ray corona, as well as allowing for testing of the newest
Compton reflection models (e.g., Murphy \& Yaqoob 2009).

%==================================================================%

\begin{acknowledgments}
This research has made use of data obtained from the \textsl{RXTE} satellite, a NASA space mission.
The authors wish to thank J.\ Wilms for reading the manuscript and providing helpful suggestions to guide the paper.
This work has made use of HEASARC online services, supported by NASA/GSFC, and the NASA/IPAC Extragalactic Database, 
operated by JPL/California Institute of Technology under contract with NASA.
The research was supported by NASA Constract NAS 5-30720 and Grant NNX09AG79G.
\end{acknowledgments}

%==================================================================%

%\bibliographystyle{apj}
%\bibliography{bible file}

%==================================================================%
\appendix

%==================================================================%

\section{A. Notes on Individual Sources}

In this section, we include notes pertaining to model fitting of each
object's spectrum, and discuss spectral complexities such as partial covering
and warm absorbers. Given the energy resolution of the PCA, all warm absorber
velocity offsets relative to systemic were kept fixed. Since
many warm absorber parameters are determined via gratings observations 
and/or using the soft X-ray band, ionization parameters and column 
densities were kept fixed.

We also compare our best-fit model parameters to those obtained by
previous investigations of broadband X-ray spectra
using \textit{RXTE}, \textit{Suzaku}, and \textit{BeppoSAX}.
We concentrate on continuum model parameters such as Compton
reflection strengths and high-energy rollovers.
Comparing our spectral model results to previous investigations is not always
straight forward, as different broadband models may have been used and the
source may display spectral variability in addition to flux variability.
Furthermore, model degeneracies may exist between
certain parameters, such as $\Gamma$ and $R$, or $\Gamma$ and \eroll.
Below, we have tried to identify 
90$\%$ %%%%%(1.5-sigma, and not 1-sigma, or 68% confidence) 
uncertainties in the literature. 

%As stated earlier, given the energy resolution of the PCA, it is not possible in most cases
%to deconvolve e.g., narrow Fe K$\alpha$ and K$\beta$ lines or broad and
%narrow components.  In most cases, the Fe line complex has been
%studied with \textit{XMM-Newton}, \textit{Chandra-HETGS}, and/or \textit{Suzaku}.

%===============================================================

\subsection{NGC~4151}    %%% see also /home/agm/xte/finalspec/NGC4151/notes.0225.4151

\textit{RXTE} PCA spectral data for NGC~4151 taken before 2001 February
were published by Markowitz \etal (2003a); however, we 
include additional data up to 2004 June as well as all available HEXTE data.
As one of the brightest Seyferts,
NGC~4151 has been studied by every major X-ray mission to date,
and is known to have a rather complex absorbed spectrum.
Indeed, fitting a simple model consisting of a power-law with no 
absorption other than the Galactic column produced a very poor fit 
($\chi^2$/dof=10324/86 with a very flat power-law photon index, $\Gamma \sim 1.2$), 
and allowing full-covering absorption by cold material in excess of the 
Galactic column offered a better, though still not quite poor fit ($\chi^2$/dof=3408/85).

We applied a partial-covering cold absorber using \textsc{zphabs}
following DeRosa et al.\ (2007), who analyzed a series of
\textit{BeppoSAX} observations of this source. $\chi^2$/dof fell to 766/84 for $N_{\rm H} \sim
9 \times 10^{23}$ cm$^{-2}$ and covering fraction $f_{\rm c} \sim 70\%$,
but there remained very large residuals below 10 keV.

Following Schurch et al.\ (2003), who analyzed \textit{XMM-Newton}
observations obtained in 2000, we then applied an additional  
full-covering cold absorber, with column density fixed at
$3.4 \times 10^{22}$ cm$^{-2}$  (see also Ogle et al.\ 2000 and Zdziarski et al.\ 2002).
DeRosa et al.\ (2007) modeled this component to
be mildly ionized, with log($\xi$)$\sim 0-1$\footnote{$\xi \equiv L_{\rm ion} n_{\rm e}^{-1} r^{-2}$, 
where $L_{\rm ion}$ is usually defined as the isotropic 1--1000 Ryd ionizing continuum luminosity, $n_{\rm e}$ is 
the electron number density, and $r$ is the distance from the central continuum source to the absorbing gas.  $\xi$ has
units of erg cm s$^{-1}$.}, but modeling this component
as cold gas was adequate to yield a good fit in our best-fit models.
This yielded our baseline model fit, with
$\chi^2$/dof = 682/83.  The best-fit parameters are listed in
Table \ref{tabbase}.  We kept the column density of the full-covering gas
as a fixed parameter: thawing it caused
several model parameters to deviate wildly from 
expected values based on previous observations, and
after adding the Compton reflection component, thawing this column
did not yield any further improvement in the fit.

After adding the Compton reflection component (see Table \ref{tabpexrav})
our best-fit values for the column density and 
covering fraction for the partial covering absorber
are broadly consistent with values found by Zdziarski et al.\ (2002) 
(for a 0.4--400 keV spectrum from near-simultaneous \textit{ASCA} and OSSE spectra)
and DeRosa et al.\ (2007). 
Our best-fit values for the covering fraction are within the ranges
given by DeRosa et al.\ (2007) and Pucetti et al.\ (2007).

Values of the Compton reflection strength $R$ in the literature when data $>$ 10 keV 
are included  %%%%% exclude R=1.95 (Schurch03, XMM <10keV)
are generally $\lesssim$ 1, and a variety of broadband
models have been used to model the complex absorption below
10 keV, affecting values of $\Gamma$. 
Given the degeneracy between $\Gamma$ and $R$, a comparison between our best-fit value
of $R$ and those in the literature is therefore not straightforward.
Furthermore, De Rosa et al.\ (2007) claim that the Compton hump was detected
in 1996 with $R \sim 2$ but disappeared by 2000/2001, so this component's
normalization relative to that of the power law may be time variable.

Adding a high-energy rollover, we found that $\chi^2$/dof fell to 132/82, for 
\eroll = $156^{+26}_{-20}$ keV, 
with $\Gamma$ flattening to $1.65 \pm 0.04$
(the best-fit values for the partial covering component
$N_\textrm{H}$ and $f_{\rm c}$ were $5.6 \pm 0.4 \times 10^{23}$ cm$^{-2}$ and $52\%$, 
respectively). However, as seen in Figure \ref{NGC4151spec}, there is a substantial
``dip'' near 100 keV. This is most likely a spurious feature related to background
subtraction, as above this energy the background spectrum flattens and the source flux falls 
with increasing energy to below $1\%$ of the background.
The majority of the decrease in $\chi^2$ when a high-energy rollover is added
is attributed to ``fitting'' this 100 keV dip, and 
so we do not believe that the evidence for a rollover is 
robust in this object.  We adopt a lower limit to \eroll of  $136$ keV,
consistent with Zdziarski et al.\ (2002) and most spectral fit results in 
DeRosa et al.\ (2007).

Analysis of a long-term, time-averaged spectrum assumes
that the form of the spectral model is valid 
on shorter time-scales. Puccetti et al.\ (2007) demonstrated that 
NGC~4151 exhibits variations in the properties of its absorber 
on both rapid (time scales of tens of ks) and long (years) time scales. They 
attributed the variations to transits along the line of sight by
gas clouds consistent with an origin in the broad line region.
However, the magnitude of the variations 
(factors of 2--5 in the column densities of the full-covering and
partial covering absorbers and factors of $\sim$20$\%$--30$\%$ in covering fraction)
is likely small enough not to invalidate the above assumption
for our long-term spectral analysis.

%===============================================================

\subsection{IC~4329a}

Previous PCA + HEXTE joint spectral fits were published by Markowitz, Reeves \& Braito (2006).
They included data only from 2003 April through 2005 October, while here we include
data up through 2007 August plus data from three earlier campaigns.
%%%The Fe line is dominated by a narrow component with only a
%%%modest broad component (\textit{XMM-Newton} analysis by Markowitz, Reeves \& Braito 2006).
%%%%% MRB2006:  Gamma = 1.894   R = 0.51+/-0.04

Following Steenbrugge et al.\ (2005b),
we included a cold absorber with a column density $N_\textrm{H}$ fixed at
$1.7 \times 10^{21}$ cm$^{-2}$ and 
two warm absorber zones, one with log($\xi$) = --1.37 
and $N_\textrm{H} = 1.3 \times 10^{21}$ cm$^{-2}$,
and the other with log($\xi$) = +1.92  and $N_\textrm{H} = 6 \times 10^{21}$ cm$^{-2}$.

Our best-fit value for $R$, $0.39 \pm 0.05$, is similar to that obtained 
by Bianchi et al.\ (2004) for joint fits to a simultaneous 
\xmm + \textit{BeppoSAX} observation in 2001 January
($R = 0.3-0.5$) and slightly lower than those obtained by Markowitz, Reeves \& Braito 
(2006) for an \xmm long-look  ($R = 0.51 \pm 0.04$)
and Perola et al.\ (2002) for two \textit{BeppoSAX} observations in 1998
($R = 0.6-0.7$).

Bianchi et al.\ (2004) reported a high-energy rollover at
130--170 keV (best-fit values depend on the model used; uncertainties were $\sim$
10--20 keV) which we do not confirm; our lower limit is 330 keV,
consistent with rollovers reported by Perola et al.\ (2002).

%%%% mention Perola 2002:  Ecut best-fit values 262 & 325 with large uncertainties 

Our best-fit value for power-law photon index $\Gamma$, $1.88 \pm 0.02$, is very close to
values reported in the aforementioned papers.

%===============================================================

\subsection{NGC~3783}

\textit{RXTE} spectral data have been previously published by Markowitz \etal (2003a),
though they only included data up until 2002 April and did not use
HEXTE data.

NGC~3783's complex warm absorber properties have been studied extensively
with gratings observations (e.g., Netzer et al.\ 2003, McKernan et al.\ 2007).
Three zones of warm absorption introduce significant
continuum curvature in the spectrum above 2 keV.
Following Reeves et al.\ (2004), we included absorbers with
$N_\textrm{H} = 1.1 \times 10^{21}$, $1.2 \times 10^{22}$, and $4.4 \times 10^{22}$ cm$^{-2}$ and 
log($\xi$) = --0.1, 2.1, and 3.0, respectively.

We clearly detected the Compton hump with a strength $R = 0.41 \pm 0.08$,
in close agreement with Markowitz \etal (2003a) as well as with
Perola et al.\ (2002) using data from the \textit{BeppoSAX} observation in 1998 June.
Using the same \textit{BeppoSAX} data but different warm absorber model codes  
and broadband models, 
Perola et al.\ (2002) and De Rosa et al.\ (2002) reported high-energy rollovers at
$156^{+37}_{-40}$ and $340^{+560}_{-107}$ keV, respectively; however,
we find no evidence for a high-energy rollover (\eroll $> 350$ keV).

%===============================================================

\subsection{NGC~5548}

\textit{RXTE} spectral data have been previously published by Markowitz \etal (2003a),
though they only included data up until 2003 January and did not use HEXTE data.

We modeled three zones of warm absorption in this source based on Steenbrugge et al.\ (2005a),
with $N_\textrm{H}$ =   $2.5 \times 10^{21}$,   $1.0 \times 10^{21}$, and $6 \times 10^{20}$ cm$^{-2}$, and
log($\xi$) = +2.3, +1.9, and --0.2, respectively.

Our best-fit value of $R$, 0.13 $\pm$ 0.04, is somewhat smaller than values reported previously, including 
Liu et al.\ (2010) for a series of \textit{Suzaku} monitoring observations obtained in 2007 ($R = 0.79^{+0.35}_{-0.32}$ 
for the summed spectrum), Perola et al.\ (2002) for a \textit{BeppoSAX} observation in 1997 ($R = 0.54^{+0.20}_{-0.13}$), 
Bianchi et al.\ (2004) for joint fits to a simultaneous \textit{BeppoSAX} + \textit{XMM-Newton} spectrum in 2001 
($R = 0.45^{+0.20}_{-0.16}$), and Markowitz \etal (2003a)for \textit{RXTE}-PCA data covering 1999--2003 
($R = 0.41^{+0.02}_{-0.08}$).  Best-fit values of $\Gamma$ in those papers are in the range 1.59--1.74.  It is interesting to 
note that the 2007 \textit{Suzaku} campaign caught the source at a relatively low flux level, with a value 
of $F_{2-10}$ for the summed spectrum of $1.9 \times 10^{-11}$ erg cm$^{-2}$ s$^{-1}$, a factor of three lower than 
our best-fit 2--10 keV flux, $6.0 \times 10^{-11}$ erg cm$^{-2}$ s$^{-1}$.  As our data cover the period from 
1996 to 2007, this is consistent with the notion that the absolute normalization of the Compton reflection component
may have remained constant while the 2--10 keV nuclear flux had dropped by a factor of $\sim$3 during the 
\textit{Suzaku} campaign in 2007 June--August, i.e., the observed reflected flux had not yet responded to the drop in continuum 
flux by 2007.  One possibility is that the bulk of the Compton-reflecting material is located at least several 
light years from the X-ray nuclear continuum source.

We do not confirm previous reports of high-energy rollovers using \textit{BeppoSAX}
(Bianchi et al.\ 2004, Perola et al.\ 2002) or \textit{Suzaku} (Liu et al.\ 2010, who
report \eroll $= 75^{+85}_{-15}$ keV),
finding a lower limit to \eroll of 260 keV.

%===============================================================

\subsection{Mkn~509}

Previous results on the warm absorbers for this object
(McKernan et al.\ 2007, Smith et al.\ 2007) indicate a negligible impact on the PCA spectrum.
The Compton reflection component is significantly detected but is weak.
Our best-fit value of $R$, $0.15 \pm 0.05$ is
somewhat lower than values reported by De Rosa et al.\ (2004)
for a variety of model fits to \textit{BeppoSAX}
spectra obtained in 1998 and 2000 ($R \sim 0.8 - 1.1$)
or Perola et al.\ (2002) for the 1998 observation ($0.58^{+0.39}_{-0.30}$).
These works also reported high-energy rollovers with best-fit values
in the range 67--115 keV; we do not find any evidence for a rollover,
with \eroll > 220 keV.

%===============================================================

\subsection{MR~2251--178}

Based on previous studies of this object's warm absorbers
(e.g, Kaspi et al.\ 2004), there is negligible impact on the PCA spectrum.
We do not find evidence for a Compton reflection hump, broadly consistent with
very low values of $R$ modeled by Orr et al.\ (2001; best-fit values of $R \sim 0.2-0.4$)
using spectra obtained from a pair of \textit{BeppoSAX} observations in 1998.

We find only marginal evidence for a rollover at
$103^{+40}_{-30}$ keV: $\chi^2$ falls by 10.1 for 1 less dof while the photon index steepens by
0.07 (from $1.63 \pm 0.02$ to $1.56 \pm 0.03$).
This rollover energy is consistent with that claimed by Orr et al (2001: best-fit values near
102--133 keV depending on the model used). However, in our spectrum,
much of this improvement in fit is likely
associated with trying to fit a small artificially narrow ``dip'' at 100 keV associated
with HEXTE background subtraction.
Ignoring the bins around this energy, adding a rollover to the model causes
$\chi^2$ to drop by only $\sim$6 for 1 less dof (significant at
the $\sim 96 \%$ confidence level according to an $F$-test).
Furthermore, there is not much improvement visually
in the data/model residuals above $\sim$50 keV.
We caution that it is unclear from our spectral modeling
whether or not the rollover is a real feature.

%===============================================================
\subsection{NGC~3516}

Markowitz \etal (2003a) have published \textit{RXTE}-PCA spectral data taken up 
through 2000 February. We include HEXTE data, as well as
additional data taken in 2000--2002 and in 2005, but this additional 
data represents an increase in good PCA exposure time of only 27$\%$ compared to 
Markowitz \etal (2003a).

This source is known to have several zones of ionized absorption
along the line of sight (see Turner et al.\ 2008 and references therein),
and the cold and ionized absorbers can display variability in column depth on both rapid
(tens of ks) and long (several years) time scales. The absorber with the most impact
on modeling of the long-term average PCA spectrum is the so-called
``heavy'' component, or ``Zone 3,'' described in Turner et al.\ (2008). We modeled this component
with a column density fixed at $2.0 \times 10^{23}$ cm$^{-2}$, log($\xi$) fixed at 
2.19,  and covering fraction left as a free parameter but expected to be 
near 50$\%$ (we find 56$\%$ in our best-fit model).  Other warm absorber components 
discussed in Turner et al.\ (2008) 
such as the ``UV absorber'' or more highly-ionized absorbers 
have negligible impact on the PCA spectrum and are ignored.

Our best-fit values of $\Gamma$ ($1.82 \pm 0.04$) and $R$ ($0.31 \pm 0.09$) for 
the reflection model are slightly lower than the best-fit values derived
by Markowitz et al.\ (2008) for the 0.3--76 keV \textit{Suzaku} spectrum 
($\Gamma = 1.97^{+0.01}_{-0.03}$ and $R = 1.8^{+0.4}_{-0.5}$ for their ``PC2'' model).
The discrepancy may be at least partially attributed to degeneracies between $\Gamma$ and $R$; 
forcing $\Gamma$ to be 1.97 in our fits, $R$ rises to $\sim$0.9.
Much of the remainder of the discrepancy may be attributed to intrinsic
source spectral variability. Specifically, it is interesting to note that
the observed 2--10 keV and 12--76 keV fluxes for the \textit {RXTE} spectrum
are $4.4 \times 10^{-11}$ (PCA) and 9.3 $\times 10^{-11}$ erg cm$^{-2}$ s$^{-1}$ (average of 
HEXTE A and B responses), respectively,
while for the \textit{Suzaku} observation $F_{2-10}$ and $F_{12-76}$ were
$2.3 \times 10^{-11}$ and 11 $\times 10^{-11}$ erg cm$^{-2}$ s$^{-1}$, respectively.
That is, the ratio $F_{12-76}$/$F_{2-10}$ was twice as high during the
\textit{Suzaku} observation, indicating a large degree of spectral variability,
and requiring a larger relative normalization for the Compton reflection component in the
\textit{Suzaku} observation.

Markowitz \etal (2003a) found lower a value
for $\Gamma$ ($\sim$1.6), but this discrepancy is likely due to the fact that
those authors did not include the partial-covering, ``heavy''
warm absorber (described by Turner et al.\ 2005)
in their modeling, affecting determination of the broadband continuum and thus
(the mildly-degenerate) photon index and Compton reflection strength.
In fact, removing the partial-covering absorber from our fits, we obtain a fit with roughly similar values of
$\Gamma$ and $R$ ($\sim$0.3) 
to Markowitz, Edelson, \& Vaughan (2003), but the fit quality is poor ($\chi^2_r \sim 3.1$).
Similarly, Bianchi et al.\ (2004), modeling simultaneous \textit{XMM-Newton} and \textit{BeppoSAX}
observations conducted in 2001, include a full-covering warm absorber
with $N_\textrm{H} \sim 1.5 \times 10^{22}$ cm$^{-2}$ in their model, but they do not model the 
partial-covering ``heavy'' component, yielding what may be a spuriously low value of 
$\Gamma$ ($\sim$1.5). 

Finally, we note that
including a high-energy rollover offers insignificant improvement to our fit, with \eroll $ > 220$ keV.

%===============================================================

\subsection{NGC~3227}

Data taken from 2000 November to 2001 May (modified Julian day (MJD) 51,850--52,050),
affected by the passage of the compact cloud cross the line of
sight (Lamer et al.\ 2003), were ignored during spectral fitting. 

Two zones of warm absorption, with $N_\textrm{H} = 1 \times 10^{21}$ and $2 \times 10^{21}$ cm$^{-2}$ 
and log($\xi$) = 1.2 and 2.9,
respectively, were included in the spectral fits following
the \textit{XMM-Newton}-RGS analysis of Markowitz et al.\ (2009).
We also included a full-covering zone of cold absorption
with $N_\textrm{H}$ as a free parameter, obtaining 
$1.5 \pm 0.9 \times 10^{21}$ cm$^{-2}$ in the best-fit model.

Markowitz et al.\ (2009) also presented joint spectral fits to
PCA + HEXTE data up to 100 keV, plus joint fits with four-channel \textit{Swift}-BAT data. 
They reported a rollover at 90 $\pm$ 20 keV, driven primarily by the
inclusion of the BAT spectrum; we find no significant
evidence for a rollover in the current spectrum (\eroll > 210 keV).

The relatively low value of $\Gamma$ we obtain, 1.69 $\pm$ 0.04,
is consistent with previous results from \textit{XMM-Newton} and \textit{RXTE}
spectral fits (Cappi et al.\ 2006).

%===============================================================
\subsection{NGC~4593}

We included one warm absorber zone, with
$N_\textrm{H} = 3 \times 10^{21}$ cm$^{-2}$ and 
log($\xi$)=2.4 (Markowitz \& Reeves 2009;
McKernan et al.\ 2003).

%% The Fe line is narrow though somewhat broadened
%% (FWHM width  ~ 10000 km/s; Brenneman et al 2007), with no additional broader component.

Using \textit{Suzaku} data, Markowitz \& Reeves (2009) modeled a Compton reflection hump
with a strength $R = 1.08 \pm 0.20$, however that particular observation occurred in 2007 December during a low-flux
state: $F_{2-10}$ was a factor of $\sim$ 3.5 lower than the historical 
long-term average (Figure \ref{lc2}).
The fact that we measure an $R$ value of $0.34 \pm 0.09$ for the long-term spectrum is consistent with the notion of
a Compton reflection component that is constant in absolute
normalization (i.e., it did not respond to the drop in continuum flux
during the \suzaku observation).
This would require that the bulk of the reflecting material is located
a minimum of $\sim$ a lt-yr from the central X-ray source, as 
the 2--10 keV flux light curve (Figure \ref{lc2}) indicates 
a decline from average 2--10 keV flux levels before MJD $\sim$ 54100
to the lower flux level during the \textit{Suzaku} observation on MJD 
MJD 54449.

Our value of $R$ is only barely inconsistent with that obtained
by Guainazzi et al.\ (1999) from the \textit{BeppoSAX} observation in 1997 
($R \sim 1.0$ with typical uncertainties of $\sim 0.5$), which caught the source at 
a typical flux level ($F_{2-10} \sim  4 \times 10^{-11}$ erg cm$^{-2}$ s$^{-1}$).
We find a photon index of $1.85 \pm 0.03$, consistent with
that reported by Brenneman et al.\ (2007) for an \textit{XMM-Newton}
long-look and Guainazzi et al.\ (1999).

There is no evidence for a rollover in this source
based on our \textit{RXTE} spectrum (\eroll > 335 keV)
or based on the \textit{BeppoSAX} observation
(\eroll > 222 keV and $\gtrsim150$ keV from
Perola et al.\ 2002 and Guainazzi et al.\ 1999, respectively).

%===============================================================

\subsection{NGC~7469}  

The warm absorbers (measured by e.g., Blustin et al.\ 2007) do not significantly impact the
modeling of the PCA spectrum and were ignored in our modeling.
Previous \textit{RXTE} spectral results were published by Nandra et al.\ (2000) and Markowitz \etal (2003a)
for the 1996 intensive monitoring campaign; our analysis also includes data from the longer-term monitoring 
campaign, from 2003 April to 2009 July.

Our best-fit value of $R$, $0.7\pm0.2$ is broadly consistent with values reported by 
Markowitz \etal (2003a) and Perola et al.\ (2002, for the
1999 \textit{BeppoSAX} observation), and slightly lower than values reported 
by De Rosa et al.\ (2002) for the \textit{BeppoSAX} observation 
for a range of spectral models (best-fit values in the range 0.9--1.8
with uncertainties as high as 0.6).

Using \textit{BeppoSAX} data, De Rosa et al.\ (2002) and Perola et al.\ (2002)
each reported a high-energy rollover;  depending on the spectral model tested,
best-fit values range from $\sim$ 140 to $\sim$260 keV with very large 
uncertainties, but we do not find evidence for a rollover in our data
(\eroll > 220 keV).

%======================================================

\subsection{3C~111}

\textit{RXTE} observations have been concentrated into five short campaigns over 1997 to 2003,
plus continuous monitoring since 2004 March. The long-term PCA light curve 
shows long-term variations by a factor of $\sim 3-4$ (Figure \ref{lc2}).

\textit{RXTE} PCA+HEXTE spectra were previously published by Eracleous, Sambruna \& Mushotzky 
(2000); the good time exposures for PCA and HEXTE were 33 and 13 ks, respectively.
Lewis et al.\ (2005) published an \textit{RXTE} observation simultaneous with an \textit{XMM-Newton} long-look
in 2001 March; good exposure times were 57 ks for the PCA and 18 ks for each HEXTE cluster.

The total X-ray absorbing column along the line of sight measured from
various X-ray missions from \textit{HEAO-1} to \textit{ASCA}
all show evidence for absorption by a column of cold gas
in excess of the Galactic column inferred from 21-cm measurements
(Reynolds et al.\ 1998). This is likely due to a molecular cloud lying
along the line of sight to 3C~111 (Bania et al.\ 1991); 
the cloud's molecular gas (\ion{H}{2} and metals/dust) will contribute to 
the total X-ray absorption but not to the measured 21 cm radio emission. 
We herein adopt a molecular hydrogen column density of 
$9 \times 10^{21}$ cm$^{-2}$ (Bania et al.\ 1991, estimated 
from CO emission measurements) for a total Galactic
nonionized hydrogen column of $1.2 \times 10^{22}$ cm$^{-2}$ 
in all fits.
Warm absorption does not significantly affect the PCA spectrum.

Previous results on BLRGs, including Wozniak et al.\ (1998) and Eracleous et al.\ 
(2000), indicated that BLRGs tend to have Compton reflection humps and Fe K$\alpha$ lines 
which are generally weak compared to those of normal Seyferts. We find no evidence for a Compton 
reflection hump in our spectrum, generally consistent with previous results from \textit{RXTE}, 
\textit{BeppoSAX}, and \textit{Ginga} observations, where
upper limits to $R$ were reported or best-fit values
of $R$ were $\lesssim$0.5 (above-mentioned references; Grandi et al.\ 2006).

We also find no evidence for a rollover, with \eroll $ > 230$ keV;
a rollover was claimed by Grandi et al.\ (2006) for a \textit{BeppoSAX} observation in 1998
($146^{+224}_{-68}$ keV).

%======================================================

\subsection{3C~120}

%%% The Fe line is a blend of both narrow and broad components (Kataoka et al 2007, using Suzaku).

No warm absorbers significantly affect the spectrum above 2 keV. 
Joint fits to PCA + HEXTE spectra have been published previously by
Gliozzi et al.\ (2003); their spectral data, obtained in 1997, contained 100 (40) ks of good exposure time
for the PCA (each HEXTE cluster) and only included data up to 50 keV in HEXTE.
Eracleous et al.\ (2000) published joint fits to the PCA + HEXTE spectrum  (up to 100 keV) 
obtained from a 2-day \textit{RXTE} observation in 1998.
Our best-fit results for the Compton reflection strength and our lack of evidence for a rollover
for our spectrum is consistent with results from those papers.
Our results are also roughly consistent with 
best-fit values for the Compton reflection strength 
obtained by Kataoka et al.\ (2007; \textit{Suzaku}) and
Zdziarski \& Grandi (2001, \textit{BeppoSAX}, and also
\textit{BeppoSAX} + \textit{CGRO}-OSSE joint fitting),
$\sim$0.4--0.6 with uncertainties typically $\sim$0.2--0.4.

Modeling a rollover causes $\chi^2$ to drop by only
3.2 for 1 less dof between the reflection and rollover models, with large changes to both $\Gamma$ (flattens by 0.2 to $\sim$1.6)
and $R$ (goes to 0), and also yields a large uncertainty in \eroll, $190^{+310}_{-70}$ keV. 
While this value of \eroll is consistent with
those reported by Zdziarski \& Grandi (2001), Grandi et al.\ (2006), and Wozniak et al.\ (1998, from
joint \textit{ASCA}+OSSE fits), the improvement in fit is not significant and the best-fit values for
$\Gamma$ and $R$ are inconsistent with previous results (e.g., 
Grandi et al.\ 2006; Gliozzi et al.\ 2003; and Zdziarski \& Grandi 2001 each
find best-fit values of $\Gamma$ closer to 1.8), and we thus
conclude that the reflection model is the best description of the data 
with a conservative estimate for the lower limit on \eroll of 120 keV. 

%======================================================

\subsection{NGC~5506}

\textit{RXTE}-PCA spectral data obtained from observations occurring from 1996 through early
1999 were published by Lamer et al.\ (2000).

The evidence for a rollover in our spectrum is not strong: $\chi^2$ falls by only 4 for 1 less dof, and
data/model residuals do not improve noticeably.
Moreover, $\Gamma$ flattens by 0.1, to 1.80, and the column density of the Compton-thin absorber
$N_\textrm{H}$ falls to $<0.9 \times 10^{22}$ cm$^{-2}$.
However, most previous studies by various X-ray missions (Bianchi et al 2003, 2004; Risaliti 2002)
indicate that $\Gamma$ is closer to 1.9--2.0 (more consistent with this source being an
``obscured Narrow Line Seyfert 1'', e.g., Dewangan \& Griffiths 2005),
and $N_\textrm{H} \sim (2-4) \times 10^{22}$ cm$^{-2}$, similar to our results when a rollover is
not included in the model. We thus conclude that the reflection model is the best 
description of the spectrum.

Most values of $R$ reported in the literature (above references, also Perola et al.\ 2002)
are similar to ours, with best-fit values of $R$ in the range 1.0--1.5.
We do not confirm the rollovers of 110 -- 140 keV ($\pm \sim 30$ keV depending on the model)
claimed by Bianchi et al.\ (2004) for a simultaneous \textit{BeppoSAX}/\textit{XMM-Newton} 
observation in 2001 February; we obtain a lower limit of 209 keV.

%======================================================

\subsection{MCG--5-23-16}

\textit{RXTE} observations have been concentrated in several campaigns
instead of sustained monitoring, but the average 2--10 keV flux within each campaign never deviates
by more than $\sim 25 \%$ from the mean 2--10 keV flux, $9.6 \times 10^{-11}$ erg cm$^{-2}$ s$^{-1}$.
Our best-fit model includes a single, full-covering cold absorber with $N_\textrm{H} = 
3.5 \pm 0.5 \times 10^{22}$ cm$^{-2}$; this is only about a factor of two higher than the column modeled
by Reeves et al.\ (2007) using \textit{Suzaku} and Braito et al.\ (2007) using \textit{XMM-Newton} and 
\textit{Chandra}-HETGS.

Weaver et al.\ (1998) published the \textit{RXTE} PCA spectrum from the 1996 November campaign 
(80 ks of good exposure time). For an inclination angle of 50$\degr$, best-fit values of $R$ are 
$0.36^{+0.24}_{-0.22}$ for \eroll fixed at 500 keV (consistent with our results), and $0.45 \pm 0.23$ for 
\eroll fixed at 200 keV. However,
Reeves et al.\ (2007) found a total Compton reflection strength of $1.1 \pm 0.2$ 
(assuming an inclination angle of 45$\degr$) using \textit{Suzaku}, a factor of $\sim$2.5 
higher than our best-fit value.
The 2--10 and 15--100 keV fluxes reported by Reeves et al.\ (2007) are virtually identical to
what we measure, $F_{2-10} = 9.6 \times 10^{-11}$ erg cm$^{-2}$ s$^{-1}$
and $F_{15-100} = 1.9 \times 10^{-11}$ erg cm$^{-2}$ s$^{-1}$. The discrepancy may be at least partially
attributed to differences in spectral modeling and degeneracies between parameters,
particularly $N_\textrm{H}$, $\Gamma$, and $R$; specifically, we find a value for 
$\Gamma$ 0.1 lower than Reeves et al.\ (2007; forcing $\Gamma$ to be 1.95 while
ignoring data $<$5 keV which influence the value of $N_{\rm H}$, $R$ rises to $\sim$ 0.75).
The value of $R$ found by Perola et al.\ (2002) is in agreement with Reeves et al.\ (2007),
given the inclination angles assumed by both papers.

Our lower limit to \eroll, 170 keV, is identical to that found by
Reeves et al.\ (2007). It is consistent with the rollover claimed by
Perola et al.\ (2002) using \textit{BeppoSAX}, $147^{+70}_{-40}$ keV,
and consistent with the results of Zdziarski \etal (1996),
who noted that spectral fits to non-simultaneous
\textit{Ginga} + OSSE data suggested a rollover near 200 keV.

%======================================================
\subsection{NGC~4507}

\textit{RXTE} observations are clustered into two campaigns, one in 1996 which accounts for
94\% of the total good exposure time, and one in 2003. 
There were three observations with \textit{BeppoSAX}, in 1997, 1998 and 1999, analyzed by Risaliti (2002).
As reported in their best-fit model with reflection (their "Model C"), from 1997 to 1999, the 
2--10 keV flux dropped by a factor of 2.1 from $1.84 \times 10^{-11}$ to $0.87 \times 
10^{-11}$ erg cm$^{-2}$ s$^{-1}$, while $R$ increased from $0.7\pm0.2$ to $2.0\pm0.5$, consistent with
the Compton hump absolute normalization staying constant despite the drop in power-law normalization.
However, from our modeling, our best-fit value of $R$ is only $0.4 \pm 0.1$ despite our best-fit 2--10 keV flux,
$1.43 \times 10^{-11}$ erg cm$^{-2}$ s$^{-1}$, lying
in that same range. Best-fit values of $\Gamma$ and $N_{\textrm{H}}$ do not differ by more than 
$\sim$ 0.2 and a factor of 1.5, respectively, between the \textit{RXTE} and \textit{BeppoSAX} models,
and so the cause of the discrepancy is not obvious. One possibility, however, is that 
the high reflection component normalization value could indicate a response to a higher past illuminating flux. 
We do not find significant evidence for a high-energy
rollover, with \eroll > 170 keV (consistent with Risaliti 2002).

%======================================================

\subsection{Cen~A}

A comprehensive spectral analysis of all \textit{RXTE} campaigns through 2009
is given by Rothschild et al.\ (in prep.).
Our best-fit model parameters are in agreement with results in that
paper (see also Rothschild et al.\ 2006 for results on \textit{RXTE} and \textit{INTEGRAL}
observations up through 2003).   
Markowitz et al.\ (2007) present analysis of the
\textit{Suzaku} observation of Cen A in 2005, covering 0.3 to 220 keV.
They present evidence for a secondary continuum emission component above 4 keV,
absorbed by a cold column with $N_{\textrm{H}} = 7.0 \times 10^{23}$ cm$^{-2}$,
though this component is not detected in the \textit{RXTE} analyses.
Neither the current nor any of the above analyses find evidence for a Compton reflection
component; previous upper limits to $R$ have usually been close to 0.05
(see also Benlloch et al.\ 2001).
A high-energy rollover is also not required by the \textit{RXTE} data; we find \eroll $ > 490$ keV,
similar to the lower limit of 400 keV found by Markowitz et al.\ (2007).
Using \textit{CGRO} spectra, Kinzer et al.\ (1995)  modeled a rollover
at 300--$\sim$700 keV (depending on flux level), while Steinle et al.\ (1998)
claimed a break near 150 keV.

%======================================================

\subsection{NGC~4945}

As pointed out by Done et al.\ (1996), \textit{CGRO}-OSSE demonstrated that
this heavily-obscured source is one of the brightest AGN in the
$>$100 keV sky. PCA+HEXTE spectral fits were published previously by
Madejski et al.\ (2000), though they only used data from the fall 1997 campaign.

We excluded data from ObsID 92118-01-40-00, observed 
on 2006 May 20, due to the presence of spectrally-steep emission below 12 keV which was
a factor of $\sim$3 times the emission for each of all other observations of NGC~4945.
The nature and origin of this emission is discussed in Appendix C.

Our spectral fit results are broadly consistent qualitatively
with the results of Itoh et al.\ (2008) for the \textit{Suzaku} observation.
In our model, the so-called ``direct'' or "transmitted" component is 
the Compton-thick-absorbed power-law which dominates above $\sim$12 keV.
A weak, cold Compton reflection component dominates from $\sim$11 keV down to 
$\sim$6 keV. Emission below 5 keV may be thermal emission from a plasma 
and/or scattered emission (e.g., Itoh et al.\ 2008); we model this emission
as a simple soft power law, but the power-law parameters 
are very poorly constrained because we have no information
below 3 keV and, given the PCA energy resolution, there is likely 
blending with the Fe line parameters.

An additional contributor to the $<$5 keV emission modeled here as a simple power-law is the ultra-luminous transient 
X-ray (ULX) source Suzaku J1305--4931, located a few arcmin to the SW of the nucleus of NGC~4945, and discovered 
in a \textit{Suzaku} observation of NGC~4945 obtained in 2006 January (Isobe et al.\ 2008). The ULX was not 
detected in a \textit{Suzaku} observation of NGC~4945 in 2005 August, nor has it been reported elsewhere, and 
so the details of the ULX's flux activity as a function of time during early 2006 are not known. However, that 
source's 0.5--10 keV flux was only $\sim 2 \times 10^{-12}$ erg cm$^{-2}$ s$^{-1}$, and the impact on the summed 
\textit{RXTE} spectrum is small, affecting only the power-law component below $\sim$10 keV. 

When we include a high-energy rollover in the model,
(the data/model residuals in Figure \ref{N4945rat} are particularly compelling in this case)
the strength of the Compton reflection hump dwindles from $R = 0.07$ to 
an upper limit of $R < 0.03$
as the curving introduced into the continuum by the high-energy rollover
accounts for a portion of the already-weak (relative to the
absorbed power-law) Compton hump. As a result of degeneracy between 
\eroll and $\Gamma$ and the fact that $\Gamma$ is already
poorly constrained due to the strong absorption, the best-fit value of
$\Gamma$ becomes rather low, $1.0\pm0.1$, when a rollover is modeled.

Itoh et al.\ (2008) and Guainazzi et al.\ (2000) each reported a high-energy rollover at 
190\err{150}{50} and $190^{+100}_{-90}$ keV, respectively; the latter is only barely inconsistent 
at the 90\% confidence level with the rollover we model.
Done et al.\ (1996) modeled the 0.6--400 keV 
spectrum constructed from non-simultaneous OSSE, \textit{Ginga} and 
\textit{ASCA} data; no strong evidence for a rollover was found.

%======================================================

\subsection{Circinus}

The good exposure time for this object is dominated by the sampling obtained in 1998.

Matt et al.\ (1999), who analyzed the spectrum obtained from the \textit{BeppoSAX} 
observation in 1998, were the first to extend the spectrum of this source out
to $>$10 keV and were the first to establish the presence of the absorbed power-law component.
With the exception of modeling soft X-ray emission lines, 
our best-fit model is similar to that of Matt et al.\ (1999):
a Compton-thick-absorbed power-law, a Compton reflection
component, a softer power-law component, and an Fe emission line.

In the case of a model with no rollover, the photon index for the soft
power law is identical to that of the base power-law,
consistent with either scattered or leaked emission.
For the model that includes the rollover component both photon indices 
are poorly constrained and subject to degeneracies.
In the context of scattered emission, we find that the
optical depth of the scattering medium is 4$\%$ or 47$\%$
when Compton reflection is included in the model  
and a high-energy rollover is excluded or included, respectively
(alternately, partial covering fractions of 96$\%$ or 53$\%$ if the
soft emission is ``leaked'').
In our best-fit model, we find roughly similar column densities
for the Compton-thick absorber and consistent $R$ values as Matt et al.\ (1999). 

We confirm the presence of the Compton-thick absorbed power-law component, e.g., 
the ``directly transmitted'' nuclear emission. A model without that component, and 
wherein the peak of the spectrum at $\sim$20 keV is explained by a dominating Compton 
reflection component, yields a poor fit ($\chi^2$/dof = 1473/63). 
Allowing for the Compton reflection component to undergo Compton-thick absorption
yields an unsatisfactory fit with $\chi^2$/dof=216/64 and with poor residuals,
and so this type of model is not discussed further.

Of all the objects in our sample, the evidence for a high energy
rollover is the strongest in this object, with $\chi^2$ dropping by $>$50 for 1
less dof.  However, given that the hard X-ray power-law undergoes
Compton-thick absorption and suffers a rollover below $\sim$ 15 keV, there 
is degeneracy between the best-fit values $\Gamma$ and \eroll, as 
shown in Figure \ref{CIRCINUSspec}. In this case we did not allow \eroll to drop below a hard limit of
40 keV in order to minimize the impact on modeling of the
Compton reflection component peaking at 20--30 keV.

Sambruna et al.\ (2001) published the 6--30 keV PCA spectrum
obtained during the observations in 2000, which were
simultaneous to the \textit{Chandra}-HETGS observation, 
the focus of their paper.
They also modeled a Compton reflection component plus absorbed power-law,
in addition to neutral and ionized Fe K lines, though
constraints on $\Gamma$ were very poor and the strength of the Compton
reflection component was not reported.

%======================================================

\subsection{NGC~7582}

In the course of our modeling, we found two models which yielded approximately equally good statistical descriptions of the data.
In the first model, "\textsc{REFL}", a single power-law component with $\Gamma = 1.79 \pm 0.10$
is absorbed by a full-covering column of Compton-thin gas with \NH $= 1.4 \pm 3\times 10^{23}$ cm$^{-2}$, and is
accompanied by a strong Compton reflection component, $R = 3.3^{+1.6}_{-0.9}$; $\chi^2$/dof = 36.3/33. 
Using the \pexrav model, values of $R$ greater than two do not necessarily have a direct geometrical meaning
but may indicate a geometry more complex than a simple reflecting slab or may indicate anisotropic beaming of the primary continuum.

The second model, ``\textsc{PC}'', includes a partial-covering Compton-thick absorber ($f_\textrm{cov} = 60\% $; 
\NH $= 3.6 \pm0.6 \times10^{24}$ cm$^{-2}$) in addition to the full-covering Compton-thin absorber 
(\NH $= 2.4 \pm0.4 \times 10^{23}$ cm$^{-2}$); $\chi^2$/dof was 40.0/32.
Adding a Compton reflection hump to the \textsc{PC} model did not yield any improvement in fit, as 
there is complete degeneracy between the strength of the Compton reflection hump and the Compton-thick absorbed power-law, 
both of which peak near 20--30 keV; this held whether or not the reflection component was modeled to be obscured 
by the same Compton-thick gas obscuring the power-law component
($\Gamma$ = 1.84$\pm0.12$; power-law normalization $A$ = 1.8\err{1.0}{0.6}$ \times 10^{-2}$ ph\,keV$^{-1}$\,cm$^{-2}$\,s$^{-1}$ at 1 keV).

In their modeling of a spectrum of NGC~7582 obtained with \textit{BeppoSAX} in 1998, Turner et al.\ (2000) 
also explored models of these two types. They favored a \textsc{PC} type model over a \textsc{REFL} type model 
on physical grounds: in the latter, 
they found the Compton hump strength $R$ to be near 6 (assuming an inclination cos$i$ = 0.45), which was incompatible with the 
small value of the EW for the Fe K$\alpha$ line (assuming that a Compton hump with a value of $R$ near 1 will be associated with
an Fe line with an EW near 150 keV, George \& Fabian 1991). In our \textsc{REFL} model, we find an Fe line EW of 331$\pm$94 eV, 
which is not incompatible with the observed Compton hump strength, and so we cannot reject this model on this basis alone (see Figure \ref{WvsR}). 
However, Bianchi et al.\ (2007) analyzed \textit{Chandra} and \textit{Hubble Space Telescope} images and 
concluded that there exists a Compton-thick torus plus large-scale Compton-thin material,
in agreement with the complex absorption structure first quantified by Turner et al.\ (2000).

A model more complex than either \textsc{PC} or \textsc{REFL} may in fact be preferred.
In the context of the \textsc{PC} model, we might reasonably expect that if there is Compton-thick absorption along the line of sight, 
there may likely be reflected emission originating in circumnuclear Compton-thick gas.
Bianchi et al.\ (2009), in their modeling of a series of four \textit{Suzaku} spectra 
of NGC~7582 obtained in 2007, include \textit{both} a partial-covering Compton-thick absorber and a Compton reflection
component. In the \textsc{PC} model as applied to the current \textit{RXTE} spectrum, we are simply not sensitive to the presence of a 
Compton reflection component; we cannot rule out the presence of a Compton reflection component with a strength 
similar to that modeled by Bianchi et al.\ (2009).

For the \textsc{REFL} model, we found no significant improvement to the fit when a high-energy rollover was modeled, obtaining
a lower limit of 200 keV.  For the \textsc{PC} model, degeneracy made the addition of a rollover unconstrainable.

%============================================

\subsection{3C~273}

3C~273 has been studied with every major X-ray observatory to date.
The X-ray spectrum typically resembles those of
Seyfert 1s with an Fe line and Compton reflection hump
likely visible at during low flux states and likely diluted by an additional X-ray continuum 
contribution from the jet (e.g., Grandi et al.\ 1997, Kataoka et al.\ 2002).

In our fits, adding a Compton reflection hump to a model 
consisting only of a power-law plus a Gaussian component
significantly improves the fit, as $\chi^2$ drops by 32 for 1 less dof. However,
the strength of the Compton hump is quite low: $R = 0.07 \pm 0.03$.

The flat photon index for our power-law component is in general
agreement with previous results, such as Kataoka et al.\ (2002), who presented
\textit{RXTE}-PCA only spectral data from monitoring
in 1996--2000, and a set of \textit{BeppoSAX} observations
analyzed by Grandi et al.\ (2006).

If a jet component does make a significant contribution
to the X-ray spectrum then a power-law rollover associated with
thermal Comptonization at 100--300 keV may be masked by the non-thermal
jet emission. We find no significant evidence for a rollover, with 
\eroll > 410 keV.

%============================================

\subsection{3C~454.3}

The \textit{RXTE} monitoring data used herein were obtained as part of a multiwavelength campaign
that occurred when the source went into a bright flaring state in 2005.
The PCA light curve (though not spectrum) is presented as a part of a multi-wavelength analysis 
in Jorstad \etal (2010); detailed discussion on the origin of the high-energy continuum emission
can be found in, e.g., that paper, Ghisellini et al.\ (2007), and Pian et al.\ (2006).
For consistency, we did not include archival \textit{RXTE} monitoring
data that did not use the offset pointing position.

%============================================

\subsection{Mkn~421}

Mkn~421 is one of the most extensively studied BL Lac objects across the electromagnetic spectrum. The 
\textit{RXTE} data used here were usually obtained as part of multi-wavelength campaigns
when the source was in outburst. Despite being a steep spectrum source, Mkn 421 is one of the brightest
AGN in X-rays when it flares, yielding high quality spectra with HEXTE. 
PCA and HEXTE spectral and light curve data during flares, and modeling of the broadband SED, have been published 
in, e.g., Giebels et al.\ (2007), Rebillot et al.\ (2006), and B{\l}a$\dot{\rm z}$ejowski et al.\ (2005).
The source's extreme levels of X-ray flux and spectral variability over a wide range of time scales have been 
well-documented previously (e.g., Cui 2004), and the average 2--10 keV flux considered here may be a factor of
3--4 less than the flux during the brightest flaring states.
We found that a simple power-law fit to the long-term
summed PCA + HEXTE spectrum yielded a poor fit, with $\chi^2$/dof  = 298/81
and poor data--model residuals. Using a broken power-law 
to allow for a model representing a slowly bending
continuum allowed for a good fit. Our results are similar to those of 
Giebels et al.\ (2007) and Foschini et al.\ (2006) for \textit{XMM-Newton} data.

%============================================

\subsection{1ES~1959+650}

The \textit{RXTE} data used here were gathered as part of 
multi-wavelength (typically including X-ray, TeV, optical, and radio)  
campaigns and have been published previously by Giebels et al.\ (2002), Krawczynski et al.\ (2004),
and Gutierrez et al.\ (2006) for the 2000, 2002, and 2003 campaigns, respectively.  

Gutierrez et al.\ (2006) demonstrated that the X-ray spectrum flattens as flux increases, with
$\Gamma$ $\sim$ 2.9 (1.6) when the X-ray flux is a factor of $\sim$3 higher (lower) than
the long-term average flux, $1.6 \times 10^{-10}$ erg cm$^{-2}$ s$^{-1}$.
Furthermore, the X-ray band lies near the peak of the synchrotron component in the broadband SED. 
Consequently, our spectral fitting found that a broken power-law model
(with $\chi^2$/dof = 47.4/62) was a better fit than an unbroken
power-law ($\chi^2$/dof = 62.6/64, with poor residuals below $\sim$8 keV).
Previous to the \textit{RXTE} programs, \textit{BeppoSAX} observed this source in 1997; the reader is referred to 
Beckmann et al.\ (2002) and Donato et al.\ (2005) for those spectral fit results.

\section{B.  Long-term Variability in the HEXTE Background}
%\begin{center}
%\textsc{Appendix A:} \\
%\textit{Long-term Variability in the HEXTE Background}\\    %%% exact title still TBD
%\end{center}

\begin{figure}
  \plotone{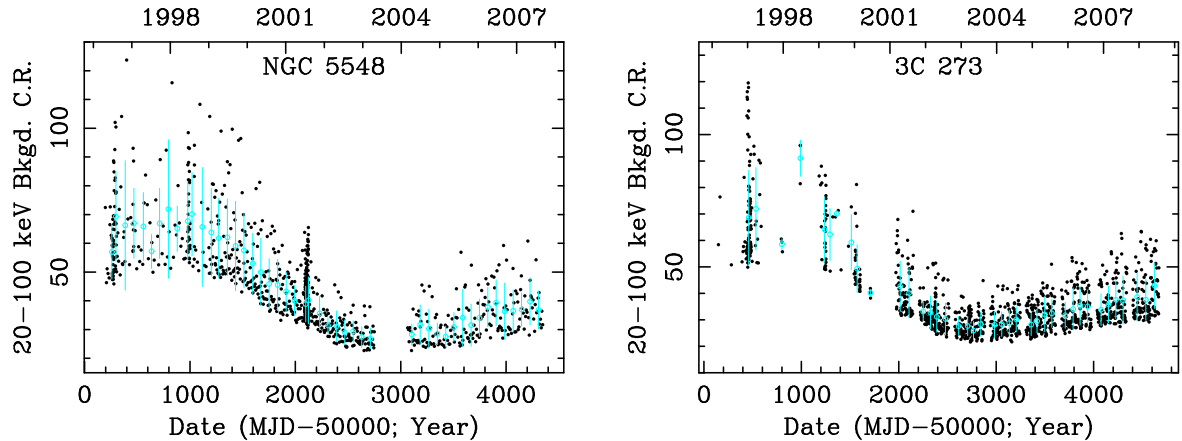}
  \caption{Light curves showing 20--100 keV HEXTE cluster B background count rates (after application of screening criteria)
for two selected targets, NGC~5548 (left) and 3C~273 (right), illustrating
the long-term decreasing trend in the average background flux 
due to the decrease in altitude, and thus particle flux intercepted by, the \rxte\ spacecraft.
Additional long-term trends, such as the increase after $\sim$2004, are likely the result of solar activity.
Blue points denote the averages over $\sim$ 80-day bins, with error bars signifying the standard deviation
of count rates within each bin.}
  \label{hexteback}
\end{figure}

In this Appendix, we summarize the long-term variability of the HEXTE
background; the reader is referred to Rothschild et al.\ (1998) and Gruber et al.\ (1989) for 
further details of the internal backgrounds of the HEXTE detectors and 
the NaI/CsI scintillators comprising the
Hard X-Ray and Low-Energy Gamma Ray Experiment (A4) aboard \textit{HEAO-1}, respectively.

The HEXTE background count rate is dominated by internal activation features.
The total count rate thus depends strongly on the particle flux intercepted by the spacecraft, 
as measured by two independent particle monitors, one for each HEXTE cluster. 
The background flux is dominated by emission lines resulting from 
short-term activation by geomagnetically trapped 
protons encountered in the SAA as well as by cosmic rays, 
and so the instantaneous background rate depends strongly
on the location of the spacecraft within 
its orbit, i.e., it is highest during and immediately after SAA passage and begins to
exponentially decay once the spacecraft exits the SAA. 
Satellite orientation with respect to the geomagnetic field 
and instantaneous geomagnetic rigidity rollover are also key factors affecting the total background
rate. Even after screening out data gathered within tens of minutes of SAA passage, 
the background count rate is still quite variable on timescales of hours to a day
and from one $\sim$1 ks duration observation to the next, depending on the closeness 
of a particular observation to the previous SAA passage.

%%%% http://mamacass.ucsd.edu/hexte/hexte_faq.html#BackgroundSpectrum

The strongest emission feature in the background is an emission feature near
30 keV due to the K lines from the Te daughters of various I decays.
From 58 to 87 keV is a blend of emission lines whose origins include
$^{121}$I decay at 58 keV and fluorescence of Pb in the collimator by both charged particles and cosmic X-rays
(from celestial sources outside the field of view, including the CXB).
At 191 keV, there is an emission feature due to decay of $^{123}$I, and there is a  
weak broad line at 154 keV due to K-escape lines from the 191 keV complex.
There is also a flat $\beta$-decay continuum component.
In addition, there is a wide range in activation half-lives
(see Table 1 in Gruber et al.\ 1989; other, weaker emission features are discussed therein).
 
However, there is an additional long-term trend in the total background flux: 
as the altitude of the \rxte\ spacecraft
has dropped from approximately 590 km in 1996 to 480 km in 2007,
the spacecraft has been intercepting smaller regions of the SAA and experiencing a relatively
stronger geomagnetic field strength. The increased geomagnetic rigidity rollover
as a function of altitude
makes it more difficult for relatively lower energy particles to get through, so
the average particle flux  
intercepted by the spacecraft has been decreasing
as documented by F\"{u}rst et al.\ (2009). 
Consequently, the average HEXTE background flux has dropped by $\sim$50$\%$
(from $\sim 12$ to $\sim 6 \times 10^{-9}$ erg cm$^{-2}$ s$^{-1}$).
Figure A1 illustrates this long-term trend by plotting the 
HEXTE cluster B background flux in the 20--100 keV band for each observation 
(after screening criteria were applied)
for two of the more consistently-sampled targets in our sample, NGC~5548 and 3C~273.

%==================================================================%

\section{C.  A Possible Transient Source Near NGC~4945}

\begin{figure}
  \epsscale{0.5}
  \plotone{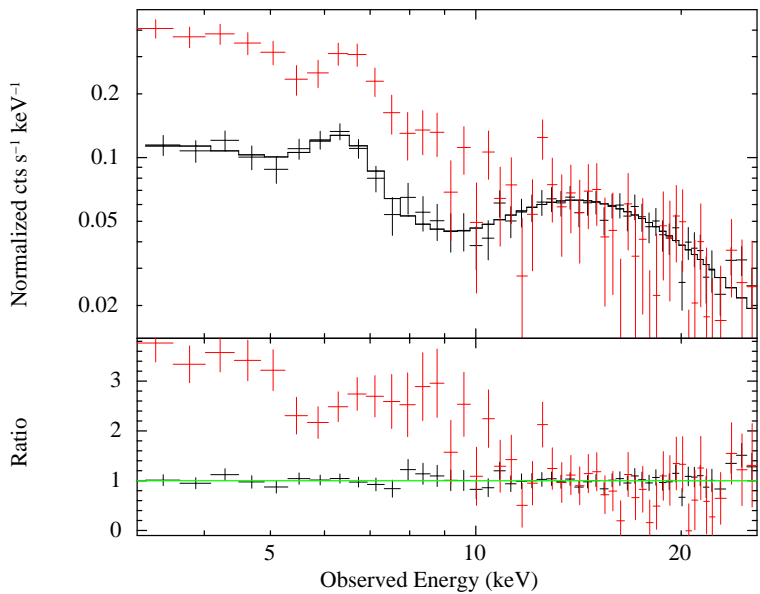}
  \caption{3--26 keV PCA data from the observation of NGC~4945. on 2006 May 20 are shown in black. 
  Red points show the summed 3--26 keV PCA spectrum from all observations in the range 2006 May 10--30, 
  excluding the observation on May 20, along with the best-fit model. In the former spectrum, below 
  $\sim$12 keV, there is steep-spectrum emission, along with a possible Fe emission line, in excess of 
  the nuclear activity and diffuse thermal emission associated with NGC~4945.}
  \label{4945flare}
\end{figure}

Data from ObsID 92118-01-40-00, observed on MJD 53875, 2006 May 20,
were not used in the summed spectrum for NGC~4945.  The 2--10 keV 
flux was found to be $2.1 \times 10^{-11}$  erg cm$^{-2}$ s$^{-1}$,
a factor of 2.5--3 greater than the flux during all other \textit{RXTE}
observations of this source, which lie in the range $(3-7) \times 10^{-12}$ erg
cm$^{-2}$ s$^{-1}$.
During this time, \textit{RXTE} was monitoring the source once every 2 days;
no other observations during this time, including those
on May 18 and 22, show evidence for an abnormally large flux. 

The spectrum for the 2006 May 20 observation (see Figure B1)
indicates the presence of an unusually steep and bright component below 12 keV, with
a line-like feature near 6 keV; the spectrum above 12 keV is roughly consistent 
with that extracted from all other observations obtained during May 10--30 (see below).
The PCA background spectra for each individual observation
during this time were examined, but nothing unusual was seen 
for any observation; furthermore, the \textit{Reuven Ramaty High Energy Solar Spectroscopic Imager}
solar flare database\footnote{http://hesperia.gsfc.nasa.gov/hessidata/dbase/hessi\_flare\_list.txt}
did not indicate any extreme solar activity within a few days of the observation.

One possibility is that the excess soft flux be due to a rapid, 
steep-spectrum flare from a transient source in the field of view. 
Non-detection of the soft emission in the observations obtained
on May 18 and 22 suggests that if this is indeed a transient flare,
its duration is limited to a maximum of 44 hr considering the 
stop/start times of the May 18 and 22 observations. 

We extracted a spectrum for NGC~4945 combining all the data 
from May 10--18 and 22--30, and applied the best-fit time-average 
model (see Table 5) to the joint PCA + HEXTE cluster B spectrum,
keeping $\Gamma_{\rm HXPL}$ and the Fe line centroid energy and
width fixed at their time-averaged values. The energy ranges 
used were 3--26 keV for the PCA and 15--100 keV for HEXTE.
A good fit was obtained ($\chi^2$/dof = 59.5/74) without the need to introduce
any additional emission components; the 
total model 2--10 keV flux was 6.8 $\times 10^{-12}$ erg cm$^{-2}$ s$^{-1}$.

We then froze all parameters and applied this model to the
PCA + HEXTE cluster B spectrum for the May 20 observation
(good exposure times after screening were 1968 s for the
PCA and 541 s for HEXTE); again, the energy ranges used were
3--26 and 15--100 keV for the PCA and HEXTE, respectively.
We added a third power-law component and a Gaussian
component to model flaring continuum and Fe K$\alpha$ line emission in excess of
that modeled for the nucleus of NGC~4945; more complex models
were not necessary to achieve a good fit.
The Fe line energy and width $\sigma$ were kept fixed at 6.4 keV and 1 eV,
respectively. We obtained a good fit ($\chi^2$/dof = 57.6/60)
with the following parameters: the photon index of the
flaring power law was 2.49\err{0.37}{0.32} and
the normalization at 1 keV was 1.17\err{0.87}{0.47} $\times 10^{-2}$  ph cm$^{-2}$
s$^{-1}$ keV$^{-1}$.
The normalization of the Fe line was $7.6\pm 5.6 \times 10^{-5}$ ph cm$^{-2}$ s$^{-1}$.
The EW of the Fe line with respect to the flaring power-law
was $600 \pm 570$ eV (taking into account uncertainties on both
the Fe line intensity and the normalization of the flaring power law). 
Removing the Fe line component from the model yielded a fit that was worse by 
$\Delta \chi^{2}$=4.5 for 1 additional dof.
Virtually identical fits were found when  
the power-law of the flaring emission was absorbed by the total
Galactic column (\NHgal$= 1.4 \times 10^{21}$ cm$^{-2}$) or unabsorbed.
The best-fit value for the 2--10 keV flux of the flaring power-law was $1.5 \times
10^{-11}$ erg cm$^{-2}$ s$^{-1}$.

We now discuss possibilities for the origin and nature of the flaring emission.
The Galactic latitude and longitude of NGC~4945 are roughly 305\degr ~and $+$13\degr,
respectively,
i.e., not close enough to either the Galactic center or anti-center
to convincingly bolster or diminish the source's likelihood of being
Galactic in origin based on sky location alone.
There are no sources in the \textit{RXTE} All-Sky Monitor source catalog in the PCA
field of view, though there 
are numerous unidentified soft X-ray sources within 1$\degr$ of the position of NGC~4945
detected with the \textit{ROSAT} Position Sensitive Proportional Counter 
(White \etal 2000) and detected in the \textit{XMM-Newton} Serendipitous
Source catalog. 

Assuming that the source is located in the host galaxy of NGC~4945, at a luminosity distance of 
11.1 Mpc, the 2--10 keV luminosity assuming isotropic emission is $2.2 \times 10^{41}$ erg s$^{-1}$, 
which is about 2/3 the 2--10 keV luminosity of the nucleus of NGC~4945 after accounting for absorption 
(see Table \ref{tabflux}).  This luminosity is slightly higher than typical X-ray luminosities inferred 
for ULX sources, $\sim 10^{37-39}$ erg s$^{-1}$ (Winter et al.\ 2006), and so we 
cannot confirm the emission as originating in a ULX on basis of luminosity alone. In a \textit{Suzaku} 
observation of NGC~4945 obtained five months earlier, in 2006 January, Isobe et al.\ (2008) reported 
the discovery of the transient ULX source Suzaku J1305--4931 located a few arcmin to the SW of the nucleus of
NGC~4945. However, that source's 0.5--10 keV flux was only $\sim 2 \times 10^{-12}$ erg cm$^{-2}$ s$^{-1}$.
As there were no \textit{Suzaku} observations of NGC~4945 in spring of 2006, it is not clear if the soft 
excess emission seen in the 2006 May 20 observation can be identified with this ULX.

Assuming instead that the source is located in our Galaxy at a distance which lies in the
range 2--15 kpc, then the 2--10 keV luminosity is in the range $(0.7 - 40) \times 10^{34}$ erg s$^{-1}$,
roughly consistent with many types of Galactic compact accreting sources. Accreting GBH systems, 
during transition states between the low/hard and high/soft spectral states, can exhibit strong flaring 
with emission that can be quantified as a power law with $\Gamma \sim 2.0--2.5$ (e.g., Pottschmidt et al.\ 2003). 
Strong, rapid (durations of hours) flaring with soft spectra in the $<$10 keV range can also be produced by 
supergiant fast X-ray transients, e.g., Sidoli et al.\ (2009, and references therein). 

\end{document}